\documentclass[12pt,a4paper]{article}
\usepackage{amssymb}
\renewcommand{\theequation}{\thesection.\arabic{equation}}
\newlength{\extraspace}
\newlength{\extraspaces}
\newcommand{\be}{\begin{equation}
\addtolength{\abovedisplayskip}{\extraspaces}
\addtolength{\belowdisplayskip}{\extraspaces}
\addtolength{\abovedisplayshortskip}{\extraspace}
\addtolength{\belowdisplayshortskip}{\extraspace}}
\newcommand{\ee}{\end{equation}}
 
\newcommand{\ba}{\begin{eqnarray}
\addtolength{\abovedisplayskip}{\extraspaces}
\addtolength{\belowdisplayskip}{\extraspaces}
\addtolength{\abovedisplayshortskip}{\extraspace}
\addtolength{\belowdisplayshortskip}{\extraspace}}
\newcommand{\ea}{\end{eqnarray}}

\newcommand{\bas}{\begin{eqnarray*}
\addtolength{\abovedisplayskip}{\extraspaces}
\addtolength{\belowdisplayskip}{\extraspaces}
\addtolength{\abovedisplayshortskip}{\extraspace}
\addtolength{\belowdisplayshortskip}{\extraspace}}
\newcommand{\eas}{\end{eqnarray*}}
 
\newcounter{subequation}[equation]
\makeatletter

\expandafter\let\expandafter
\reset@font\csname reset@font\endcsname

\def\subeqnarray{\arraycolsep1pt
    \def\@eqnnum\stepcounter##1{\stepcounter{subequation}%
        {\reset@font\rm(\theequation\alph{subequation})}}
\jot5mm     \eqnarray}

\def\subarray{\arraycolsep1pt
    \def\@eqnnum\stepcounter##1{\stepcounter{subequation}%
        {\reset@font\rm(\alph{subequation})}}
\jot5mm     \eqnarray}

\makeatother

\newcommand{\newsection}[1]{
\vspace{15mm}
\pagebreak[3]
\addtocounter{section}{1}
\setcounter{equation}{0}
\setcounter{subsection}{0}

\setcounter{footnote}{0}
\addcontentsline{toc}{section}
{\protect\numberline{\arabic{section}}{#1}}
 
\begin{flushleft}
{\large\bf \thesection. #1}
\end{flushleft}
\nopagebreak
\medskip
\nopagebreak}
 
\newcommand{\newsubsection}[1]{
\vspace{1cm}
\pagebreak[3]
\addtocounter{subsection}{1}

\addcontentsline{toc}{subsection}
{\protect\numberline{\thesection.\arabic{subsection}}{#1}}
 
\noindent{ \bf \thesection.\arabic{subsection} #1}
\nopagebreak
\vspace{2mm}
\nopagebreak}

\newcommand{\newappendix}[1]{
\vspace{15mm}
\pagebreak[3]
\addtocounter{section}{1}
\setcounter{equation}{0}
\setcounter{subsection}{0}

\addcontentsline{toc}{section}
{\protect\numberline{\thesection}{#1}}

\renewcommand{\theequation}{\Alph{section}.\arabic{equation}}
\begin{flushleft}
{\large\bf Appendix \Alph{section}: #1}
\end{flushleft}
\nopagebreak
\medskip
\nopagebreak}

\newcommand{\NP}[1]{Nucl.\ Phys.\ {\bf #1}}
\newcommand{\PL}[1]{Phys.\ Lett.\ {\bf #1}}

\newcommand{\R}{\mbox{\rm I\hspace{-.4ex}R}}

\newcommand{\bra}{\langle}
\newcommand{\ket}{\rangle}
\newcommand{\ra}{\rightarrow}

\newcommand{\rra}{\ \longrightarrow \ }

\newcommand{\is}{ &\! =\! & }
\newcommand{\nonum}{\nonumber \\[1.5mm]}
\newcommand{\sspace}{\makebox[1cm]{ }}
\newcommand{\bspace}{\makebox[2cm]{ }}
\newcommand{\nspace}{\!\!\!\!\!\!\!\!\!\!}

\newcommand{\inv}{^{-1}}

\newcommand{\lb}{\lambda}
\newcommand{\eps}{\epsilon}

\newcommand{\dd}{{\partial}}

\newcommand{\fbar}{{\overline{f}}}

\newcommand{\sbar}{{\overline{s}}}
\newcommand{\lbar}{{\overline{l}}}

\newcommand{\cD}{{\cal D}}

\newcommand{\cH}{{\cal H}}
\newcommand{\cL}{{\cal L}}

\newcommand{\cO}{{\cal O}}

\newcommand{\nl}{\mbox{[\hspace{-.35ex}[}}
\newcommand{\nr}{\mbox{]\hspace{-.32ex}]}}

\newcommand{\hbbar}{{\overline{h}}}
\newcommand{\rhobar}{{\overline{\rho}}}
\newcommand{\phibar}{{\overline{\phi}}}
\newcommand{\sigmabar}{{\overline{\sigma}}}
\newcommand{\lbbar}{{\overline{\lambda}}}
\newcommand{\gamhat}{\widehat{\gamma}}

\renewcommand{\d}{\mbox{\tiny \sl D}}
\renewcommand{\b}{\mbox{\tiny \sl B}}

\begin{document}

\begin{titlepage}
%
\renewcommand{\thefootnote}{\fnsymbol{footnote}}
\mbox{}
\vspace{1.5cm}

\begin{center}
{\Large \bf Renormalization and asymptotic safety}\\[4mm]
{\Large \bf in truncated quantum Einstein gravity}
\vspace{2.5cm}

{\large Max Niedermaier}
\\[10mm]
{\small\sl Laboratoire de Math\'{e}matiques et Physique Th\'{e}orique}\\ 
{\small\sl CNRS/UMR 6083, Universit\'{e} de Tours}\\
{\small\sl Parc de Grandmont, 37200 Tours, France}%
\footnote{permanent address; e-mail: {\tt max@phys.univ-tours.fr}} 
\\[1mm]
{\small\sl and}
\\[1mm]
{\small\sl Max-Planck-Institut f\"{u}r Physik}\\
{\small\sl 80805 Munich, Germany}
\vspace{1cm}

{\bf Abstract}
\end{center}

A perturbative quantum theory of the 2-Killing vector
reduction of general relativity is constructed. Although
non-renormalizable in the standard sense, we show that to
all orders of the loop expansion strict cut-off independence
can be achieved in a space of Lagrangians differing only by a
field dependent conformal factor. In particular the Noether
currents and the quantum constraints can be defined as finite
composite operators. The form of the field dependence in the conformal
factor changes with the renormalization scale and a closed formula is obtained
for the beta functional governing its flow. The flow possesses
a unique fixed point at which the trace anomaly is shown to vanish.
The approach to the fixed point adheres to Weinberg's ``asymptotic safety''
scenario, both in the gravitational wave/cosmological sector and
in the stationary sector.
\vfill

\setcounter{footnote}{0}
\end{titlepage}

\tableofcontents
\newpage
\newsection{Introduction and survey}

The hope that quantum Einstein gravity, although perturbatively
non-renormalizable \cite{Weinberg,tHVelt74,GorSagn86,Ven92} 
is in some sense renormalizable beyond perturbation theory underlies 
most approaches toward constructing such a theory. Explicitly this 
hope is expressed in S.~Weinberg's ``asymptotic safety'' scenario
\cite{Weinberg}. Implicitly however it also underlies the modern 
background independent approaches, such as the dynamical triangulations 
approach \cite{Ambj} or the canonical quantization program of 
A.~Ashtekar and its ramifications. In these approaches the 
renormalizability problem seems to reappear in specific, not foreseeable 
features like the tendency of random geometries to degenerate 
\cite{Ambj,spin02}, the singular support of diffeomorphism invariant 
measures \cite{FL01}, or the scarcity of semi-classical states 
\cite{Thie02}. The hope that these problems can 
eventually be overcome in one or more of these approaches has 
recently been revived by the results of M.~Reuter and O.~Lauscher 
\cite{Reuter98,Reuter02a,Reuter02b}, reporting non-trivial though not 
yet compelling evidence for the viability of the original 
asymptotic safety scenario. See also \cite{Souma99,FalkOdi98,PercPern02}. 
The goal of the present work is to gain more detailed insight into 
the renormalizability issue in a more manageable, truncated version 
of the theory, corresponding to all spacetimes with two Killing vectors.

The 2-Killing vector reduction has already been widely used as a 
laboratory for studying quantum aspects of general relativity,
see e.g.~\cite{Kuch71,AshPie96a,KorNic96,KorSam98,qernst}. In 
particular for selfinteracting cylindrical gravitational waves 
an in principle exact `bootstrap' quantization was proposed in 
\cite{qernst}. In that framework the issue of (non-)renormalizability 
is by-passed. Since it is a reduced phase space quantization based on a 
bootstrap principle, however, some important issues cannot 
be addressed. In particular one would like to understand 
the link to the perturbative divergencies and their bookkeeping, 
the status of the quantum constraints and their algebra,
and the projection onto the physical state space. In other 
words one would like to have a more-or-less conventional quantum 
field theoretical framework in which a Dirac quantization program 
could be implemented. This is what we set out to initiate here.

The reduced phase turns out to be equivalent to that of two-dimensional 
(2D) gravity non-minimally  coupled via a `radion' field $\rho$ to a 
2D matter system. The radion field is related to the determinant of the 
internal metric, the matter system is a noncompact ${\rm O}(1,2)$ 
nonlinear sigma-model. This means the matter fields are maps 
$n = (n^0,n^1,n^2)$ from a 2D manifold into the hyperboloid 
$H_+ = \{n = (n^0,n^1,n^2) 
\in  \R^{1,2}\,|\, n\cdot n = (n^0)^2 - (n^1)^2 - (n^2)^2~=~1, \,n^0 >0\}$. 
A sign $\eps = \pm 1$ will distinguish the 
two main situations, where either both Killing vectors are spacelike 
($\eps =+1$) or one is spacelike and the other timelike ($\eps =-1$). 
Accordingly the 2D metric $\gamma_{\mu\nu}$ will have signature $1 -\eps$; 
without (much) loss of generality we shall always assume it to be 
conformally flat $\gamma_{\mu\nu} \sim e^{\sigma} \eta_{\mu\nu}$,
and describe the dynamics in terms of $\sigma$. The reduced phase space 
is then characterized by the equations of motion and the symplectic 
structure following from the flat space action 
$S = \int \!d^2 x L$, with 
\begin{equation}
L(n, \rho, \sigma) = -\frac{1}{2 \lb}[ \rho 
\dd^{\mu} n \cdot \dd_{\mu} n + \dd^{\mu} \rho \dd_{\mu}
(2 \sigma + \ln \rho)]\;,\quad n\cdot n = 1\,,
\label{i1} 
\end{equation} 
where $\lb$ is Newton's constant per unit volume of the internal space.
As explained in appendix A both sectors $\eps =\pm 1$ can be described 
by the action (\ref{i1}) but the 4D spacetime interpretation of the
fields is very different.   
Further, to account for the original 2D diffeomorphism invariance 
the weak vanishing of the hamiltonian and the diffeomorphism 
constraints $\cH_0 \approx 0,\,\cH_1 \approx 0$, has to be imposed. 
The latter are given by $\cH_0 = T_{00}$ and $\cH_1 = T_{01}$,
if $T_{\mu\nu}$ denotes the classical energy momentum tensor 
derived from (\ref{i1}). The trace $T^{\mu}_{\;\mu}$ vanishes on-shell.

Suppose we now want to embark on a Dirac quantization of the 
system. That is the vector $n$ as well as $\rho, \sigma$ are promoted 
to independent quantum fields whose dynamics is governed by the 
action (\ref{i1}). The indefiniteness in its $\rho,\sigma$ part
reflects the stability problem related to (or replacing)  
the ``conformal factor'' problem of 4D quantum
Einstein gravity \cite{MazMot}. The associated negative norm states hopefully 
will decouple from the physical state space, defined schematically 
as the `kernel' of the constraints. The constraints ought 
to be defined as composite operators through a generalized 
action principle. Clearly the key issue to be 
addressed then is that of the renormalizability of an action functional 
that is motivated by (\ref{i1}) and its symmetries and extended by 
suitable sources needed to define composite operators.     
In contrast to nonlinear sigma-models without coupling to 
gravity we find that the quantum field theory based on (\ref{i1}) 
is {\it not} ultraviolet renormalizable in the standard sense.

Renormalizability in the standard quantum field theoretical sense 
typically presupposes that the bare and the renormalized 
(source extended) action have the same functional form, only the 
arguments of that functional (fields, sources, and coupling constants) 
get renormalized. Though the bare action is motivated by 
the classical one it can be very different from it. In any case the 
form of the (bare=renormalized) action functional is meant to be known 
before one initiates the renormalization. The 2-Killing vector reduction 
has been known for some time to be 1-loop renormalizable \cite{dWGNR92}. 
Here we confirm this result, but we also find 
that the system is not renormalizable in the above sense beyond 1-loop.%
\footnote{The reduced system thus accurately portrays the features of 
full quantum Einstein gravity \cite{tHVelt74,GorSagn86,Ven92}.} 
The reason is that the dependence on the dimensionless `radion' 
field $\rho$ is not constrained by any Noether symmetry. Thus it 
can -- and does -- enter the counter terms in a different way as in the 
bare action, no matter how the latter is chosen. 

The solution we propose 
is to renormalize the theory in a space of Lagrangians differing 
by an overall conformal factor that is a function of $\rho$ 
\cite{PTletter}. More precisely we show that to {\it all orders} in 
the loop expansion nonlinear field renormalizations exist such 
that for any prescribed bare $h_B(\,\cdot\,)$ there exists
a renormalized $h(\,\cdot\,)$ such that       
\begin{equation}
\frac{h_B(\rho_B)}{\rho_B} L(n_B,\rho_B,\sigma_B) = 
\frac{h(\rho)}{\rho} L(n,\rho,\sigma)\;,\quad
\mbox{but} \quad h_B(\,\cdot\,) \neq h(\,\cdot\,) \,.
\label{i2}
\end{equation}
A subscript `${}_B$' denotes the bare fields while the plain 
symbols refer to the renormalized ones. The fact that 
$h_B(\,\cdot\,)$ and $h(\,\cdot\,)$ differ marks the deviation from 
conventional renormalizability; it holds with one notable exception
described in appendix C. In order to be able to perform explicit 
computations we show this result in a specific computational scheme: 
Dimensional regularization, minimal subtraction and the
covariant background field expansion. We expect however that 
the main features remain valid in a regularization independent 
analysis along the lines of \cite{BBBC88}; c.f.~section 3.  
We made no attempt to address the infrared problem here because, 
guided by the analogy to the abelian sector \cite{ERwaves}, we 
expect it to disappear upon projection onto the physical state space.

In view of (\ref{i2}) the function $h(\,\cdot\,)$ plays 
the role of a generalized (renormalized) coupling, which 
is ``essential'' in the sense of \cite{Weinberg}. As such 
it is subject to a flow equation $\mu \frac{d}{d\mu} \hbbar 
= \beta_h(\hbbar/\lb)$, where $\mu$ is the renormalization scale 
and $\mu \,\ra \,\hbbar(\,\cdot\,,\mu)$ is the `running' coupling 
function. Remarkably $\beta_h(h)$ can be obtained in closed form 
and is given by 
\begin{equation}
\beta_{h}(h/\lb) = - \rho \dd_{\rho} 
\left[ \frac{h(\rho)}{\lb} \int_{\rho}^{\infty} 
\frac{du}{u} \frac{h(u)}{\lb} \beta_{\lb} 
\Big(\frac{\lb}{h(u)} \Big) \right]\,. 
\label{i3}
\end{equation}
Here $\beta_{\lb}(\lb)$ is the conventional (numerical) beta 
function of the ${\rm O}(1,2)$ nonlinear sigma-model without
coupling to gravity, computed in the minimal subtraction scheme. 
$\beta_{h}(h)$ can thus be regarded as a ``gravitationally
dressed'' version of $\beta_{\lb}(\lb)$, akin to the phenomenon
found in \cite{KlKoPo93}. The $\hbbar$-flow turns out to have a
unique {\it UV stable fixed-point} $h^{\rm beta}$, satisfying 
$\beta_h(h^{\rm beta})=0$. This establishes Weinberg's asymptotic 
safety scenario in the truncation considered.

We proceed by showing that the energy momentum tensor can be 
defined as a finite composite operator $\nl T_{\mu\nu}\nr$ by adding a
judiciously chosen improvement term $\Delta T_{\mu\nu} =
(\dd_{\mu} \dd_{\nu} - \eta_{\mu\nu} \dd^2)(f(\rho) + f_0 \,\sigma)$.
After renormalization the latter is characterized by a function 
$f(\,\cdot\,)$ and a constant $f_0$. Scale changes then trigger a 
non-autonomous inhomogeneous flow equation 
for a running $\fbar(\,\cdot\,,\mu)$. Clearly a necessary condition for 
the quantum constraints  $\nl T_{00}\nr$ and $\nl T_{01}\nr$  
to have a sufficiently large `kernel' on the
state space generated by $\rho,\dd_{\mu} \sigma, n^i \dd_{\mu} n^j$
is that the trace anomaly $\nl T^{\mu}_{\;\;\mu} \nr$ 
vanishes modulo contributions proportional to the equations of motion 
operator. One might expect that when $\hbbar$ becomes stationary also 
$\fbar$ becomes stationary and defines the proper improvement potential 
at the fixed point. This is indeed the case, moreover
\begin{equation}
\mu \frac{d}{d \mu} \hbbar = 0 = \mu \frac{d}{d \mu} \fbar
\quad \Longleftrightarrow \quad \nl T^{\mu}_{\; \mu} \nr =0 \,,
\label{i4}
\end{equation}
to all loop orders. That is, the trace anomaly of the improved energy 
momentum tensor vanishes precisely at the fixed point of the functional 
flow.

The article is organized as follows: In the next section we motivate 
the rescaled Lagrangians in (\ref{i1}) via Weyl transformations in 
the classical theory and discuss heuristic aspects of the quantization 
procedure in both $\eps =\pm 1$ sectors. Section 3 provides the counter 
terms through a reformulation as a Riemannian sigma-model.
The main renormalization architecture is laid out in section 4 
at a fixed renormalization scale $\mu$. Variation of $\mu$ induces  
flow equations whose fixed point structure is investigated subsequently.
Section 6 establishes the link to the vanishing of the trace anomaly
alluded to. We conclude with some directions for future research.

\newpage
\newsection{Putting conformal factors into action} 

For the 4D Einstein-Hilbert Lagrangian we adopt the normalization 
\be
L(G) = -\frac{1}{\lb^{(4)}} \sqrt{-\det G} \,R^{(4)}(G)\;,
\label{LEH}
\ee
with the conventions detailed in appendix A. Inserting the 
generic form (\ref{lineel}), (\ref{Mn}) of a metric with two
Killing vectors it turns into the Lagrangian of the truncated 
theory we wish to explore: 
\be
L_{\eps}(n, \rho, \sigma) = -\frac{1}{2 \lb}\rho \sqrt{\gamma} 
\gamma^{\mu\nu} [ \dd_{\mu} n \cdot \dd_{\nu} n + 
\eps \rho^{-2} \dd_{\mu} \rho \dd_{\nu} \rho] 
- \frac{\eps}{\lb} \rho \sqrt{\gamma} R^{(2)}(\gamma) +
\frac{2 \eps}{\lb} \sqrt{\gamma}\gamma^{\mu\nu}
\nabla_{\mu}\!\nabla_{\nu} \rho \,. 
\label{Lgamma}
\ee
Here all fields are functions of the non-Killing coordinates $(x^0,x^1)$ 
only. The coupling constant $\lb > 0$ is Newton's constant
per unit volume of the internal space. Further $\nabla_{\!\mu}$ is 
the covariant derivative with respect to $\gamma_{\mu\nu}$, which we 
assume to be diffeomorphic to $\eta_{\mu\nu} e^{\sigma}$ throughout. 
The 2D curvature is normalized such that $R^{(2)}(e^{\sigma} \eta) 
= - e^{-\sigma} \dd^2 \sigma$. The last term in (\ref{Lgamma}) 
is a total derivative which of course could be discarded in the action.
We keep it here because of its nontrivial interplay with 
conformal transformations of the 4D metric and the 2D metric. 

The induced `matter' part $-\eps \dd_{\mu} n\cdot \dd_{\nu}n = 
(\dd_{\mu} \Delta \dd_{\nu} \Delta + \eps \dd_{\mu}B \dd_{\nu}B)/\Delta^2$ 
is the pull-back of the canonical metric on the hyperboloid 
$H_{\eps}$ in (\ref{hyp}). For $\eps =-1$ the induced `matter' fails to 
have good positivity properties. However there exists a dual (classically
equivalent) Lagrangian $L^{\d}_{\eps}$ such that
\ba
&& L^{\d}_{\eps}(\eta(\phi_{\d})) = \eps L_+(\phi_{\d})\,,  
\nonum
&& \eta (\Delta, B_{\d}, \rho, \sigma) = \Big(\frac{\rho}{\Delta},
B_{\d}, \rho, \sigma + \frac{1}{2} \ln \rho - \ln \Delta\Big) \,,
\sspace \eta^2 = {\rm id}\,. 
\label{LDDi}
\ea
Here $\dd_{\mu} B_{\d} = - \eps \rho \Delta^{-2} \eps_{\mu\nu} \dd^{\nu} B$,
on-shell, and $\phi_{\d} = (\Delta, B_{\d}, \rho, \sigma)$ denotes the 
collection of fields. Thus also for the stationary sector, $\eps =-1$, 
can one use an action with a definite `matter' part, just that one of
fields is (on-shell) nonlocally related to the one appearing in the line 
element. The transition from $L_{\eps}$ to $L^{\d}_{\eps}$ is an instance of 
abelian T-duality in the sense of Buscher \cite{Buscher}; 
the involution $\eta$ is closely related to the 
Kramer-Neugebauer involution \cite{KN}; c.f.~appendix A and section 2.2.    
Apart from the heuristic discussion in section 2.2 we shall exclusively 
use the Lagrangian $L_{\eps}$, keeping in mind that $L_+$ can be used 
to describe both the gravitational wave/cosmological sector ($\eps =+1$) 
and the stationary axisymmetric sector $(\eps =-1$). The case $L_-$ 
will be carried along, mainly in order to highlight the differences.  

In summary the reduced Lagrangian is that of a 2D gravity theory 
non-minimally coupled via $\rho$ to a non-compact ${\rm O}(1,2)$ 
nonlinear sigma model. It is important that the step from (\ref{LEH}) 
to (\ref{Lgamma}) is a ``symplectic reduction'', i.e.~both the reduced 
Einstein equations and the symplectic structure induced from the 4D 
theory coincide with the ones derived from the action 
$S_{\eps} = \int \! d^2 x \,L_{\eps}$.

\newsubsection{4D and 2D Weyl transformations -- generalized Ernst systems}

The rescaled Lagrangians instrumental for the renormalization 
process via (\ref{i2}) have a natural classical counterpart: 
If $G_{MN}(x)$ denotes a generic 4D metric with two Killing vectors 
(in adapted coordinates) the relevant structure are Weyl transformations
of the form  
\be
G_{MN}(x) \rra \exp\omega(\rho(x)) \,G_{MN}(x)\,,
\label{rhoWeyl}
\ee
where $x$ are the non-Killing coordinates and $\rho(x)$ is the 
`radion' field related to the determinant of the internal metric.
Weyl transformations where the spacetime dependence of the 
conformal factor enters through a function of a scalar
field are frequently used in scalar-tensor theories of 
gravity and in the dimensional reduction of gravity theories; 
see e.g.~\cite{FGN98} for a review.

Clearly a $x$-dependent rescaling of the 4D metric,
$G_{MN}(x) \ra e^{\omega(x)} \,G_{MN}(x)$, maps one metric with
two Killing vectors onto another one. In the parameterization 
(\ref{lineel}) this amounts to the simultaneous replacement 
$\gamma_{\mu\nu} \ra e^{\omega} \gamma_{\mu\nu}$, $\rho 
\ra e^{\omega} \rho$, while $\gamma_{\mu \nu} \ra e^{\omega}
\gamma_{\mu\nu},\, \rho \ra \rho$ are 2D Weyl transformations. 
From (\ref{Lgamma}) one computes 
\ba
\mbox{2D}: \sspace L_{\eps}(n, \rho, \omega + \sigma) \is 
L_{\eps}(n,\rho,\sigma) +  
\frac{\eps}{\lb} \rho \sqrt{\gamma} \gamma^{\mu\nu} 
\nabla_{\mu}\! \nabla_{\nu} \omega\;,
\nonum
\mbox{4D}: \;\,\quad L_{\eps}(n, e^{\omega} \rho, \omega +\sigma) \is 
e^{\omega} L_{\eps}(n,\rho, \sigma) + \frac{6\eps}{\lb} 
\sqrt{\gamma} \gamma^{\mu\nu} e^{\omega/2} 
\nabla_{\mu} (\rho \nabla_{\nu} e^{\omega/2})\;. 
\label{Lconf} 
\ea 
The first equation implies in particular that the action is invariant 
under a restricted class of (`conformal') 2D Weyl transformations 
$\gamma_{\mu\nu} \ra e^{\omega} \gamma_{\mu\nu}$ with 
$\nabla^2 \omega =0$. This also explains why the 
trace $T^{\mu}_{\;\;\mu}$ vanishes only 
on-shell. The 2D Weyl transformations can also be used
to generate non-vacuum solutions of Einstein's equations with 
two Killing vectors from vacuum solutions \cite{Wain}. In the 
present context 4D Weyl transformation of the form (\ref{rhoWeyl})
turn out to be important.

Consider the following generalization of the Lagrangian 
(\ref{Lgamma}) or (\ref{action2}) 
\ba
L_{hab} \is \frac{1}{2 \lb}h(\rho) \sqrt{\gamma} 
\gamma^{\mu\nu} [- \dd_{\mu} n \cdot \dd_{\nu} n + 
\rho^{-2}a(\rho) \dd_{\mu} \rho \dd_{\nu} \rho] 
+ \frac{1}{2\lb} f(\rho) \sqrt{\gamma} R^{(2)}(\gamma)
\nonum 
&\simeq & \frac{1}{2\lb} h(\rho) [ - \dd^{\mu} n \cdot \dd_{\mu} n
+ a(\rho) \rho^{-2} \dd^{\mu} \rho \dd_{\mu} \rho + 
2 b(\rho) \rho^{-1} \dd^{\mu} \rho \dd_{\mu}\sigma]\,.
\label{Labgamma}
\ea
Here we introduced arbitrary functions $h(\rho), a(\rho)$ and $b(\rho)$ 
such that
\be
f(\rho) = 2 \int^{\rho} \frac{du}{u} h(u) b(u)\;,
\label{phi_hb}
\ee
and omitted total derivative terms. The constraints read
\be
\lb T^{hab}_{\pm\pm} = 
h(\rho)[ - \dd_{\pm} n \cdot \dd_{\pm} n + \rho^{-2} a(\rho) 
(\dd_{\pm}\rho)^2 ] + \dd_{\pm} \sigma \dd_{\pm} f
- \dd_{\pm}^2 f\,.
\label{abconstr}
\ee
while $\lb T^{hab}_{+-} = \dd_+ \dd_- f$ vanishes on account 
of the equations of motion $\dd^{\mu}\dd_{\mu} f =0$. 

The classical systems (\ref{Labgamma}) -- (\ref{abconstr}) 
are related to the original one in two ways: First for 
$a(\rho), b(\rho)$ given by specific expressions in terms 
of $h(\rho)$ the former is the 4D Weyl transform of the latter, 
with $e^{\omega} = h(\rho)/\rho$ \cite{PTletter}. Indeed, in 
conformal gauge this amounts to the substitution
\be
\mbox{4D}: \sspace \rho \rra h(\rho)\,,
\quad \sigma \rra \sigma + \ln[h(\rho)/\rho]\;.
\label{L4Dsubst}
\ee
Discarding total derivatives one finds 
\ba
&& L_{\eps}(n, h, \sigma+\ln h/\rho) \simeq L_{hab}  \quad \mbox{with} 
\nonum
&& a(\rho) = -\eps[3 (\rho \dd_{\rho} \ln h)^2 -2 
\rho \dd_{\rho} \ln h]\;,
\quad b(\rho) = -\eps \rho \dd_{\rho} \ln h\,.
\label{L4Dconf}
\ea
Specifically for $h(\rho) = \rho^p$ the functions $a(\rho), b(\rho)$ 
again reduce to (shifted) constants: $a = -\eps (3p^2 -2p),\; b= -\eps p$. 

Secondly, for generic $a(\rho)$ and $b(\rho)$ unrelated to $h(\rho)$ 
one can use field redefinitions to simplify the Lagrangian 
(\ref{Labgamma}) such that it differs from the one in (\ref{i1}) 
only by an overall factor $h(\rho)/\rho$. The explicit transformation 
is a by-product of the considerations in section 3.1; we thus postpone 
its description to Eq.~(\ref{coord_change3}). The class of Lagrangians  
$L_{hab}$ in (\ref{Labgamma}) with constant $a,b$ turns out to be the 
appropriate setting for the renormalization.
Before turning to this, however, some heuristics is called for.

\newsubsection{Quantization -- heuristics}

Morally speaking one would like to make sense out of the 
functional integral
\ba
&& (a) \;\;\int_{\rm 2-Killing} \cD G \,e^{i S[G]}\sspace \mbox{or} 
\nonum
&& (b)\;\;\int \cD n \cD \rho \cD \sigma \, 
e^{i d^2 y \int \!d^2 x \,L_{\eps}(n,\rho,\sigma)}
\quad \mbox{with}\quad \cH_0 \simeq \cH_1 \simeq 0\,. 
\label{q1}
\ea
In the first version the intended functional integral ranges over 
all 4D Lorentzian metrics with two Killing vectors of fixed 
signature and fixed unit volume $d^2 y$ of the internal space. $S[G]$ is 
the Einstein-Hilbert action. Of course (a) cannot be expected to be 
``the same'' as ``first quantizing and then truncating.'' Also, but not only, 
because the latter lacks a precise meaning so far. However in the context
of the asymptotic safety scenario the truncation should preserve the 
existence and the qualitative features of a fixed point: If the full 
theory is assumed to have an UV stable fixed point every truncation that 
preserves the presumed `ferromagnetic' or `anti-screening' 
\cite{FalkOdi98,Reuter02b} nature of the self-coupling should likewise
display such a fixed point (but not vice versa). Simply `freezing' 
the fluctuations transversal to the Killing orbits should meet this 
condition. The search for an UV stable fixed point in (a) and (b) 
thus provides an important self-consistency test for the 
asymptotic safety scenario.  

In version (b) of the functional integral part of the 4D reparameterization 
invariance has been fixed by choosing the parameterization (\ref{lineel}) of 
$G_{MN}$ adapted to the Killing vectors and the residual 2D 
diffeomorphism invariance has been partially fixed at the 
expense of the constraints. In this version one aims at a Dirac
quantization of the system; it is the one which we 
investigate in the bulk of the paper. 

The reason for spelling out both intentions (\ref{q1}) is that 
reasonable implementations of (a) always lead to versions of (b) where 
the `matter' part of the reduced Lagrangian corresponds to a sigma-model
with a Riemannian (rather than pseudo-Riemannian) target space,
i.e.~the hyperboloid $H_+$ in (\ref{hyp}). The resulting functional
integral in (b) can then be interpreted either as modeling a subsector of 
4D Lorentzian quantum gravity or as a subsector of 4D Euclidean quantum gravity.

For $\eps =+1$ this conclusion is obvious. Both the original action 
(\ref{action2}) and the dual action (\ref{Ldual}) have 2D `matter' sectors
based on the hyperboloid $H_+$. Both follow directly from a 4D line
element, namely (\ref{Elineel}) and (\ref{Dlineel}), respectively. 
The base space of the 2D reduced actions has Minkowski signature
and the functional integral (b) has its usual quantum field theoretical meaning. 
Since the Lagrangians are Poincar\'{e} invariant and we confine ourselves to a 
perturbative study a Wick rotation is legitimate, technically
convenient, but not essential. Looking back at (\ref{Elineel}) or (\ref{Dlineel})  
one sees that a Wick rotation of $x^0$ gives the 4D line
elements a Riemannian signature, $(-,-,-,-)$. The reduced Lagrangians 
can in both versions be written in the form 
\be
L_E(n, \rho, \sigma) = \frac{1}{2 \lb} \left[ \rho 
\frac{(\dd_{\mu}\Delta)^2 + (\dd_{\mu} B)^2}{\Delta^2} - 
\dd_{\mu} \rho \dd_{\mu}(2 \sigma + \ln \rho) \right]\,.
\label{q2} 
\ee
This is readily checked to coincide with the result of the 
2-Killing reduction of the Einstein-Hilbert 
action for Riemannian metrics of signature $(+,+,+,+)$. 
As expected, the ${\rm O}(1,2)$ part is manifestly positive,
while the indefiniteness of the $\dd_{\mu} \rho \dd_{\mu} \sigma$ 
part (manifest by diagonalization) is reflects the stability problem 
related to (or replacing) the ``conformal factor'' problem 
\cite{MazMot}. The exponential now looks more appealing, 
$\exp( - d^2 y \!\int \!d^2 x \,L_E)$, but in a perturbative context 
it is equivalent to the original $\exp(i \,d^2 y \!\int \!d^2 x \,L_+)$
of (b).

In contrast, when starting from the stationary axisymmetric  
line element, i.e.~Eq.~(\ref{Elineel}) or (\ref{Dlineel}) with 
$\eps =-1$, the 2D matter sector in the reduced actions is indefinite.  
However the 2D base space is already Euclidean, so that 
$\exp(i\, d^2 y \!\int \!d^2 x \,L_-)$ 
has no immediate quantum field theoretical interpretation.
Such an interpretation can be restored in two ways. One proceeds 
by analytic continuation and the other  by dualization.
Both are merely heuristic but both give rise to a 2D `matter' 
sigma-model based on the hyperboloid $H_+$. In the first argument 
one notes that upon analytic continuation in the Killing time $y^2$ the 
functional integral regains a quantum field theoretical meaning. However 
the line element (\ref{Elineel}) then is complex, unless one 
performs an additional replacement $B \ra i B$ (or restricts
attention to the abelian $B =0$ sector). Alas, doing both 
replacements $y^2 \ra i y^2$ and $B \ra iB$ gives back (\ref{q2}).     
This argument is somewhat unsatisfactory because replacing 
a real variable in a functional integral by a purely imaginary one
can hardly be expected to be legitimate. A safe conclusion 
would be that no conclusion can be drawn based on the actions
obtained by direct reduction in the stationary axisymmetric sector.

The second argument is better and proceeds by dualization. As outlined 
in appendix A, applying an abelian T-duality transformation to the 
Lagrangian $L_-$ in (\ref{Lgamma}) leads to the classically equivalent 
Lagrangian (\ref{Ldual}) where the spins live on $H_+$. Adapting the 
familiar functional integral argument \cite{Buscher} to the case at hand 
this comes about as follows. Consider 
\ba
&& \int \cD \Delta \cD V \cD B_{\d} \,\exp\{ 
i \,d^2 y \! \int \! d^2 x \,L^{\rm gauge} \}\,, 
\nonum
&& 
L^{\rm gauge} = \frac{1}{2\lb}\left[ \rho \Delta^{-2} 
(- (\dd_{\mu} \Delta)^2 + V_{\mu}^2 ) + 2 B_{\d}\, 
\eps^{\mu\nu} \dd_{\mu} V_{\nu} - \rho^{-1}(\dd_{\mu} \rho)^2
- 2  \dd_{\mu} \rho \dd_{\mu} \sigma \right]\,.
\label{gaugeint}
\ea   
The Lagrangian can be viewed as a gauged version of $L_-$ by decomposing 
$V_{\mu}$ as $V_{\mu} = A_{\mu} + \dd_{\mu} B$, with $\delta B = \lb(x),\,
\delta A_{\mu} = - \dd_{\mu} \lb(x)$. The overall `$i$' stems from 
the original (heuristic) functional integral (a) over all 4D Lorentzian 
metrics with one timelike and one spacelike Killing vector. Integrating over 
the Lagrange multiplier field $B_{\d}$ implements the 
$\eps^{\mu\nu} \dd_{\mu} V_{\nu} =0$ constraint, i.e.~$V_{\mu} 
= \dd_{\mu} B$ locally, and hence (in the gauged interpretation after 
gauge fixing) gives back the original reduced functional integral. 
On the other hand one can perform the Gaussian integral over $V_{\mu}$ 
which (after dropping a divergent measure factor) yields 
\be
\int \cD \Delta \cD B_{\d} \,\exp\{ i\, d^2 y  \!\int \!d^2 x \,L^{\d}_- \} =
\int \cD \Delta_{\d} \cD B_{\d} \,\exp\{ -i\, d^2 y \!\int \!d^2 x \,L_+ \}\,,
\label{dualint}
\ee
using (\ref{LDD}) in the second step.
As noted in (\ref{LDdirect}) the dual action is no longer directly related 
to a 4D line element and is therefore unaffected by the analytic continuation 
in the Killing time $y^2$ that restores a standard quantum field theoretical 
interpretation of (b). The discrepancy between the result (\ref{Ldual}) 
for $\eps =-1$ and the standard duality formula (where no sign flip occurs in the 
relevant component of the target space metric $g^{\d}_{\b\b} = +1/g_{\b\b}$)  
arises because in \cite{Buscher, Alvdual} the overall `$i$'    
in (\ref{gaugeint}) is replaced with an overall `minus' (as required 
by Euclidean quantum field theory) and $B_{\d}$ with $i B_{\d}$. We conclude 
that after dualization the formal 
functional integral relevant for the stationary axisymmetric subsector 
is likewise one governed by a sigma-model Lagrangian whose `matter' 
part corresponds to the  Riemannian symmetric space $H_+$ rather than 
the pseudo-Riemannian $H_-$ entering via the direct reduction. Due to the 
formal nature of the argument (and the fact that both systems are not 
renormalizable in the standard sense) one cannot expect that 
the orginal and the dual theory (based on $L_-$ and $L_+$, respectively)
are equivalent. One is thus free to take the dual version (\ref{dualint}) 
as the starting point for 
the construction of a perturbative quantum theory.

Our main goal will therefore be to construct a finite perturbative measure 
based on the exponentiated Lagrangian $\exp( i \int d^2 x L_+)$ or 
$\exp(- \int d^2 x L_E)$. For $\eps =+1$ this Lagrangian can be taken 
to be the one (\ref{action2}) obtained by direct reduction. For $\eps =-1$ 
it should be interpreted as the dualized Lagrangian in (\ref{Ldual}), (\ref{LDD});
for simplicity we drop the subscripts `${\d}$' on the fields throughout. 
However one should keep in mind that the fields in the $L_+$ Lagrangian when  
viewed as describing the stationary axisymmetric sector have a very 
{\it different} 4D spacetime meaning than in the $\eps =+1$ sector. 

In order to highlight the differences to the $H_-$ case we carry it along 
in the version  $\exp( - \int d^2 x L_-)$, where the quantum field theoretical 
meaning is restored while giving up a link to 4D quantum gravity.      
Of course the $H_-$ system can still be interpreted as 2D Euclidean 
(conformally flat) quantum gravity non-minimally coupled to a hyperbolic 
sigma-model.

For the Ernst-like systems so far only quantizations of the reduced 
phase space have been investigated, see e.g.~\cite{KorNic96,KorSam98,qernst}.
As stressed in the introduction a number of important issues 
can however only be addressed in a Dirac quantization program. 
A Dirac approach also allows one to decompose 
the full problem into simpler subproblems: (i) Construction of a 
finite perturbative measure for the basic Lagrangian. (ii) Extension
by sources according to the composite operators aimed at -- which should 
include at least the constraints and the Noether currents. 
(iii) Projection onto the physical state space. In this article we 
focus on steps (i) and (ii). Concerning (i) we find that although the 
quantum systems based on (b) in (\ref{q1}) are {\it not} renormalizable 
in the standard quantum field theoretical sense, they can be 
rendered renormalizable in the `conformal' sense outlined in the 
introduction. This also allows one to define composite operators in a
systematic way. In particular, quantum versions of the 
constraints can be defined, although the actual 
construction of physical states and proving the absence of 
negative norm states will still be difficult.
Throughout we address only the ultraviolet aspects because it 
seems that infrared problems will disappear upon projection 
onto the physical state space. This picture is suggested by the 
situation in the abelian systems, where (hopefully) Dirac quantization 
is equivalent to a reduced phase space quantization, and the latter 
is free of infrared problems \cite{ERwaves}. In other words 
we hope that an infrared cutoff in the nonabelian systems 
can be removed after projection onto the physical state 
space. In the following we thus concentrate on steps 
(i) and (ii), i.e.~on perturbatively defining an UV finite quantum 
theory based on the (source extended) action (\ref{i1}) 
or its generalization $L_{hab}$ in (\ref{Labgamma}).

\newpage
\newsection{Formulation as a Riemannian sigma-model}

For the reasons explained above we aim at a Dirac quantization 
of the system, promoting also $\rho$ and $\sigma$ to independent 
quantum fields. As always the raw material for a perturbative 
quantization are the counter terms. It is convenient to interpret 
the generalized Ernst systems (\ref{Labgamma}) as Riemannian 
sigma-models in the sense of Friedan \cite{Frie85}. Taking advantage 
of the vast literature on these systems one gets the 
counter terms almost for free. To preclude a
misconception, however, let us stress that the `renormalization 
architecture' built from these counter terms 
will be different from the one used for Riemannian sigma-models.

\newsubsection{Conformal geometry of the target space}

The Ernst system can be treated as a Riemannian sigma-model by promoting 
$\rho$ and $\sigma$ to extra coordinates on a 4-dim.~auxiliary
target space. To this end set
\ba
&& \phi^1 = \Delta\;,\quad \phi^2 = B\;,\quad \phi^3 = \rho\;,
\quad \phi^4 = \sigma\;,
\nonumber \\[4mm]
&& \widehat{g}_{ij}(\phi) = \left( \begin{array}{cccc} 
\eps \rho/\Delta^2 & 0 & 0 &0 \\
0 & \rho/\Delta^2 & 0 &0\\
0 & 0 & a/\rho & b \\
0 & 0 & b &0 \end{array} \right)\;.
\label{Tmetric}
\ea
For $a =b=-\eps$  then $\frac{1}{2\lb }\widehat{g}_{ij}(\phi)\, 
\dd \phi^i \dd \phi^j$ reproduces the Lagrangian of 
(\ref{action2}). In the following we keep $a \in \R$ and 
$b \neq 0$ as parameters, first as a check on the convention-independence,
and second because they might turn into coupling constants.

For the rest of this section we now study the geometry of this
4D target space. First note that $\widehat{g}_{ij}(\phi)$ has signature 
$(\eps, +,+,-)$. Further $\widehat{g}_{ij}(\phi) d\phi^i d\phi^j$ has 
the structure of a ``warped product'' of the hyperboloid $H_{\eps}$ 
with $\R^{1,1}$. Indeed
\be
\widehat{g}_{ij}(\phi) d\phi^i d\phi^j = 
e^{u^+}[\eps ds^2_{H_{\eps}} + 2b du^+ du^-]\;,\sspace
u^+ = \ln \rho \;,\quad u^- = \sigma + \frac{a}{2b} \ln \rho\;.
\label{ydef}
\ee
The scalar curvature is $R(\widehat{g}) = - 2 \eps/\rho$, so that 
$\rho$ plays the role of a `radion' field parameterizing the curvature
radius of the target manifold.%
\footnote{Note that although $H_{\eps}$ has constant curvature $-2$ 
for both signatures $\eps = \pm 1$ of the Killing vectors, the 
curvature of the 4D target space depends on $\eps$.}

Next we determine the Killing symmetries of the metric (\ref{Tmetric}). 
One readily finds that (\ref{Tmetric}) admits the following Killing 
vectors: ${\bf t}_-:= \dd_{\sigma}$ and ${\bf e},\, {\bf h},\,{\bf f}$, 
where
\be
\begin{array}{lll}
{\bf e} = \dd_B \;,\quad & {\bf h} = 2(B \dd_B + \Delta \dd_{\Delta}) \;,
\quad & {\bf f} = (-B^2 + \eps \Delta^2) \dd_B - 2 B \Delta \dd_{\Delta}\;,
\\[2mm]
[{\bf h},{\bf e}] = -2 {\bf e}\;,\quad &
[{\bf h},{\bf f}] = 2 {\bf f}\;,&
[{\bf f},{\bf e}] = {\bf h}\;,
\label{Ksl2}
\end{array}
\ee
generate the isometries of the hyperboloid $H_{\eps}$. In addition to 
these proper Killing vectors (\ref{Tmetric}) admits two
conformal Killing vectors 
\ba
{\bf t}_+ &=& \rho \dd_{\rho} - \frac{a}{2b} \dd_{\sigma}\;,\nonum
{\bf d} &=& -\rho\ln \rho \,\dd_{\rho} + 
\Big(\sigma + \frac{a}{b} \ln \rho \Big) \dd_{\sigma} 
= - u^+ {\bf t}_+ + u^- {\bf t}_-\;.
\label{Konf1}
\ea  
Together with ${\bf t}_- \!= \!\dd_{\sigma}$ they generate the algebra
of isometries of $\R^{1,1}$, i.e.~$[{\bf t}_+, {\bf t}_-]=0$ and 
$[{\bf d}, {\bf t}_{\pm} ] = \pm {\bf t}_{\pm}$.~(Presumably 
there exists a relation to the conformal symmetries
in \cite{JulNic96}.) Further ${\bf t}_{\pm}$ are null vectors, 
$\widehat{g}_{ij}(\phi) {\bf t}_{\pm}^i {\bf t}_{\pm}^j =0$.  
For later use we also note the finite transformations and the 
scaling properties of the line element $ds^2 = 
\widehat{g}_{ij}(\phi) d\phi^i d\phi^j$: 
\begin{subeqnarray}
e^{-\ln \Lambda \, {\bf t}_+} \;&:& \;\;\,(\rho,\sigma)\;\; 
\rra \big( \Lambda^{-1} \rho, \sigma + \frac{a}{2 b} \ln \Lambda \big)\,,
\sspace \,ds^2 \rra \Lambda^{-1}\, ds^2\,,
\\
e^{-\ln \Lambda \, {\bf d}} \;&:& \,(u^+,u^-) 
\rra ( \Lambda u^+, \Lambda^{-1} u^-)\,, 
\bspace ds^2 \rra \rho^{\Lambda-1}\, ds^2\,,
\label{Kconf2}
\end{subeqnarray} 
with $u^{\pm}$ as in (\ref{ydef}). 

Conversely one can now ask what is the most general form of a 
target space metric compatible with these symmetries. This will dictate in 
what subspace of 4D Riemannian metrics the renormalization flow 
can move. Fixing a coordinate system adapted to the Killing  
vectors proper, one finds that the generic form of the metric admitting 
in addition the above conformal Killing vectors is 
\be
g_{ij}(\phi) = h(\rho) \left( \begin{array}{cccc} 
\eps/\Delta^2 & 0 & 0 &0 \\
0 & 1/\Delta^2 & 0 &0\\
0 & 0 & a(\rho)/\rho^2 & b(\rho)/\rho \\
0 & 0 & b(\rho)/\rho & 0 \end{array} \right)\;,
\label{Tabmetric}
\ee
for some functions $h(\rho),\,a(\rho),\,b(\rho)$. The corresponding
Lagrangian is the one anticipated in Eq.~(\ref{Labgamma}). The vanishing 
of $g_{44}$ in (\ref{Tabmetric}) ensures that the curvature of 
the lower $2 \times 2$ block vanishes; correspondingly the 
isometries of $\R^{1,1}$ (acting 
like conformal Killing vectors on the full metric) are still 
present. Indeed, the two conformal Killing vectors now read
\ba
&\nspace & {\bf t}_+ = \frac{b}{b(\rho)}\rho \dd_{\rho} - 
\frac{b a(\rho)}{2 b(\rho)^2}\, \dd_{\sigma} \,,\nonum
&\nspace & {\bf d} = - \left\{ \frac{\rho}{b(\rho)} 
\int^{\rho} \frac{du}{u} b(u) \right\} \dd_{\rho} + 
\left\{ \sigma + \int^{\rho} \frac{du}{u} \frac{a(u)}{2 b(u)} 
+ \frac{a(\rho)}{2 b(\rho)^2} \int^{\rho} \frac{du}{u} b(u)
\right\} \dd_{\sigma}\,.
\ea
The scale factors in the conformal Killing equations are 
\ba
\cL_{{\bf t}_+} g_{ij} \is \frac{b\rho\dd_{\rho} \ln h}{b(\rho)}
\;g_{ij}\;,\nonum
\cL_{{\bf d}} g_{ij} \is -\frac{\rho\dd_{\rho} \ln h}{b(\rho)} 
\int^{\rho} \frac{du}{u} b(u) \;g_{ij}\,. 
\ea
One can also check that ${\bf t}_+,\,{\bf d}$, and ${\bf t}_- = \dd_{\sigma}$ 
continue to generate the isometries of $\R^{1,1}$.

Of course in (\ref{Tabmetric}) one is still free to perform 
coordinate transformations in the non-Killing coordinate 
$\rho$. If we insist that the ${\bf t}_-$ Killing vector 
continues to act like the $j=4$ coordinate derivative the allowed 
residual transformations are
\be
\rho \rra \widetilde{\phi}^3(\rho) = \tilde{\rho}\;,
\sspace    
\sigma \rra \widetilde{\phi}^4(\rho) + \sigma = \tilde{\sigma}\;.
\label{coord_change1}
\ee
Spelling out 
$$
g_{ij}(\phi) = \frac{\partial \widetilde{\phi}^k}{\partial \phi^i}
\frac{\partial \widetilde{\phi}^l}{\partial \phi^j}
\,\widetilde{g}_{kl}(\widetilde{\phi})\,,
$$
and solving for $\widetilde{\phi}^3(\rho),\,\widetilde{\phi}^4(\rho)$
one obtains
\begin{subeqnarray}
&& \widetilde{h}(\widetilde{\phi}^3(\rho)) = h(\rho) \;,
\\
&& \int^{\widetilde{\phi}^3(\rho)} \frac{du}{u} 
\widetilde{b}(u) = 
\int^{\rho} \frac{du}{u} b(u)\,,
\\
&& 
\widetilde{\phi}^4(\rho) = \int^{\rho} \frac{du}{2 u} 
\frac{a(u)}{b(u)} \left[1 -  
\frac{\widetilde{a}(\widetilde{\phi}^3(u))}{a(u)}
\bigg(\frac{b(u)}{\widetilde{b}(\widetilde{\phi}^3(u))}\bigg)^2
\right]\,.
\label{coord_change3}
\end{subeqnarray}
These relations can be utilized in several ways. One can use 
(\ref{coord_change3}a) to bring $h(\rho)$ into a prescribed form;
then $\widetilde{\phi}^3(\rho)$ is fixed and can no longer help 
to simplify $b(\rho)$. Alternatively one can use  
(\ref{coord_change3}b) to bring $b(\rho)$ into a prescribed 
form, in which case $\widetilde{\phi}^3(\rho)$ is fixed by this 
requirement and can no longer be used to simplify $h(\rho)$. 
In both cases $a(\rho)$ can largely be changed at will by means of 
(\ref{coord_change3}c). We shall adopt the second option and adjust
$b(\rho)$ to be a nonzero constant $b$. Likewise $a(\rho)$ 
is adjusted to be some constant $a$. Summarizing, in an adapted 
coordinate system the generic form of the target space metric 
compatible with the above (conformal) Killing 
vectors is 
\be
g_{ij}(\phi) = \frac{h(\rho)}{\rho} \,\widehat{g}_{ij}(\phi)\,,
\label{Tgmetric}
\ee
where $\widehat{g}_{ij}(\phi)$ is given by (\ref{Tmetric}) with 
$b\neq 0,\,a\in \R$. The main modification compared to the 
initial situation is the $\rho$-dependent scale factor $h(\rho)/\rho$.     
The corresponding Lagrangian is (\ref{Labgamma}) with constant 
$a$ and $b \neq 0$. 

Even within the class of metrics (\ref{Tgmetric}) some residual 
transformations (\ref{coord_change1}) are possible. Let 
$\widetilde{h}(\widetilde{\rho})$ and $\widetilde{a},\widetilde{b} \in \R$
parameterize a metric of this form. Then (\ref{coord_change1})
with 
\be
\widetilde{\phi^3}(\rho) = \rho^p\;,\quad p := b/\widetilde{b}\,,
\sspace \widetilde{\phi}^4(\rho) = 
\frac{1}{2b}[a - \widetilde{a} p^2] \ln \rho
\label{coord_change4}
\ee
maps it onto a metric of the same form, with constants $a,b$ 
and $h(\rho) = \widetilde{h}(\rho^p)$. In particular
one can map $\tilde{h}(\tilde{\rho}) = \tilde{\rho}^{\tilde p}$
with $\tilde{p} >0$ onto $h(\rho) = \rho^{p \tilde{p}}$, where the 
new power may be negative. Qualitatively this exchanges the role 
of small and large $\rho$ in the asymptotics of $h(\rho)$.

It is instructive to convert the above target space symmetries into
current identities. To this end consider first a generic infinitesimal 
diffeomorphism $\phi^j \ra \phi^j + v^j(\phi)$, $g_{ij} \ra 
g_{ij} - \cL_v  g_{ij}$. The invariance of $L = \frac{1}{2 \lb} g_{ij}(\phi) 
\dd \phi^i \dd \phi^j$ can be expressed as 
\be 
\dd^{\mu}\Big[ \frac{1}{\lb}g_{ij}(\phi) v^i(\phi)\dd_{\mu} \phi^j \Big] 
- \frac{1}{2\lb} \cL_v g_{ij}(\phi) \dd^{\mu} \phi^i \dd_{\mu} \phi^j +
\frac{\delta S}{\delta \phi^i} v^i(\phi) =0 \;. 
\label{diffward1}
\ee
Here  
\be
\cL_v g_{ij} = v^m \frac{\dd}{\dd \phi^m} g_{ij} + 
\frac{\dd v^m}{\dd \phi^i} g_{j m} +
\frac{\dd v^m}{\dd \phi^j} g_{i m}\;,
\label{Lie}
\ee
is the Lie derivative with respect to the vector field $v^m(\phi)$.   
The quantity $\lb J_{\mu}(v) = g_{ij}(\phi) v^i(\phi) \dd_{\mu}
\phi^j$ may be interpreted as a ``diffeomorphism current''. If $v^i(\phi)
\dd_i$ is a Killing vector of $g_{ij}(\phi)$ the identity
(\ref{diffward1}) simply expresses the conservation of the associated 
Noether current; so for our (\ref{Tgmetric}) there are four Noether 
currents, associated with ${\bf e}, {\bf h}, {\bf f}$, and 
${\bf t}_-$. The latter reads 
\be
J_{\mu}({\bf t}_-) = \frac{b}{\lb} \dd_{\mu} \left( \int^{\rho} 
\frac{du}{u} h(u) \right) = \frac{b}{\lb} h \dd_{\mu} \ln \rho\,. 
\label{sigma_curr}   
\ee
For the O$(1,2)$ Noether currents often the vector basis 
is more convenient
\ba
&& J_{\mu}^i = \frac{\eps}{\lb}h (n \times \dd_{\mu} n)^i\;,  
\sspace \mbox{where}
\nonum 
&& J_{\mu}^0 = \frac{\eps}{2}[J_{\mu}({\bf f}) -J_{\mu}({\bf e})]\,,
\quad  
J_{\mu}^1 = \frac{\eps}{2} J_{\mu}({\bf h})\,,
\quad 
J_{\mu}^2 = \frac{\eps}{2}[J_{\mu}({\bf f}) +J_{\mu}({\bf e})]\,.
\label{O12curr}
\ea
More interestingly also the two conformal Killing vectors 
give rise to the on-shell identities:
\ba
\nspace \dd^{\mu} J_{\mu}({\bf t}_+) \is \rho \dd_{\rho} \ln h \cdot L\;,
\quad \;\sspace  \;J_{\mu}({\bf t}_+) = 
\frac{b}{2 \lb}h(\rho) \dd_{\mu}(2 \sigma + \frac{a}{b} \ln \rho)\,,
\nonum
\nspace \dd^{\mu} J_{\mu}({\bf d}) \is -\ln \rho \cdot \rho \dd_{\rho} \ln h 
\cdot L\;,\quad\, J_{\mu}({\bf d}) = \frac{b}{\lb}h(\rho)
(\sigma \dd_{\mu} \ln \rho - \ln \rho \dd_{\mu} \sigma)\,. 
\label{cward1}
\ea
Observe that there are only two choices for $h(\rho)$ for which the 
Lagrangian is a total divergence on-shell: $h(\rho) \sim \rho^p$ and 
$h(\rho) \sim \ln\rho$. (Since ${\bf t}_+$ and ${\bf d}$ are the only 
conformal Killing vectors of the target space metric there can be no 
other identities of this form.) The case $h(\rho) \sim \rho^p$ 
illustrates that even after the conformal gauge in (\ref{Labgamma}) 
has been chosen, the system still `remembers' the link to the 2D 
diffeomorphism invariance. (Recall that in a diffeomorphism 
invariant theory the Lagrangian can always be written as a 
total divergence on-shell.)  
In the quantum theory the identities (\ref{cward1}) can be converted 
into Ward identities which help to characterize the quantum theory.

\newsubsection{Background field expansion and non-renormalization 
of \boldmath{$\xi^3$}}

The covariant background field method which we shall employ in 
the quantum theory involves decomposing the fields $\phi = 
(\Delta,B,\rho,\sigma)$ into a classical background field configuration 
$\varphi$ and a formal power series in the quantum fields $\xi$ whose 
coefficients are functions of $\varphi$. The series is defined in terms 
of the geodesic curve $[0,1] \ni s \ra \gamma^j(s)$ from the point 
$\varphi = \gamma(0)$ to the (nearby) point $\phi= \gamma(1)$, where $\xi^j = 
\frac{d}{ds} \gamma^j(s)|_{s=0}$ is the tangent vector at 
$\varphi$. E.g.~to second order in $\xi$ one has $\phi^j = \varphi^j 
+ \xi^j -\frac{1}{2}\Gamma(\varphi)^j_{\;\;kl} \xi^k \xi^l + 
O(\xi^3)$, where $\Gamma(\varphi)^j_{\;\;kl}$ is the 
metric connection evaluated at the point $\varphi$ in target 
space, i.e.~at the background field configuration. Generally we shall 
write $\phi^j(\varphi;\xi)$ for this series, and refer to 
$\phi,\,\varphi$, and $\xi$ as the full field, the background field, 
and the quantum field, respectively. For our target space 
metric (\ref{Tgmetric}) no major simplification occurs 
with one important exception: The geodesic equation for 
the 3-component $\gamma^3(s)$ decouples from the others and 
can be solved in closed form. The solution emanating from 
$\varphi$ with tangent vector $\xi$ reads
\be 
\xi^3 \frac{h(\varphi^3)}{\varphi^3} s = 
\int_{\varphi^3}^{\gamma^3(s)} \frac{du}{u} h(u)\;.
\label{3geodesic}
\ee
In particular $\phi^3 \!=\! \rho \!= \!\gamma^3(1)$ depends only on 
$\varphi^3$ and $\xi^3$, the first terms being 
$\rho = \varphi^3 + \xi^3 + [1/\varphi^3 - \dd_3 \ln h(\varphi^3)] 
(\xi^3)^2 + O((\xi^3)^3)$. This feature turns out to lead 
to a crucial simplification in the renormalization 
analysis.

We refer to appendix B for an outline of the renormalization 
of generic Riemannian sigma-models. In our quantum theory 
Eq.~(\ref{Tgmetric}) gives the renormalized 
target space metric from which the $g$-dependent counter 
tensors in (\ref{Rcounter1}) are computed. In addition to 
these coupling/source renormalizations also the quantum 
fields $\xi^j$ are renormalized in a nontrivial way. The 
transition from the bare fields $\xi_{\b}^j$ to the renormalized 
ones $\xi^j$ is governed by Eq.~(\ref{xiren1}). For the  
target space geometry (\ref{Tgmetric}) this is still true, 
with the notable exception of $\xi^3$:
\be
\xi_{\b}^3 = \xi^3 \quad \mbox{to all loop orders.}
\label{xiren3}
\ee    
This arises through the combination of the following facts:
(i) From (\ref{3geodesic}) we know that the inverse normal 
coordinate expansion for $\xi^3$ depends only on $\varphi^3$ and 
$\phi^3-\varphi^3$, i.e. $\xi^3 = \xi^3(\varphi^3;\phi^3 - \varphi^3)$.
(ii) The operator $Z(g)-1$ from which the renormalization
$\xi_{\b}(\xi)$ is computed is a scalar differential operator 
without constant terms built from the covariant derivative 
$\nabla_{\!i}$ and the curvature tensors of $g_{ij}$, both 
referring to the full field $\phi^j$.  
(iii) A covariant tensor $z_{i_1\ldots i_n}(g)$ 
of arbitrary rank built from curvature tensors and their covariant 
derivatives vanishes if $i_k =4$ for one or more $k=1,\ldots,n$. 
This can be seen to be a consequence of the flatness of the 
lower $2 \times 2$ block of the target space metric (\ref{Tgmetric}). 

The verification of (\ref{xiren3}) then is straightforward. 
By (ii) a typical monomial in $Z(g)-1$ is of the form $z^{i_1 \ldots i_n}(g) 
\nabla_{i_1} \ldots \nabla_{i_n}$, with $n \geq 2$. By (i) it 
acts on a function of $\rho= \phi^3$ only. One easily checks that 
then only the $z^{3 \ldots 3}(g)\nabla_3^n$ term contributes.
However on account of (iii) this vanishes, which proves 
(\ref{xiren3}).

The importance of (\ref{xiren3}) lies in the fact that composite  
operators $H$ that are arbitrary functions of (the renormalized 
full field) $\rho = \phi^3$ do {\it not} require renormalization, i.e.
\be
\nl H(\rho) \nr = \mu^{d-2} H(\rho)\,,
\label{frho_nonren}
\ee
where $\nl\, \cdot \,\nr$ is the normal product defined in 
Eq.~(\ref{normalproducts}). Hence, up to the trivial $\mu$-prefactor    
the function $H$ and the composite operator can be identified. 

Of course the same is not true for functions depending solely on 
one of the other fields $\Delta,\,B$ or $\sigma$. In particular 
the associated quantum fields $\xi^1,\,\xi^2$ and $\xi^4$ are 
renormalized in a nontrivial way. For example, taking advantage 
of (\ref{xiren3}) one finds from (\ref{xiren2}) 
\be
\xi^i_{\b} = \xi^i \left[ 1 + \frac{1}{2 -d} \left( 
\frac{\lb}{2\pi} \frac{\eps}{3 h(\rho)} + O(\lb^2) \right) + \ldots 
\right], \quad i =1,2\,,
\label{xiren12}
\ee
and a similar more complicated expression for $\xi_{\b}^4$.

\newsubsection{Structure of the metric counter terms}

The (conformal) Killing vectors studied in section 3.1 also 
constrain the form of the coupling/source counter tensors
in (\ref{Rcounter1}). Here we consider specifically the metric 
counter tensors $T_{ij}(g)$ since they will be of immediate 
importance. Similar arguments however apply to the other 
purely $g$-dependent counter terms $Z^V(g), N(g), Z(g)$ and 
$\Psi(g)$. We shall discuss them separately when needed. 
Here recall the notation $T^{(\nu,l)}_{ij}(g)$, $l \geq \nu$, 
for the $l$-loop $\nu$-th order pole term in the metric counter 
term. For the moment we only need two generic properties 
of them. First, they are built from the curvature tensors of 
$g_{ij}$ and their covariant derivatives. Second they can be chosen 
to transform as as $T_{ij}^{(\nu,l)}(\Lambda^{-1}g) = 
\Lambda^{l-1} T_{ij}^{(\nu,l)}(g)$, $\Lambda \in \R$, under 
constant rescalings of the metric.

The first property implies that the geometry of the hyperboloid 
$H_{\eps}$ remains intact. Further -- related to the flatness 
of the lower $2\times 2$ block in (\ref{Tgmetric}) -- all 
covariant $4$-components of the Riemann tensor and its covariant 
derivatives vanish. A detailed analysis shows that 
the counter tensors must then be of the form 
$T^{(\nu,l)}_{ij}(g) \sim {\rm diag}(\eps/\Delta^2, 1/\Delta^2,\, *,\, 0)$,
where both the proportionality factor and the $i\!=\!j\!=\!3$ component are 
functions of $\rho$ only. The latter are further constrained by the  
scaling property of the counter tensors. Combined with (\ref{Kconf2}) 
it poses the condition
\be
T^{(\nu,l)}_{ij}(\widehat{g}) d\phi^i d\phi^j\bigg|_{
\phi^i \ra \exp(- \ln \Lambda\,{\bf t}_+)  \phi^i} = 
\Lambda^{l-1} T^{(\nu,l)}_{ij}(\widehat{g}) d\phi^i d\phi^j\;,
\label{Tcounter2}
\ee  
on the counter tensors computed from $\widehat{g}_{ij}(\varphi)$. 
They must therefore be of the form \newline $\rho^{-l} {\rm diag} 
(\eps \rho/\Delta^2, \rho/\Delta^2, a_l/\rho,0)$ for some
constants $a_l$. (This structure was already noted in \cite{Satoh}.) 
For a generic $h(\rho)$ in (\ref{Tgmetric}) only the  $i\!=\!j\!=\!3$ 
component of the counter tensor can contain $\rho$-derivatives, so that 
the before mentioned proportionality factor must be $\zeta_l\, h^{1-l}$,
for some number $\zeta_l$. 

Later on we need only the $\nu=1$ counter 
tensors, for which we also have information about the $\zeta_l$ through 
the flat space O$(1,2)$ sigma-model. In summary we conclude that  
to all loop orders the counter tensors $T_{ij}^{(1,l)}(g)$
are constrained to be of the form 
\be
T_{ij}^{(1,l)}(g) = \frac{1}{h^{l-1}}\, {\rm diag} 
\left( \frac{\eps\,\zeta_l}{\Delta^2},\,\frac{\zeta_l}{\Delta^2},\,
\frac{S_l(h)}{\rho^2},\, 0 \right)\,,\sspace \forall\, l \geq 1\,,
\label{Tcounter1}
\ee
where $S_l(h)$ is a differential polynomial in $h$ that 
is invariant under constant rescalings of $h$ and vanishes for 
constant $h$. The numbers $\zeta_l$ are the  
counter term coefficients of the flat space hyperbolic sigma-model 
with 2D target space $H_{\eps}$ computed (for $l >2$) in the minimal 
subtraction scheme. Using (\ref{Rcounter2}) and (\ref{Tgmetric})
the first three are explicitly given by: 
\be
\label{Chtensors}
\begin{array}{lcll}
T^{(1,1)}_{ij}(g) &\;:\quad & \zeta_1 = -\eps\;,
&\quad S_1(h) = 
- (\rho\dd_{\rho})^2 \ln h + \frac{1}{2} 
(\rho \dd_{\rho} \ln h)^2 \;,\\[2mm]
T^{(1,2)}_{ij}(g) &\;:\quad & \zeta_2 = 1/2\;, &\quad S_2(h) = 0\,,
\\[2mm]
T^{(1,3)}_{ij}(g) &\;:\quad & \zeta_3 = -\eps 5/12\;, 
&\quad S_3(h) = 
-\frac{1}{4}(\rho\dd_{\rho})^2 \ln h + \frac{1}{12} 
(\rho \dd_{\rho} \ln h)^2 \;.
\end{array}
\ee
Modulo signs the coefficients $l \zeta_l$ indeed coincide with the known 
beta function coefficients in the O$(3)$ nonlinear sigma model computed 
in minimal subtraction \cite{BrezHik78,Hikami81,Wegner86,Wegner89}. The 
sign pattern is induced by the combination of the overall sign in the 
Lagrangian and the signature of the hyperboloid $H_{\eps}$.

The form of the metric counter tensors in the Ernst-like systems thus 
is highly constrained, which will be important later on. 
Nevertheless one sees that the counter terms (\ref{Tcounter1}) are 
{\em not} of the form (\ref{Tgmetric}) but differ by inverse 
powers of $h(\rho)$. One might still hope to achieve field 
theoretical renormalizability by: (i) allowing for non-linear 
field redefinitions, (ii) promoting the constants $a,b$ (and possibly 
the coefficient of a $\dd^{\mu} \sigma \dd_{\mu} \sigma$ term) 
to couplings that might get renormalized, and (iii) allowing the 
renormalized target space metric to be of the generic form 
(\ref{Tgmetric}). We state without proof that (i) 
is indeed enough to absorb the  one-loop counter term, but 
beyond one loop even the combination of (i),(ii) and (iii)
is not sufficient to ensure field theoretical renormalizability.

\newsection{Renormalization by preserving the Killing symmetries} 

Anticipating these lessons we formulate in the following 
a renormalization procedure that is only slightly weaker than quantum 
field theoretical renormalizability. For want of a better term 
we shall refer to it as {\em conformal renormalizability}. The term is 
meant to indicate that although the functional form of the Lagrangian
cannot be preserved identically in the renormalization process, 
it can be maintained up a a field dependent conformal factor in 
a way that preserves all, in particular the {\it conformal} Killing 
vectors of the original target space metric. We show this in 
the framework described above to all orders in the loop exapansion
and borrow again various results from the renormalization of Riemannian 
sigma-models. To get started recall that in this setting field theoretical 
renormalizability amounts to the condition that the bare and the 
renormalized target space metric have the same functional form: 
$g_{ij}^{\b}(\,\cdot\,) = g_{ij}(\,\cdot \,)$. Since this cannot  
be achieved for the Ernst-like systems we relax the condition 
as follows:

\newsubsection{Conformal renormalizability}

Motivated by the form of the counter tensors we allow for a change 
of the target space metric by a singular $\rho$-dependent prefactor. 
Explicitly we assume the transition to be of the form 
\be
g_{ij}^{\b}(\phi) = \mu^{d-2} \,g_{ij}(\phi)\left[
1 + \frac{1}{2-d} \sum_{l \geq 1} \Big(\frac{\lb}{2\pi} \Big)^l 
\,H_l(\rho) + \ldots \right]\,,
\label{RBmetric}
\ee
with the functions $H_l(\rho)$ to be adjusted. The arguments on {\it both} 
sides of (\ref{RBmetric}) are the renormalized fields and the dots 
indicate higher pole contributions. Here we anticipate that the 
parameters $a,b$ do not get renormalized. The same holds for the 
coupling $\lb$ which therefore merely plays the role of a loop counting 
parameter. The renormalized metric $g_{ij}(\phi)$ is of the 
form (\ref{Tgmetric}), where the function $h$ is part of the 
specification of the quantum theory. We shall return to the issue of 
adequately fixing $h$ later. Technically it is advantageous to 
leave $h$ unspecified and to formulate the renormalization procedure 
for generic $h$.

In doing so we assume that the bare and the renormalized 
fields are related by 
\be
\phi_{\b}^j = \phi^j + \frac{1}{2 -d} 
\sum_{l \geq 1} \Big( \frac{\lb}{2\pi} \Big)^l \phi_l^j(\phi) 
+ \ldots \,.
\label{phiren}
\ee
The only requirement on the functions $\phi_l^j(\phi)$ is that 
they do not contain derivatives of the fields $\phi^j$. 
In a generic Riemannian sigma-model the bare metric value is expanded 
in terms of the renormalized value as 
\be 
g_{ij}^{\b}(\phi_{\b}) = \mu^{d-2} \left[ g_{ij}(\phi)
+ \frac{1}{2-d} \sum_{l \geq 1} \Big( \frac{\lb}{2\pi}\Big)^l 
\,T_{ij}^{(1,l)}(g)+ \ldots\right]\;,
\label{gbare}
\ee
where we only displayed the counter tensors for the 
simple poles; c.f.~Eqs.~(\ref{Rcounter1}), (\ref{nulpieces}). 
Combining (\ref{RBmetric}), (\ref{phiren}) 
and (\ref{gbare}) one arrives at the finiteness conditions
\be
\cL_{\phi_l} g_{ij}+  H_l(\rho) g_{ij} 
= T_{ij}^{(1,l)}(g)\;,\sspace l \geq 1\,,
\label{Hfinite}
\ee
where $\cL_v g_{ij}$ is the Lie derivative (\ref{Lie}), $T_{ij}^{(1,l)}(g)$ 
are the counter terms in (\ref{Tcounter1}), and all quantities are evaluated 
at the renormalized fields. The $\rho$-dependence in the $H_l(\rho)$ term 
marks the difference to a renormalization in the quantum field theoretical 
sense, see e.g.~\cite{Hungary96}. 
The structure (\ref{Tcounter1}) of the counter tensors implies
that $H_l(\rho)$ and $\phi_l^j(\rho)$ scale as 
\be
H_l(\rho) \rra \Lambda^{-l} H_l(\rho)\;,\sspace 
\phi_l^j(\rho) \rra \Lambda^{-l} \phi_l^j(\rho)\;,
\label{Hxiscaling}
\ee 
under $h(\rho) \ra \Lambda h(\rho)$.

The finiteness condition (\ref{Hfinite}) is easily seen to imply 
that the fields $\Delta, B$ are at most multiplicatively renormalized.
For $\phi_l^3(\phi)$ and $\phi^4_l(\phi)$ one has to allow for a non-trivial
$\rho$-dependence and this also turns out to be sufficient. Thus 
we assume $\phi_l^3(\phi)= \phi_l^3(\rho)$ and $\phi^4_l(\phi)=\phi_l^4(\rho)$.
The finiteness condition then amounts to a pair of coupled 
differential equations for $\phi^3_l(\rho)$ and $H_l(\rho)$. 
The solutions are
\be
\phi^3_l(\rho) = -\zeta_l\,\rho \int_{\rho_l}^{\rho} \frac{du}{u} 
\frac{1}{h(u)^l}\;,\quad \;\;\;
H_l(\rho) = -\frac{1}{h} \rho \dd_{\rho}( h \phi_l^3/\rho)\;,
\quad \forall l \geq 1\,.
\label{ren_sol1}
\ee
The integration constants $\rho_l$ are fixed by requiring that
$\phi^3_l(\rho)$ does not contain a term linear in $\rho$, 
i.e.~$\rho_l = \infty$. One reason why this is a natural choice 
is that the coefficient of the linear term always has to match that 
in a renormalization of $\lb$, and both are redundant in the sense that 
they are not useful for the absorption of counter terms. We thus set 
both to zero which leads to the above criterion for fixing 
$\rho_l$. The same criterion will be recovered below 
from another viewpoint. With (\ref{ren_sol1}) known the solutions 
for $\phi_l^4(\rho)$ can be obtained by a simple integration 
and read
\be
\phi_l^4(\rho) = -\frac{a}{2b \rho}\phi_l^3(\rho) 
+ \frac{1}{2b} \int_{\rho_l}^{\rho} \frac{du}{u}\, 
\frac{S_l(h)(u)}{h(u)^l} + d_l\;.
\label{ren_sol2}
\ee
Again integration constants $d_l = \phi_l^4(\rho_l)$, appear 
which can be put to zero without loss of generality in the following 
sense: 

The point to observe is that whenever $\phi_l^j(\rho)$ contains an additive 
contribution proportional to a conformal Killing vector ${\bf v}^j$ of 
$g_{ij}(\phi)$, one can trade it for an additive contribution 
$\frac{1}{2} \nabla_{\!j} {\bf v}^j$ 
to $H_l(\rho)$. In other words the solution of the finiteness condition
(\ref{Hfinite}) contains an ambiguity in that certain pieces can be 
shuffled from the Lie derivative term to the term proportional to the 
metric. The ambiguity is linked to and parameterized by the conformal
Killing vectors of $g_{ij}$. The conformal Killing vector ${\bf t}_+$ 
in the $(\Delta,B,\rho,\sigma)$ coordinates is a linear combination 
of $(0,0,\rho,0)$ and the Killing vector $(0,0,0,1)$ generating 
translations in $\sigma$. Clearly the ambiguities induced by such 
linear combinations via the above mechanism just correspond to 
the arbitrariness in the integration constants in 
$\phi_l^3(\rho), \,\phi_l^4(\rho)$. Effectively the 
above criterion to fix the integration constants thus amounts to 
removing any part in $\phi_l^j$ proportional to the ${\bf t}_+$ 
conformal Killing vector. After allowing $\phi_l^4$ to also depend 
linearly on $\sigma$ the conformal Killing vector ${\bf d}$ 
could similarly be used to add multiples of $\ln\rho \cdot  
\rho \dd_{\rho} \ln h $ to $H_l(\rho)$.

This completes the solution of the finiteness condition 
(\ref{Hfinite}). The crucial renormalization is that of the scale 
factor in (\ref{RBmetric}) where, subject to the above specifications, 
the $H_l(\rho)$ are uniquely determined functionals of $h$.
Since the residues of the higher order poles are determined 
by those of the first order poles this structure will carry 
over to the entire divergent part of (\ref{RBmetric}), (\ref{phiren}),
and (\ref{gbare}). Together these renormalizations guarantee the 
existence of a well-defined renormalized action. In section 4.3
we describe how this extends to the renormalization of 
correlation functions.

Before taking this up let us evaluate the $l=1,2$ renormalization 
functions $H_l(\rho)$ and $\phi^3_l(\rho)$ for the Ernst system 
proper, where $h(\rho) =\rho$. One gets $H_1(\rho) = 0$, 
$H_2(\rho) = 1/(4\rho^2)$ and $\phi^3_1(\rho) = -\eps$, 
$\phi^3_2(\rho) = 1/(4\rho)$. In particular this means the Ernst 
system is renormalizable in the conventional sense at the 1-loop level
but beyond that only in the above `conformal' sense. 
More generally one finds 
\be
H_1(\rho) \equiv 0\quad \;\mbox{iff}\; \quad h(\rho) \sim \rho^p 
\quad \mbox{with} \quad \left\{ 
\begin{array}{ll} p> 0 & \mbox{and} \;\;\rho_1 = \infty\,,\\
                  p< 0 & \mbox{and} \;\;\rho_1 = 0\,.
\end{array} \right.
\label{fixp1}
\ee
Thus, also if $h$ is a generic power of $\rho$ the system 
remains strictly renormalizable at the 1-loop level. 
One can also verify that apart from the constant 
there are no other $h$ functions with that property.  
The $p>0$ and the $p<0$ sector can in principle be related by a 
field redefinition of the form (\ref{coord_change4}) with 
$b/\tilde{b} <0$. Both sectors turn out to be equivalent;
we consider the $p>0$ sector throughout.

A bonus of $h(\rho) \sim \rho^p$ is that the 
1-loop field renormalizations are gradients of a `potential'.
For $p >0$ one finds 
\ba
&\nspace & \phi_1^j = \left( 0,0,-\frac{\eps}{p}\rho^{-p+1} ,
\frac{2 a \eps -p^2}{4 b p} \rho^{-p} + d_1\right)\;,
\nonum
&\nspace & \phi_1^j = -\dd^j \Phi_1\;, \quad  
\Phi_1(\rho,\sigma) = \frac{1}{4p}[2 \eps a + p^2]\, 
\ln \rho -\frac{b d_1}{p}\, \rho^p + \frac{\eps b}{p}\,\sigma\,,
\label{fixp3}
\ea 
where $d_1$ is the integration constant entering through the solution 
of (\ref{ren_sol2}) and in $\Phi_1$ an irrelevant additive constant 
has been omitted. 
The constant $d_1$ corresponds to the before mentioned ambiguity 
in the solution of the finiteness condition associated with the 
Killing vector ${\bf t}_- = \dd_{\sigma}$, and can be set to zero.

Eqn.~(\ref{fixp3}) is also convenient to discuss the relation to 
the observation of de Wit et al \cite{dWGNR92} that the 1-loop counter term 
in the Ernst system is a total divergence on-shell (with respect to 
the base space). In view of the diffeomorphism identity (\ref{diffward1})
this is equivalent to $R_{ij} = \cL_{\phi_1} g_{ij}$, for 
some field renormalization vector $\phi_1^j(\phi)$. Since $H_1(\rho) =0$
for $h(\rho)=\rho$ this of course is in agreement with our result. 
If the 4D target space was compact one could also infer from
a general theorem by Bourguinion (reviewed in \cite{Polch}) that 
$\phi_1^j$ is the gradient of a scalar without actually 
computing it. In the case at hand the target space is non-compact 
(with non-zero curvature) and the fact that $\phi_1^j$ nevertheless 
comes out to be the gradient of a scalar is non-trivial. It is also 
crucial for the physics of the system in that $\phi_1^j \sim \dd^j \Phi_1$ 
is a necessary condition for conformal invariance \cite{HullTown86}.
We should also mention that a class of Riemannian sigma-models 
with a target space of Minkowski signature and a null Killing vector
has been studied by Tseytlin \cite{Tseyt93a,Tseyt93b}. 
However both the setting and the results are not directly related
to ours.

\newsubsection{Essential couplings through finite quantum deformations}

So far $h(\rho)$ has been treated as $\lb$-independent. Motivated by 
the analogy to a generalized coupling we now allow it to be of 
the form%
\footnote{I am indebted to P.~Forg\'acs for suggesting this.}
\be
h(\rho, \lb) = h_0(\rho) + \frac{\lb}{2\pi} h_1(\rho) + 
\Big(\frac{\lb}{2\pi} \Big)^2 h_2(\rho) + \ldots \;.
\label{hlb}
\ee
As indicated we use $h_0(\rho)$ to denote a 
$\lb$-independent prefactor in (\ref{Tgmetric}) and $h(\rho,\lb)$ for one 
of the form (\ref{hlb}). Here (the renormalized=bare) $\lb$ serves 
as the loop counting parameter. The `adjustable' functions 
$h_l=h_l[h_0],\,l\!\geq \!1$, are regarded as (local) functionals 
of $h_0$. We shall assume $h_0(\rho) \sim \rho^p, \,p\!>\!0$, 
throughout as this ensures 1-loop renormalizability. Moreover we 
consider only genuine deformations where not all of the $h_l(\rho)$, 
$l\! \geq \!1$, are again proportional to $\rho^p$.

In addition to being technically rather natural the deformation (\ref{hlb}) 
has a profound physical significance. In Weinberg's terminology 
\cite{Weinberg} it replaces the ``inessential'' coupling $h_0(\rho) 
\sim \rho^p$ by an ``essential'' coupling function $h(\,\cdot\,,\lb)$. 
Roughly speaking an inessential coupling is one whose flow is 
affected by field redefinitions and which may continue 
to run even at a fixed point. A simple test is to compute the 
variation of the Lagrangian with respect to the bare quantity.
If it comes out a total divergence modulo the equations of motion
the quantity is an ``inessential coupling''. Applied to the 
Einstein-Hilbert action this criterion disqualifies Newton's
constant as an inessential coupling \cite{Weinberg}. It is only
if one includes a cosmological constant term and/or higher order 
curvature scalars that the ratios of their prefactors become 
``essential couplings'' -- in the space of which one can search for 
a fixed point \cite{Reuter02a,Reuter02b}. In the context of 
the 2-Killing reduction the deformation (\ref{hlb}) achieves 
precisely the same: Since $h \frac{\dd}{\dd h} L = L = - \lb 
\frac{\dd}{\dd \lb} L$ both the reduced Newton  
constant $\lb$ and $h(\,\cdot\,)$ will be inessential if 
$L$ can be written as a total divergence on-shell. Going back 
to Eqs.~(\ref{cward1}) one sees that this is the case iff 
$h(\rho) \sim \rho^p$ or $h(\rho) \sim \ln \rho$. Since we 
insist on 1-loop renormalizability and exclude trivial deformations 
in (\ref{hlb}) the set of functions $h(\,\cdot\,,\lb)$ qualifies as an 
{\it essential coupling}. In particular the space of these functions
is the appropriate arena to search for a fixed point. We shall
take up the search in section 5. Our immediate concern though is
to determine the impact the functions $h_l,\,l \!\geq \!1$, 
have on the renormalization of the system(s).

We begin by examining the effect of (\ref{hlb}) on the solution
(\ref{ren_sol1}) of the finiteness condition. In a renormalizable 
quantum field theory finite coupling redefinitions correspond 
to a change of scheme. Their impact can be studied simply 
by substituting into the solution computed in the original scheme 
and re-expanding in powers of the loop counting parameter. Since in the 
case at hand non-algebraic manipulations were involved in arriving at 
(\ref{ren_sol1}) we made sure that such a substitution procedure 
is legitimate also here by going back to the finiteness conditions. 
That is we used (\ref{hlb}) in the renormalized metric and the counter 
tensors from the beginning and expanded in powers of $\lb$ to arrive 
at a modified set of finiteness conditions (\ref{Hfinite}), which 
is then solved as before.  

The general finiteness condition for the $\lb$-dependent quantities 
is obtained along the same lines as before, i.e.~by combining 
the general identity
\ba
&& g^{\b}_{ij}(\phi_{\b}) \dd^{\mu}\phi_{\b}^i \dd_{\mu} \phi_{\b}^j =
g^{\b}_{ij}(\phi) \dd^{\mu}\phi^i \dd_{\mu} \phi^j +
\frac{1}{2-d}\,\cL_{\Xi(\phi,\lb)} g^{\b}_{ij}(\phi) 
\,\dd^{\mu}\phi^i \dd_{\mu} \phi^j \;,
\nonum
&& \phi^j_{\b} = \phi^j + \frac{1}{2-d}\, \Xi^j(\rho,\lb) \;,
\sspace \Xi^j(\rho,\lb) := \sum_{l \geq 1} \Big(\frac{\lb}{2\pi} \Big)^l
\phi_l^j(\rho)\;,
\label{xigRB}
\ea
with the relation between the bare and the renormalized metric 
functional (\ref{RBmetric}) specific for the Ernst system 
\be
g_{ij}^{\b}(\phi) = \mu^{d-2} \,g_{ij}(\phi,\lb)\left[
1 + \frac{1}{2-d} H(\rho,\lb) + \ldots \right]\,,
\quad H(\rho, \lb) := \sum_{l \geq 1} \Big(\frac{\lb}{2\pi} \Big)^l 
\,H_l(\rho)\;.
\label{RBlbmetric}
\ee
The renormalized metric here depends on $\lb$ through
a prefactor $h(\rho,\lb)$ of the form (\ref{hlb}). Again the 
arguments on {\it both} sides of (\ref{RBlbmetric}) are the 
renormalized fields. The finiteness condition obtained from 
(\ref{xigRB}), (\ref{RBlbmetric}) reads 
\be
\cL_{\Xi(\rho,\lb)} g_{ij}(\phi,\lb) + H(\rho,\lb) g_{ij}(\phi,\lb) 
= \lb T_{ij}^{(1)}(g(\phi,\lb)/\lb)\;,
\label{Hlbfinite}
\ee 
with $T^{(1)}_{ij}(g)$ from Eqs.~(\ref{Tcounter1}), (\ref{nulpieces}). 
If one now expands in powers of $\lb$ the $l=1$ equation coincides with 
(\ref{Hfinite}) but the $l \geq 2$ equations are modified. The solutions 
$H_l(\rho)$, $\phi_l^j(\rho)$ will depend on $h_0, \ldots, h_{l-1}$. 
We won't need the explicit form of the modified $l\! \geq \!2$ finiteness
conditions because eventually their solution turns out to  
coincide with that of the substitution procedure, despite the 
fact that non-algebraic manipulations are involved.
Strictly speaking this holds only if the integration constants
$\rho_l$ are assumed to be equal, otherwise 
some trivial ambiguities have to be taken into account.

The result can be summarized by saying that simply substituting
$h$ for $h_0$ and re-expanding in powers of $\lb$ produces the 
correct solution of the modified finiteness conditions.
For illustration let us quote the explicit 
three-loop results for the solution of the finiteness condition
for  $h(\rho,\lb)$ of the form (\ref{hlb}): 
The one loop solutions are unchanged, 
i.e.~are given by (\ref{ren_sol1}) with $h(\rho) = h_0(\rho)$. 
The two and three loop coefficients are modified according to 
\begin{subeqnarray}
\phi_2^3(\rho) & = & \rho \int_{\rho}^{\infty} 
\frac{du}{u h_0(u)^2} [ \zeta_2 - \zeta_1 h_1(u)]\,,
\\
\phi_3^3(\rho) & = & \rho \int_{\rho}^{\infty} 
\frac{du}{u h_0(u)^3} [ \zeta_3 - 2 \zeta_2 h_1(u) - 
\zeta_1 (h_2 h_0 - h_1^2)(u) ]\,,
\\
H_2(\rho) &=& - \frac{1}{h_0} \rho \dd_{\rho}
\Big(\frac{h_0 \phi^3_2}{\rho}\Big) 
- \phi_1^3(\rho) \dd_{\rho}\Big(\frac{h_1}{h_0} \Big)\,, 
\\
H_3(\rho) &=& - \frac{1}{h_0} \rho \dd_{\rho}
\Big(\frac{h_0 \phi^3_3}{\rho}\Big) 
-\phi_2^3(\rho) \dd_{\rho}\Big(\frac{h_1}{h_0} \Big) 
+ \phi_1^3(\rho) \dd_{\rho} \Big\{\frac{1}{2} 
\Big(\frac{h_1}{h_0} \Big)^2 - \frac{h_2}{h_0} \Big\}\,. 
\label{Hxichange}
\end{subeqnarray} 
Here we took $\rho_l = \infty$ for all the 
integration constants. Observe that the new solutions obey the 
scaling (\ref{Hxiscaling}) if $h_l$ is assigned scaling dimension 
$1\!-\!l$. For the existence of the integrals in (\ref{Hxichange}a,b)
only a mild constraint on the large $\rho$ asymptotics is needed,
for example 
\be
\frac{h_l}{h_0^{l+1}} = o\Big(\frac{1}{\ln^{1+p} \rho} \Big)\,,\;\;
p>0 \quad 
\mbox{for} \quad \rho \ra \infty\,,\quad  l\geq 1\,,
\label{hasympt1}
\ee
with $1/h_0 \sim o(1/\ln^{1+p}\rho)$ is sufficient. As stated earlier we 
specifically take $h_0(\rho) \sim \rho^p,\,p>0$, throughout as this 
ensures 1-loop renormalizability. We also assume that derivatives 
are well-behaved, e.g.~$\dd_{\rho} h = O(h)$, for large $\rho$, etc.

Having justified the the substitution procedure we can 
use (\ref{ren_sol1}) to obtain closed expressions for $H(\rho,\lb)$ 
and $\Xi^j(\rho,\lb)$ valid also for $h(\rho,\lb)$ of the form
(\ref{hlb}). The counter terms can be written as 
\be
\lb T^{(1)}_{ij}(g/\lb) = {\rm diag}\left(
\frac{\eps h}{\Delta^2} B_{\lb}\Big(\frac{\lb}{h}\Big),\;
\frac{h}{\Delta^2} B_{\lb}\Big(\frac{\lb}{h}\Big),\;
\frac{h}{\rho^2}S(\rho,\lb),\; 0\right)\,,
\label{Tlambda} 
\ee
where $S(\rho,\lb)  = \sum_{l \geq 1} (\frac{\lb}{2\pi})^l h^{-l} 
S_l(h)$. Further 
\be
B_{\lb}(\lb) = \sum_{l \geq 1} \zeta_l \Big(\frac{\lb}{2\pi} \Big)^l
= \int_0^{\lb} \frac{d s}{s^2} \beta_{\lb}(s) \;,
\sspace  
\beta_{\lb}(\lb) = \lb^2 \frac{\partial}{\partial\lb} 
\sum_{l \geq 1} \zeta_l \Big(\frac{\lb}{2\pi} \Big)^l \,,
\label{Blb}
\ee
is related to the beta function of the flat space O$(1,2)$ sigma-model 
with target space $H_{\eps}$, computed in the minimal subtraction 
scheme. The solutions of (\ref{Hlbfinite}) come out as 
\ba 
\label{HXi}
H(\rho,\lb) \is - \frac{1}{h(\rho,\lb)} 
\rho \dd_{\rho} \left[h(\rho,\lb) \frac{\Xi^3(\rho,\lb)}{\rho}\right]\,.
\nonum
\Xi^3(\rho,\lb) \is \rho \int_{\rho}^{\infty}\frac{du}{u} 
B_{\lb} \!\left( \frac{\lb}{h(u,\lb)} \right)\,,
\\[2mm]
\Xi^4(\rho,\lb) \is -\frac{a}{2 b \rho} \Xi^3(\rho,\lb) + 
\frac{1}{2b} \int^{\rho} \frac{du}{u} S(u,\lb)\;,
\nonumber
\ea
where again $\rho_l \!=\! \infty,l\! \geq \!1$, was assumed, and  
in $\Xi^4$ a $\lb$-dependent integration constant was absorbed into the 
lower integration boundary. These expressions generalize 
(\ref{ren_sol1}), (\ref{ren_sol2}).

The condition (\ref{hasympt1}) also implies that $h(\rho,\lb)$ 
grows for $\rho \ra \infty$ at least like $\rho^p\,,p>0$, though 
off-hand it could grow much faster. The counter term
tensor (\ref{Tlambda}) thus has a finite and universal limit
\be
\lb T^{(1)}_{ij}(g/\lb) \rra 
\frac{\lb}{2\pi}{\rm diag}
\left(\eps\frac{\zeta_1}{\Delta^2}, \frac{\zeta_1}{\Delta^2}
\frac{S_1(\rho)}{\rho^2},0 \right)\quad \mbox{for} \quad \rho \ra \infty\,,
\label{Tlambda_asympt} 
\ee
where the subleading terms are down by a power of $1/h$. 
Geometrically $R(g) = -2\eps/h(\rho)$ is the scalar curvature of 
the metric (\ref{Tgmetric}) so that the limit (\ref{Tlambda_asympt})
corresponds to weak curvature. (For $p<0$ the relevant limit would 
be $\rho \ra 0$.) Since (\ref{Tlambda_asympt}) coincides with the 
1-loop counter term one can read off the asymptotic solution 
$H_{\infty}, \Xi_{\infty}^j$ of the finiteness condition from the 
1-loop results. There is no reason to introduce an ad-hoc $\lb$-dependence 
into the asymptotics $h_{\infty}(\rho)$ of $h(\rho)$. Assuming it to be 
$\lb$-independent enforces $h_{\infty}(\rho) = h_0(\rho) \sim \rho^p$ 
and the solution triggered by (\ref{Tlambda_asympt}) is 
\be 
H_{\infty} =0\,,\sspace 
\Xi^j_{\infty} = \frac{\lb}{2\pi}\phi_1^j\,,
\label{HXi_p_asympt} 
\ee  
with $\phi_1^j$ given by (\ref{fixp3}).  

\newsubsection{Renormalized currents}

In the previous sections we have been concerned with the 
renormalization of the basic Lagrangian. Of course eventually one 
is interested in constructing correlation functions of suitable composite
operators, where the Lagrangian (co-)determines the perturbative 
measure. Since for the Ernst-like systems (\ref{L4Dconf}) the 
classical observables are built from the Noether currents and $h$ 
it seems natural to primarily aim at renormalizing their correlation 
functions. In addition the constraints ought to be constructed 
as composite operators. This can be achieved 
by including suitable {\it local} sources in the Lagrangian
such that after renormalization the composite operators can be 
obtained by functional differentiation. The sources of course 
are likewise subject to renormalization
and the problem consists in showing that they can be included 
in a way that preserves the ``conformal renormalizability'' of the 
system in the sense introduced above. Technically it is again 
convenient to borrow results from the renormalization of 
Riemannian sigma-models. We shall use the results and the 
notations of appendix B throughout.  
 
In a first step we determine the analogue 
of the source-extended Lagrangian (\ref{Lsource}) appropriate 
for the Ernst-like systems. In the purely metric part a 
local source can be included by making $h(\rho)$ explicitly 
$x$-dependent: $h(\rho) \ra h(\rho;x)$. This manifestly 
preserves the (conformal) symmetries of the target space 
metric (\ref{Tgmetric}), in particular $\overline{\dd}_{\mu} 
g_{ij} = \overline{\dd}_{\mu}\!\ln h \, g_{ij}$. The vector 
sources should evidently respect the ${\rm O}(1,2)$ symmetry. 
For the moment we are only interested in renormalizing the 
four Noether currents (\ref{O12curr}) and (\ref{sigma_curr}) 
possibly multiplied by functions of $\rho$. 
We thus introduce an ${\rm O}(1,2)$ vector source 
$\omega_{\mu}^j(\rho;x),\,j=0,1,2$, and a scalar source 
$\omega_{\mu}(\rho;x)$. As indicated both are vectors on the 
base space, functions of $x$ and functionals of $\rho$. 
The corresponding source term is $\omega_{\nu}\cdot J_{\mu} + \omega_{\nu} 
J_{\mu}({\bf t}_-)$ and replaces $V_{\nu i} \dd_{\mu} \phi^i$ 
in (\ref{Lsource}). Since we describe the Ernst-like systems 
in terms of an action on a flat base space there is in principle 
no need to minimally couple the system to an external 
background metric $\gamhat_{\mu\nu}$. In practice though a source 
term $R^{(2)}(\gamhat) \Phi$ is a convenient tool to generate 
an improvement term for the energy momentum tensor (i.e.~to 
the constraints). Anticipating that such an improvement term
is needed later, we minimally couple the original system to a
fiducial background metric $\gamhat_{\mu\nu}$ on the base space
and include a source term $R^{(2)}(\gamhat) \Phi$. 
It is not hard to see (c.f.~section 6) that the interpretation of 
$\Phi$ as a potential for the improvement term constrains it 
to be of the form $\Phi = f(\rho) + f_0 \sigma$, 
for a constant $f_0$. Finally $F$ mainly serves as a tool to 
determine the generalized ``wave function renormalizations'' and 
will be chosen accordingly. Including in it a term quadratic in the 
fields also provides an infrared regulator. In summary we arrive at 
the following source-extended Lagrangian for the Ernst-like systems 
\ba
\lb L(G; \phi) \is \frac{h(\rho;x)}{2 \rho} 
\widehat{g}_{ij}(\phi)\gamhat^{\mu\nu}  
\dd_{\mu} \phi^i \dd_{\nu} \phi^j + 
\gamhat^{\mu\nu} [\omega_{\nu}\cdot J_{\mu} + \omega_{\nu} 
J_{\mu}({\bf t}_-)]
\nonum  
&+& \frac{1}{2} R^{(2)}(\gamhat)[f(\rho) + f_0\, \sigma] + F(\phi)\,. 
\label{LEsource}
\ea    
Here $G$ now stands for the collection of generalized couplings/sources 
$\{h,\omega_{\mu},f,F\}$. 

As a simple illustration let us compute the ``wave function'' 
renormalization of the spin fields $n^j,\,j=0,1,2,$ to lowest order.  
In the present framework they are treated as composite operators  
$n^j = n^j(\Delta,B)$ according to (\ref{nspins}). Taking as scalar source 
$F(\phi) = \omega(\rho;x)\, l_j n^j(\Delta,B)$, for a singlet $\omega(\rho;x)$ 
and a constant ${\rm O}(1,2)$ vector $l_j$ one finds from 
(\ref{Rcounter1}) and  Eq.s~(\ref{frho_nonren}), (\ref{phiren}), 
(\ref{fixp3}) 
\be 
n_{\b}^j = n^j \left[ 1 - \frac{1}{2 -d} \,\frac{\lb}{2\pi} 
\left( \frac{\eps}{h(\rho)} + \dd_{\rho} \ln \omega(\rho)\,  \phi^3_1(\rho) 
\right) + \ldots \right]\,. 
\label{nspin_ren} 
\ee  
In the decoupling limit of constant $h(\rho) = 1/\lb_0$ and constant 
source $\omega$ this correctly reproduces the leading wave function 
renormalization of the spins in the ${\rm O}(1,2)$ sigma-model 
without coupling to gravity.   

Next we consider the renormalization of the Noether currents. In a 
renormalizable quantum field theory conserved Noether currents are not
renormalized in the sense that the coupling and field renormalizations
are enough to render them finite and conserved as composite 
operators. The goal in the following is to derive an analogous
result for the Ernst-like systems which are renormalizable 
only in the broader ``conformal'' sense.   
In order to explain the result we write $J_{\mu}^i(h;\phi)$ 
for the ${\rm O}(1,2)$ Noether currents (\ref{O12curr}) and 
$J_{\mu}(h;\rho)$ for the Noether current (\ref{sigma_curr}) associated 
with the $\sigma$-translations. Then the following ``non-renormalization''
results hold to all orders in the loop expansion:
\ba
&& \nl J_{\mu}^i(h;\phi)\nr = J^i_{\mu}(h_{\b};\phi_{\b})\,,\quad 
i=0,1,2\,,
\nonum
&& \nl J_{\mu}(h;\rho) \nr = \mu^{d-2} J_{\mu}(h;\rho)\,.
\label{curr_nonren}
\ea   
The normal products are defined in Eq.~(\ref{normalproducts});
in general evidently the functional form of a dimension 1 
operator will change under renormalization. For the ${\rm O}(1,2)$ 
Noether currents however this is not the case and (\ref{curr_nonren}) 
states that they can be rendered finite and conserved as 
composite operators by renormalizing the fields and the generalized
coupling $h(\,\cdot\,)$. The even stronger result for $J_{\mu}(h;\rho)$ 
is of course related to (\ref{frho_nonren}).

To show (\ref{curr_nonren}) we employ the consistency 
conditions (\ref{diffconsist}) entailed by the diffeomorphism Ward 
identity. Combined with (\ref{Rcounter1}) the first relation 
implies $[g_{ij} + T_{ij}(g)]{\bf v}^j = Z^V(g)_i^j {\bf v}_j$, for a 
Killing vector ${\bf v}^j$. Further, as noted in section 3.3, the 
Killing vectors are eigenvectors of the metric counter terms $T_{ij}(g)$,
in the case of ${\bf t}_-$ with vanishing eigenvalue. By the above 
consistency condition this carries over to $Z^V(g)$: 
\ba
Z^V(g)_i^j ({\bf t}_-)_j \is ({\bf t}_-)_i \,,
\nonum
Z^V(g)_i^j {\bf v}_j \is \left[ 1 + \frac{1}{2-d} 
B_{\lb}\Big(\frac{\lb}{h} \Big) + \ldots \right] 
{\bf v}_i\,,\quad {\bf v} = {\bf e},{\bf h}, {\bf f}\,,
\label{Z_Killings}
\ea  
where (\ref{Tlambda}) has been used. The second equation in 
(\ref{diffconsist}) implies for a Killing vector 
\be 
{\bf v}_i \cdot \frac{\dd Y}{\dd V_i^{\mu}} = 
{\bf v}^i Z^V(g)_i^j V_{\mu j} - Z({\bf v}^i V_{\mu i}) + 
{\bf v}^i N_i^{\; jk}(g) \overline{\dd}_{\mu} g_{jk}\,.
\label{Y_Killings}
\ee       
The last term on the right hand side is readily seen to vanish, 
for the other two we recall that the sources $V_{\mu i}$ 
relevant for the Noether currents are proportional to the 
respective Killing vector. The proportionality factor is a 
function of $\rho$ and $x$ only and can be pulled out of the 
differential operators in $Z^V(g)$ and $Z(g)$. In the end 
it is set to zero and kills the right hand side of (\ref{Y_Killings}).
Thus only the counter terms in (\ref{Z_Killings}) remain.
For $J_{\mu}(h;\rho)$ this directly gives the second equation in 
(\ref{curr_nonren}). 
The counter terms for the ${\rm O}(1,2)$ currents are obviously 
the ones associated with the metric renormalization in  
(\ref{gbare}), but they can also be interpreted as a generalized 
coupling renormalization via 
\be 
h_{\b}(\rho_{\b}) = \mu^{d-2} h(\rho) \left[ 1 + \frac{1}{2-d} 
B_{\lb}\Big(\frac{\lb}{h} \Big) + \ldots\right]\,.
\label{hBrhoB}  
\ee
This follows from Eq.~(\ref{RBh}) below, the $\rho_{\b}$ renormalization in 
(\ref{xigRB}), (\ref{HXi}), and (\ref{Hlbfinite}). The purely 
$(\Delta,B)$ dependent part of the currents remains unrenormalized 
and one arrives at the first equation in (\ref{curr_nonren}).

\newpage
\newsection{Flow equations}

The renormalization until here was performed at a fixed normalization 
scale $\mu$. Changing the scale leaves the bare quantities unaffected
but the renormalized ones have to compensate for it by carrying
a $\mu$-dependence. It turns out that both the function 
$h(\,\cdot\,,\lb)$ and the fields $\rho,\,\sigma$ are subject to 
nontrivial flow equations. The former is analogous to the running 
coupling in an ordinary quantum field theory. The latter is 
induced by the $h$-dependence of the renormalized fields and 
generalizes the concept of an anomalous dimension matrix.

\newsubsection{Gravitationally dressed beta function} 

Recall that the function $h$ in (\ref{Tgmetric}) could 
be prescribed at will and constituted part of the specification 
of the quantum theory. The same is true for $h(\,\cdot\,,\lb)$ of the 
generic form (\ref{hlb}); in order not to clutter the notation 
we will often suppress the $\lb$-dependence in the following.   
As $\mu$ changes the functional form of $h$ can in general 
not be maintained. Rather $h(\,\cdot\,)$ has to become a function 
$\hbbar(\,\cdot\,,\mu)$ of the normalization scale $\mu$ which 
is analogous to the running coupling in an ordinary field theory. 
Of course all functions connected by varying $\mu$ must be regarded 
as equivalent and do not define different theories.

A natural way to define a beta function for $h(\rho)$ is to 
interpret (\ref{RBlbmetric}) as a relation between the bare and 
the renormalized scale factor
\be
h^{\b}(\rho) = \mu^{d-2} h(\rho,\lb) \left[1 + \frac{1}{2-d} H(\rho,\lb) 
+ \ldots\right]\,,
\label{RBh}
\ee
allowing now the renormalized $h$ to be of the generic form (\ref{hlb}).
Following the usual procedure yields the beta function
\be
\lb \beta_{h}(h/\lb) = (2-d) h(\rho) - h(\rho) \int \!du\, h(u) 
\frac{\delta H(\rho,\lb)}{\delta h(u)}\,,
\label{hbeta1}
\ee
where we suppress the $\lb$-dependence of $h$. 
Since the functional derivative in (\ref{hbeta1}) will 
frequently reappear we introduce the shorthand
\be
\dot{X}(\rho) := \int du \,h(u) \frac{\delta X(\rho)}{\delta h(u)}\;,
\label{Xdot}
\ee
for a functional $X(\rho)= X[h](\rho)$ of $h(\rho)$. Observe that for any 
differential or integral polynomial $X_l$ in $h$ which is homogeneous of 
degree $l$, the functional derivative (\ref{Xdot}) just measures the degree, 
$\dot{X}_l = l X_l$.%
\footnote{As a warning let us add that this simple rule applies 
only if $h$ is unconstrained. For example in taking the `$\,\cdot\,$'
derivative of Eq.~(\ref{XiPhi}) one has to take into account that 
$\rho$ is functionally dependent on $h$.}
In particular the `$\,\cdot\,$' derivatives
of the solution (\ref{HXi}) of the finiteness condition will be needed
frequently and come out as 
\ba
\label{HXidot}
\dot{H} \is - \frac{1}{h} \rho \dd_{\rho} \left[ \frac{h}{\rho}
\dot{\Xi}^3\right]\,,
\nonum
\dot{\Xi}^3 \is - \rho \int_{\rho}^{\infty} \frac{du}{u} 
\frac{h(u)}{\lb} \beta_{\lb} \left( \frac{\lb}{h(u)} \right) \,,
\\[2mm] 
\dot{\Xi}^4 \is - \frac{a}{2b \rho} \dot{\Xi}^3 + \frac{1}{2 b} \int^{\rho} 
\frac{du}{u} \dot{S}(u,\lb)\;.
\nonumber
\ea 
In $\dot{\Xi}^4$ we absorbed a $\lb$-dependent additive constant 
into the lower integration boundary and used 
$\dot{S}(\rho,\lb)  = -\sum_{l \geq 1} (\frac{\lb}{2\pi})^l l h^{-l} 
S_l(\rho)$, as $\dot{S}_l(\rho) =0$. For the $\beta_h(h)$ function 
(\ref{hbeta1}) this yields explicitly 
\be
\lb \beta_{h}(h/\lb) = (2-d)h - \rho \dd_{\rho} 
\left[ h \int_{\rho}^{\infty} \frac{du}{u} \frac{h(u)}{\lb} \beta_{\lb} 
\Big(\frac{\lb}{h(u)} \Big) \right]\;. 
\label{hbeta2}
\ee
Interestingly $\beta_h(h)$ comes out to be a total $\ln \rho$-derivative. 
Further the functional beta function for $h$ 
is completely determined by the conventional beta function in 
(\ref{Blb}) of the ${\rm O}(1,2)$ sigma-model in flat space, and thus 
can be viewed as a ``gravitationally dressed'' version of the latter. 
A similar concept was (in a somewhat different context and at 
the 1-loop level) employed in \cite{KlKoPo93}, from which we borrow 
the term. See also \cite{PSZ97}. Eq.~(\ref{hbeta2}) is a structural 
result valid to all loop orders. The corresponding flow equation is 
\be
\mu \frac{d}{d \mu} \hbbar  
= \lb \beta_{h}(\hbbar/\lb)\quad \mbox{with} \quad
\hbbar(\rho,\mu_0) = h(\rho)\,.
\label{hsflow}
\ee

Before discussing its properties let us briefly comment on 
the relation of $\beta_h(h)$ to the tensorial beta function 
in (\ref{Gflow}). To find the relation one operates with 
$1 - \int\!\mbox{\small\it h} \frac{\delta}{\delta h}$ on both sides of 
(\ref{Hlbfinite}). Using $\dot{T}_{ij}^{(1,l)}(g) = 
(1-l) T_{ij}^{(1,l)}(g)$ this gives
\be
(d-2) g_{ij} + \lb \beta_{ij}(g/\lb) = - h [ h^{-1} \lb T_{ij}^{(1)}
(g/\lb)]^{{\displaystyle\dot{}}} 
= - \dot{H} g_{ij} - \cL_{\dot{\Xi}} g_{ij}\,.
\label{ts_beta}
\ee
One sees that the piece proportional to $g_{ij}/h$ gives back 
the $\beta_h(h)$ function, while the Lie derivative term is induced
by the nonlinear field renormalizations. However 
a naive transcription of the tensorial flow equation (\ref{Gflow}) 
would {\it not} give rise to consistent equations for the individual 
metric components. This is because in parameterizing the 
transition between the bare and the renormalized quantities via 
equations (\ref{RBmetric}), (\ref{phiren}), we also decided to 
treat $h$ as a generalized coupling on which the 
nonlinear field renormalizations depend on. For a generic Riemannian 
sigma-model on the other hand only redefinitions (\ref{diff10}) 
of fields are considered that are independent of the renormalized 
metric. Indeed, even if one would accept the odd feature that the 
coordinates on the target space are co-determined by the metric 
tensor (and vice versa) it would be impossible to disentangle the combined 
$\mu$-dependence in $g_{ij}(\phi[g](\mu),\mu)$ with respect 
to the `moving' coordinates $\phi^j[g](\mu)$.

Next let us verify that (\ref{hsflow}) gives sensible answers in 
two simple special cases. The first one is the decoupling limit 
where $h$ equals a constant. For constant $h$ the 
$\dd^{\mu} n \cdot \dd_{\mu} n$ 
term in the action based on (\ref{Tgmetric}) decouples from 
the $\dd^{\mu} \ln \rho \,\dd_{\mu} (\sigma + \frac{1}{2}\ln \rho)$ 
term. The field redefinitions of the latter do not effect the former 
and one expects to recover the ordinary beta function for a sigma 
model with the 2D hyperbolic target space (\ref{hyp}). 
This is indeed the case provided (\ref{hsflow}) is 
interpreted as a flow equation for the proportionality 
factor. (Recall that for the Ernst-type systems $\lb$ does not get 
renormalized and, at fixed $\mu$, just serves as a loop counting 
parameter while for constant $h$ the coupling does get renormalized and is 
conceptually distinct from the loop counting parameter). 
For $h = {\rm const} = 1/\lb_0$ the $\beta_h(h)$ function of 
Eq.~(\ref{hbeta2}) evaluates to 
\be
\beta_h(h/\lb) \bigg|_{h = 1/\lb_0} = 
\frac{1}{(\lb \lb_0)^2} \beta_{\lb}(\lb \lb_0)\,.
\label{beta_hlb} 
\ee
As expected the coupling $\lb_0$ only occurs in the combination
$\lb \lb_0$ and thus can serve in itself as a loop counting 
parameter. Putting $\lb$ equal to unity the flow equation
(\ref{hsflow}) becomes
\be
h(\rho) = \frac{1}{\lb_0}: \sspace 
\mu \frac{d\lbbar_0}{d\mu} = - \beta_{\lb}(\lbbar_0) = 
- \frac{\lbbar_0^2}{2\pi}\left(-\eps + 
\frac{\lbbar_0}{2\pi} \right) + \ldots\,,
\label{lbflow}
\ee
which is the correct flow equation for a flat space ${\rm O}(1,2)$ 
sigma-model with target space $H_{\eps}$. As usual the dependence on the 
renormalized coupling $\lb_0$ enters the solution 
of (\ref{lbflow}) only through the initial condition $\lbbar(\mu_0) = 
\lb_0$, for some $\mu_0$.

Another special case is the abelian subsector where $B\equiv 0$. 
This amounts to deleting the second row and column in the target space 
metric (\ref{Tgmetric}) upon which its scalar curvature vanishes.
Repeating the previous computations one finds that the counter tensors 
are of the form
\ba
&& T^{(1,l)}_{ij}(g) = {\rm diag}(0, \rho^{-2} h^{1-l}(\rho) 
S_l(\rho) ,0)\,,\sspace {\rm with}\nonum
&& S_1(h) = -\frac{1}{2} (\rho \dd_{\rho})^2 \ln h 
+ \frac{1}{4} (\rho \dd_{\rho} \ln h)^2\,,\quad 
S_2(h) = S_3(h) =0\,.
\label{ERcounter}
\ea
The solution of the general finiteness condition (\ref{Hlbfinite}) 
then is 
\be
\Xi^3(\rho) = C(\lb) \rho\;,\sspace
H(\rho)= -C(\lb) \rho\dd_{\rho}\ln h\,,
\label{ERren}
\ee
for some integration constant $C(\lb)$, while $\Xi^4(\rho)$ in 
(\ref{HXi}) retains its form. The formula (\ref{hbeta1}) for the 
functional beta function still applies and gives $\beta_h(h) \equiv 0$
(at $d=2$), to all loop orders. This is gratifying because in 
a reduced phase space quantization the abelian system is 
non-interacting and can be renormalized simply by normal ordering. 
The results (\ref{ERren}) also illustrate again the features discussed after 
Eq.~(\ref{ren_sol2}): $\Xi^3(\rho)$ is proportional to the $\rho \dd_{\rho}$ 
conformal Killing vector and $H(\rho)$ is the associated conformal factor. 
A renormalization of $\rho_B$ is not enforced by the counter terms;
putting $C(\lb) =0$ gives $\rho_B = \rho$ and $H \equiv 0$.    
On the other hand $\sigma$ does get renormalized, although 
(assuming $S_l(h) =0$ for all $l\!>\!1$, and putting the integration 
constants to zero) only by a 1-loop contribution. Taking $h(\rho) =\rho^p$
one has $\phi_1^4(\rho) = - \frac{p}{8 b} \rho^{-p}$.

An initially puzzling feature of $\beta_h(h)$ is that it comes a total 
$\ln \rho$-derivative. Restoring the interpretation of $\rho =\rho(x)$ 
as a field on the 2D base space, however, it has a
natural interpretation: An immediate consequence of (\ref{hbeta2}) 
is that (putting $d=2$) contour integrals of the form
\be
\int_C dx^{\mu} \dd_{\mu} \ln \rho \, \hbbar(\rho,\mu) \,,
\label{hcontour}
\ee
are $\mu$-independent for any closed contour $C$ in the base space.
They are thus invariants of the flow and can be used to discriminate
the inequivalent quantum theories (redundantly) parameterized by 
$h_l[h_0]$. With the initial condition $\hbbar(\rho,\mu_0) = h(\rho)$
the $\mu$-independence of (\ref{hcontour}) is equivalent to 
$\dd^{\mu}[(\hbbar - h) \dd_{\mu} \ln \rho] =0$. On the other 
hand the (classical and quantum) equations of motion for $\rho$ with respect 
to the $h$-modified action are just $\dd^{\mu}(h \dd_{\mu} \ln \rho) =0$.
Combining both we find that the significance of $\beta_h(h)$ being a 
total $\ln\rho$-derivative is that this feature preserves the 
equations of motion for $\rho$ under the $\mu$-evolution
of $\hbbar(\,\cdot\,,\mu)$:  
\be
\int^{\rho} \frac{du}{u} h(u) \quad \mbox{harmonic} 
\quad \Longrightarrow \quad 
\int^{\rho} \frac{du}{u} \hbbar(u,\mu) \quad \mbox{harmonic}\,.
\label{hmuharmonic} 
\ee
This provides an important consistency check as (\ref{hmuharmonic}) 
is also required by the non-re\-nor\-malization of the $\dd_{\sigma}$ 
Noether current, c.f.~(\ref{curr_nonren}). 

Finally let us consider the $\rho \ra \infty$ limit of the flow 
equation (\ref{hsflow}). On account of the reasoning leading to 
Eq.s~(\ref{HXi_p_asympt}) one will want the 
initial $h(\rho) = \hbbar(\rho,\mu_0)$ to have a $\lb$-independent 
asymptotics $h_{\infty}(\rho) \sim \rho^p$. From (\ref{HXi_p_asympt}) 
it then follows that $\beta_h(h) \ra 0$ for $\rho \ra \infty$. 
For $\rho \ra \infty$ 
the $\hbbar$-flow therefore freezes: $\hbbar(\rho,\mu) \sim 
h_{\infty}(\rho)$ for all $\mu$. This guarantees that the 
functional flow is solely driven by the counter terms and not by
artifacts. Subject to these asymptotic boundary conditions 
the $\hbbar$-flow is unambiguously defined. In the next section 
we proceed to determine its fixed points and study the linearized 
flow in its vicinity.

\newsubsection{Fixed point function and linearized flow} 

The stationary points of the flow (\ref{hsflow}) 
can be computed by converting the vanishing condition for the 
$\beta_h(h)$ function (\ref{hbeta2}) into a differential equation.
It reads 
\be
\frac{\lb}{2\pi} \rho \dd_{\rho} h = 
C(\lb) h^2 \;\frac{h}{\lb}
\beta_{\lb}\left(\frac{\lb}{h}\right)\,,
\label{betaode}
\ee 
for some $C(\lb) = \sum_{l \geq 0} C_l (\frac{\lb}{2\pi})^l$ with
constant $C_l$. This can be solved recursively for 
$h_0,h_1,$ etc. We denote the solutions by $h^{\rm beta}_l(\rho)$.
The minimal solution corresponding to a $\lb$-independent 
$C(\lb) = C_0 = p/\zeta_1$ is 
\be 
h^{\rm beta}(\rho,\lb) = 
\rho^{p} - \frac{\lb}{2\pi} \frac{2\zeta_2}{\zeta_1} 
- \Big(\frac{\lb}{2\pi}\Big)^2 \frac{3\zeta_3}{2 \zeta_1} \rho^{-p}
+ \ldots\,,
\label{betaminsol}
\ee
Switching on the constants $C_1,C_2$, etc produces 
deformations of the functions $h^{\rm beta}_l(\rho)$.
In order to compute them we expand (\ref{betaode}) in powers 
of $\lb$ to find 
\ba
\rho \dd_{\rho} h_0  \is  C_0 \zeta_1 h_0 \,, 
\nonum
\rho \dd_{\rho} h_1  \is  C_0 \zeta_1 h_1 + 
C_1 \zeta_1 h_0 + 2 C_0 \zeta_2\,,
\nonum
\rho \dd_{\rho} h_2  \is  C_0 \zeta_1 h_2 + 
C_1 \zeta_1 h_1 + C_2 \zeta_1 h_0 + \frac{3 C_0 \zeta_3}{h_0} 
+ 2 C_1 \zeta_2\,,
\label{hbeta_sol1}
\ea
etc. From them the solutions $h_0, h_1, h_2$, etc are computed recursively.
One finds $h_0(\rho) = \rho^p/\lb_0$, with $p= \zeta_1 C_0$,
and a normalization constant $\lb_0$. Further  
\ba 
h_1^{\rm beta} \is -\frac{2 \zeta_2}{\zeta_1} + \frac{C_1}{C_0} 
h_0 \ln h_0\,,
\nonum
h_2^{\rm beta} \is -\frac{3\zeta_3}{2\zeta_1} h_0^{-1} 
+ \frac{C_2}{C_0}h_0 \ln h_0  
+ \frac{1}{2}\Big(\frac{C_1}{C_0}\Big)^2 h_0 \ln^2 h_0\,, 
\label{hbeta_sol2}
\ea 
where trivial additive terms proportional to $h_0$ have been 
omitted. Generally $h_l,\,l\!\geq \!1$, is a function of 
$h_0$ containing $l$ deformation parameters,
$C_k/C_0$, $k=1,\ldots, l$. The significance of these parameters 
can be seen from the $\rho \ra \infty$ limit, where
the curvature radius of the target space manifold approaches zero.
In this limit $h^{\rm beta}(\rho,\lb) = 
h_{\infty}(\rho,\lb)[-\frac{2\zeta_2}{\zeta_1} + O(h_0^{-1})]$, with
\be
\frac{h_{\infty}(\rho,\lb)}{h_0(\rho)} =  1+ \frac{\lb}{2\pi} \frac{C_1}{C_0} 
\ln h_0 + \Big( \frac{\lb}{2\pi} \Big)^2\left( \frac{C_2}{C_0}\ln h_0 
- \frac{1}{2} \Big(\frac{C_1}{C_0} \Big)^2 \ln^2 h_0\right) + 
O(\lb^3)\,.   
\ee
In other words, the $C_k,\,k\geq 1$, switch on a $\lb$-dependence 
of the $\rho \ra \infty$ asymptotics that is not enforced 
by the counter terms;~c.f.~Eq.~(\ref{Tlambda_asympt}) 
and the subsequent discussion. Putting them to zero therefore 
is a natural extension of the minimal subtraction scheme used
throughout, whereby one recovers the minimal solution 
(\ref{betaminsol}). In this sense the fixed point (\ref{betaminsol}) 
is unique. The leading quantum correction has the 
scheme independent coefficient $-2 \zeta_2/\zeta_1 = \eps$. 
Finally we cannot resist mentioning the resemblance of (\ref{betaminsol}) to 
the ``least coupling'' form of the dilaton functional proposed in 
\cite{PolyDam94a,PolyDam94b}.

For later use let us also prepare an integrated form of 
(\ref{betaode}) 
\be
- \frac{2\pi}{\lb}\frac{\dot{\Xi}^3}{\rho} =
\widetilde{C}(\lb) + C^{-1}(\lb) \frac{1}{h}\,,
\label{Xi3_Ctilde}
\ee
which off-hand gives rise to an integration constant $\widetilde{C}(\lb)$. 
However the boundary conditions for $\rho \ra \infty$ fix the latter 
to vanish. This is because with setting discussed in section 4
$h(\rho,\lb)$ grows at least like $\rho^p,\,p\!>\!0$.
Taking the $\rho \ra \infty$ limit of (\ref{Xi3_Ctilde}) 
then enforces $\widetilde{C}(\lb)=0$, as asserted.

Next we consider the linearization of the flow equation (\ref{hsflow}) 
around the fixed point function $h^{\rm beta}$. In a renormalizable 
quantum field theory the linearized flow for the essential couplings  
encodes information about the critical manifold and the rate of 
approach to it. In our case even the linearized flow equation is an 
integro-differential equation. Since the lowest order term 
is fixed by strict renormalizability an appropriate parameterization is  
\be
\hbbar(\rho,\lb) = h^{\rm beta}(\rho,\lb) + \frac{\lb}{2\pi} \sbar_1(\rho)
+ \Big(\frac{\lb}{2\pi} \Big)^2 \sbar_2(\rho) + \ldots, 
\label{sbaransatz}
\ee
where the $\sbar_l(\rho)$ are functions of $\rho$ and $\mu$ which 
at fixed $\mu$ vanish for $\rho \ra \infty$. This boundary condition
adheres to the ``freezing'' of the full, non-linear $\hbbar$-flow
at $\rho = \infty$. By (\ref{hbeta_sol2}) we also know that 
$\mu$-independent solutions of the linearized flow equations
would have to involve linear combinations of $h_0 \ln^k h_0, 
\,k \geq 0$, which would again switch on an artificial 
$\lb$-dependence of the $\rho \ra \infty$ asymptotics. We 
conclude that the properly defined linearized $\hbbar$-flow    
does not admit ``zero-modes''. 

We assign to the functions $\sbar_l$ a scaling dimension $1-l$ in order to 
match the scaling dimensions of the $h_l^{\rm beta}$ under constant 
rescalings of $h_0 = \rho^p$. Decomposing the linearization of
(\ref{hsflow}) into homogeneous pieces results in a recursive system 
of integro-differential equations for the $\sbar_l,\,l \geq 1$.
We first anticipate the result and then comment on its 
derivation and its significance. The result is that for all 
$l\geq 1$,  
\be
\sbar_l(\rho,\mu)\rra 0 \quad \mbox{for} \quad 
\left\{ \begin{array}{ll} 
\eps =+1 \;\;\,& \mbox{and}\;\;\;\,\mu/\mu_0 \ra \infty\, \\
\eps =-1 \;\;\,& \mbox{and}\;\;\;\,\mu/\mu_0 \ra  0\,.
\end{array} \right.
\label{s1sol2}
\ee
For $\sbar_1$ this arises as follows. The defining equation is 
\be
2 \pi \mu \frac{d}{d \mu} \sbar_1 =
p \zeta_1 \rho^p \int_{\rho}^{\infty} \frac{du}{u^{2 p + 1}} \sbar_1(u)
- \frac{\zeta_1}{p} \rho^{1-p} \dd_{\rho} \sbar_1\,.
\label{s1flow}
\ee
It admits a simple generic solution parameterized by a function $r_1$ 
of one variable via
\be
\sbar_1(\rho,\mu) = \rho^p \int_{\rho^p}^{\infty} 
\frac{du}{u} r_1\Big(u - \frac{\zeta_1}{2\pi}\ln \mu/\mu_0 \Big)
\;.
\label{s1sol1}
\ee
Here $r_1(u)$ has to decay sufficiently fast for $u \ra \infty$
to ensure that the integral converges and vanishes faster than 
$1/\rho^p$ for large $\rho$. As long as the flow variable in the 
argument of (\ref{s1sol1}) is positive this translates  
into decay properties as a function of $\mu$. However since 
$\zeta_1 = -\eps$ the sign $\eps = \pm 1$ makes a crucial difference. 
Under the conditions stated one obtains (\ref{s1sol2}). For $\sbar_l$,
$l >1$, the derivation of (\ref{s1sol2}) employs a recursive solution 
formula and then proceeds by induction; the proof is 
somewhat technical and will be described elsewhere. 

The behavior (\ref{s1sol2}) suggests ultraviolet stability of  
the fixed point for $\eps =+1$ and infrared stability for $\eps=-1$. 
Notably this is {\it opposite} to the stability pattern of the 
renormalization flow in ${\rm O}(1,2)$ models without coupling to 
gravity, where the $\eps =+1$ system is not asymptotically free.  
Its nonrenormalizable modification though, obtained by truncation of Einstein 
gravity, is asymptotically safe! In view of the discussion in section 2.2 it is 
gratifying to see that precisely for the $L_+$ Lagrangian is the coupling flow
locally driven toward the fixed point in the ultraviolet. 
This is because, as argued in section 2.2, {\it only} for $L_+$ does the 
functional integral in (\ref{q1}b) plausibly model the 
truncated 4D quantum gravity in (\ref{q1}a). This holds irrespective 
of the signature of the Killing vectors though for the stationary sector 
an additional dualization is needed. The existence of an 
ultraviolet fixed point in this non-renormalizable theory therefore 
is in the spirit of Weinberg's ``asymptotic safety'' scenario.
To quote from \cite{Weinberg}: ``A theory is said to be asymptotically 
safe if the `essential' coupling parameters approach a fixed-point 
as the momentum scale of their renormalization point goes to infinity.''
In a sense elucidated in appendix C the fixed-point can also be 
regarded as ``non-Gaussian''. (Perhaps it is thus not accidental that 
1-loop perturbation theory in the full 4D theory does not seem to display the 
anti-screening phenomenon \cite{Dono02}.) Moreover one would expect 
that this feature of the truncated theory is a necessary condition for 
full quantum Einstein gravity to have a non-trivial UV stable fixed point.

On the other hand the present results cannot be subsumed literally 
into the original asymptotic safety scenario: In the case at hand 
the space of Lagrangians in which the flow moves has 
no preferred parameterization in terms of (infinitely many) numerical
parameters. This is because the bare and renormalized 
$h$-functions are related in a nonlinear and nonlocal way (in field space), 
-- a feature one might expect to occur whenever a dimensionless 
scalar unprotected by symmetries (like a dilaton or the 4D conformal factor) 
is involved. In particular one cannot classify numerical coupling vectors by 
their eigenvalues with respect to the gradient-matrix of the beta 
function and the $\mu$-dependence of the linearized flow will 
not always be power-like. For example for $n\!>\!1$ a choice 
$r_1(u) \sim e^{-u} u^n$ in (\ref{s1sol1}) induces a power-like decay in $\mu$ 
while $r_1(u) \sim u^{-n}$ induces a log-like decay in $\mu$.   
One might try to classify the functions $r_1$ by the rate of decay 
they induce in (\ref{s1sol2}) but it is unclear how to `count' them. 
In summary, the result (\ref{s1sol2}) is in the spirit of the 
asymptotic safety scenario but there is no obvious way to define the 
dimension of the critical manifold.

\newsubsection{Flow equations for the fields and 
``gravitational undressing''} 

Recall from (\ref{xigRB}) the relation between the bare and the 
renormalized fields 
\be
\phi^j_{\b} = \phi^j + \frac{1}{2 -d} \Xi^j(\rho,\lb) 
+ O\Big(\frac{1}{(2-d)^2}\Big)\,,
\label{field_flow1}
\ee
where $\Xi^1 = \Xi^2 =0$, while $\Xi^3,\,\Xi^4$ have been computed 
in (\ref{HXi}) and depend on $h$. Since the bare 
fields are $\mu$-independent the renormalized fields $\phi^j$ have 
to carry an implicit $\mu$-dependence through $h$. 
(This is analogous to the situation in an ordinary multiplicatively 
renormalizable quantum field theory, where the 
coupling dependence of the wave function renormalization 
induces a compensating $\mu$-dependence of the renormalized fields 
governed by the anomalous dimension function.) 
From (\ref{field_flow1}) and the $\hbbar$-flow (\ref{hsflow}) 
one derives 
\be
\mu \frac{d}{d \mu} \rhobar =  - \dot{\Xi}^3[\hbbar](\rhobar)\,,
\bspace 
\mu \frac{d}{d \mu} \overline{\sigma} = 
- \dot{\Xi}^4[\hbbar](\rhobar)\,,
\label{field_flow2}
\ee
where $\dot{\Xi}^3[\hbbar],\,\dot{\Xi}^4[\hbbar]$ refer to 
(\ref{HXidot}) with the solution of (\ref{hsflow}) inserted 
for $h$. Note that, conceptually, the problems decouple:
One first solves the autonomous equation (\ref{hsflow}) to obtain 
the coupling flow $\mu \ra \hbbar(\,\cdot\,,\mu)$ which is then used to 
specify the right hand side of the $\rhobar$-flow equation
whose solution in turn determines the $\sigmabar$-flow. 
For a given solution $\hbbar$ let 
$\phibar^j$ denote the moving field vector and let $\overline{g}_{ij}$ 
be the target space metric with $\hbbar$ as a prefactor. 
The distance-squared traveled along the flow is obtained by integrating
\ba
\overline{g}_{ij}(\phibar)\,\mu \frac{d\phibar^i}{d\mu} \,
\mu\frac{d\phibar^j}{d\mu}  
\is \mu\frac{d}{d\mu} \left(\int^{\rhobar} \frac{du}{u} 
\hbbar(u,\lb) \int_u^{\infty} \frac{dv}{v} \dot{S}[\hbbar](v,\lb)
\right) 
\nonum
\is  \mu\frac{d}{d\mu} \Big( - \frac{\lb}{4\pi} \ln \rhobar^p 
+ O(\lb^2) \Big) \,.
\label{RGdist1}
\ea
For a given $\hbbar$-flow it manifestly depends only on the 
initial and final $\rho$ configuration. In the decoupling limit 
of constant $h$ the right hand side vanishes identically and the 
field flow describes null geodesics.   

Generally however only the leading terms in $\rhobar,\sigmabar$ are 
readily accessible. Since we insist on having $h(\rho,\lb) = 
\rho^p + O(\lb)$, the leading term on the right hand side of 
(\ref{field_flow2}) is given by $\frac{\lb}{2\pi} \phi_1^j(\rhobar)$, 
$j=3,4$, with $\phi_1^j$ from (\ref{fixp3}). The solution is 
\ba
\rhobar \is \rho + \frac{\zeta_1}{p} \frac{\lb}{2\pi} \rho^{-p+1} 
\ln \mu/\mu_0 + O(\lb^2)\,,
\nonum
\overline{\sigma} \is \sigma +  \frac{\lb}{2\pi}
\left[d_1 - \frac{2 a \zeta_1 + p^2}{4 b p} \rho^{-p} \right]
\ln \mu/\mu_0 + O(\lb^2)\,.  
\label{field_flow3}
\ea 
In contrast the higher orders are difficult to control.

Although essential for the consistency of the formalism the 
moving fields $\rhobar(\mu)$ and $\sigmabar(\mu)$ are ``inessential'' 
couplings in the sense of \cite{Weinberg}. Recall from section 
4.2 that the flow of an inessential coupling is effected by field 
redefinitions and may continue to run even at a fixed point. 
For $\rhobar(\mu)$ and $\sigmabar(\mu)$ this is almost tautological 
but it is important for the interpretation of the results: 
One may observe from (\ref{field_flow3}) that the 
flow pattern of $\rhobar$ is {\it opposite} to that of the 
linearized $\hbbar$-flow in (\ref{s1sol2}). For example for $\eps =+1$ 
the value of $\rhobar$ is decreasing with increasing $\mu/\mu_0$. 
Thus if one was to identify $\rhobar$ with an essential 
coupling its flow would drive it out of the perturbative 
`large $\rho$' regime in the ultraviolet. However $\rhobar$ and 
$\sigmabar$ cannot be regarded as couplings for at least four
reasons: 
(i) First they meet the defining criterion for being ``inessential'' 
discussed in section 4.2. 
(ii) They continue to run at the fixed point $h^{\rm beta}$. 
(iii) They are still functions on base space and 
their value at some point $x$ has no intrinsic meaning. 
(iv) The $x$-dependence is such that the currents 
(\ref{sigma_curr}), (\ref{cward1}) with $\rhobar, \sigmabar$ 
inserted fail to be conserved in general.

Properties (i) and (iii) are obvious. Feature (ii) is present 
already to lowest order in (\ref{field_flow3}); a closed expression 
for the $\rhobar$-flow at the fixed point is given in 
Eq.~(\ref{field_flow8}) below. To see (iv) we combine 
(\ref{hsflow}) and (\ref{field_flow2}) to obtain 
\be
\mu \frac{d}{d \mu}\left( \int_{\rho}^{\rhobar} 
\frac{du}{u} \hbbar(u,\mu) \right) 
= - \hbbar(\rho,\mu) \frac{\dot{\Xi}^3[\hbbar](\rho)}{\rho}
=\frac{\lb}{2\pi} \frac{\zeta_1}{p} + O(\lb^3)\,,
\label{field_flow6}
\ee
where the right hand side is independent of $\rhobar$. As indicated 
its leading term is also $\mu$-independent so that one recovers 
(\ref{field_flow3}). On the other hand one can decompose the integral
as $\int_{\rho}^{\Lambda} + \int_{\Lambda}^{\rhobar}$ with some ($x$- and 
$\mu$-independent) constant $\Lambda$. The left hand side of 
(\ref{field_flow6}) then splits into two terms the first of which 
is a harmonic function in $x$ by (\ref{hmuharmonic}), while the right 
hand side in general is not. For a generic initial $h$ we thus find: 
\be
\int^{\rhobar} \frac{du}{u} \hbbar(u,\mu) \quad 
\mbox{is not harmonic}\,.
\label{lbarflow3}
\ee 
This demonstrates (iv) for the current (\ref{sigma_curr}); a similar 
analysis could be made for the others. In view of (i)--(iv) we 
may safely conclude that $\rhobar$ and $\sigmabar$ are ``inessential'' 
couplings.

Nevertheless the field flow (\ref{field_flow2}) is crucial
for the consistency of the formalism. This is highlighted by 
the pattern that emerges if one inserts $\rhobar$ into the 
first argument of the running $\hbbar(\,\cdot\,,\mu)$: 
Specifically we consider the combination
\be 
\lbar(\mu) := \frac{1}{\hbbar(\rhobar, \mu)}\;
\quad \mbox{with} \quad \lbar(\mu_0) = \frac{1}{h(\rho)}\,,
\label{field_flow4}
\ee
which depends on the value of $h(\rho(x))$ -- and hence on $x$ -- 
parametrically through the initial condition. Either by rewriting 
(\ref{field_flow6}) or by  
direct computation from (\ref{hsflow}) and (\ref{field_flow2})
one finds that $\lbar$ satisfies (putting $d=2$ for simplicity)  
\be 
\mu \frac{d}{d \mu}\lb \lbar = - \beta_{\lb}(\lb \lbar)\,.  
\label{lbarflow1}
\ee
This is the usual flow equation for the flat space ${\rm O}(1,2)$ 
sigma-model. The gravitationally dressed functional flow 
(\ref{hsflow}) has been `undressed'! This occurs independent of the 
form of the initial $h(\,\cdot\,,\lb)$ and may be interpreted as 
an {\it equivalence principle} for 2D quantum gravity non-minimally coupled 
to sigma-models: {\it By using a scale dependent `clock field' 
$\rhobar(x;\mu)$ the effect of 2D quantum gravity on matter can 
locally be undone.}  

Technically this occurs because in defining $\lbar(\mu)$ in 
(\ref{field_flow4}) we study the flow of the {\it numerical value} of 
$h(\rho)$ with respect to a `comoving' coordinate system in field space. 
Since both the $\hbbar$- and the $\rhobar$-flow were induced by 
splitting a set of $\rho$-modified ${\rm O}(1,2)$ counter terms  
(\ref{Tcounter1}) according to the principle of conformal renormalizability,
it is plausible that the relative flow encoded in $\lbar(\mu)$ is 
governed by (\ref{lbarflow1}). In fact the flow equation (\ref{lbarflow1}) 
can also directly be obtained from Eq.~(\ref{hBrhoB}), which highlights that 
the counter terms driving $\lbar(\mu)$ are those relating the 
value of $h_{\b}(\rho_{\b})$ to the value of $h(\rho)$.  
Of course in itself (\ref{lbarflow1}) is of little use, as one 
is really interested in the flow equation of $\hbbar(\,\cdot\,,\mu)$ 
with respect to a {\it fixed} set of field coordinates. In other 
words the complexity of the function flows (\ref{hsflow}) and 
(\ref{field_flow2}) has to be addressed 
because one needs to disentangle the $\mu$-dependence of the function 
$\hbbar(\,\cdot\,,\mu)$ from the $\mu$-dependence of its first 
argument. 

Ignoring the fact that $\lbar$ depends on $x$, one might 
be tempted to interpret it as a running coupling. As with $\rhobar$ 
and $\sigmabar$ this would have the discomforting consequence that 
the flow $\mu \ra \lbar(\mu)$ has characteristics opposite to that of 
the linearized $\hbbar$-flow (\ref{s1sol1}), (\ref{s1sol2}).
However $\lbar$ must again be considered as an {\it inessential coupling}.
One cannot directly apply the variational criterion discussed 
in section 4.2 because $\lbar$ and $\rhobar$ are not independent;
so the variation of the Lagrangian with respect to $\lbar$ would
be cumbersome to study. However there are indirect arguments that 
safely identify $\lbar$ as an inessential coupling. Most importantly 
it continues to flow even after $\hbbar(\,\cdot\,,\mu)$ reached the 
fixed point. Since the flow (\ref{lbarflow1}) is the same for any 
initial $h(\,\cdot\,)$ one can obtain an explicit formula by 
evaluating it for $h^{\rm beta}$. This trivializes the 
$\hbbar$-flow and the $\mu$-dependence is entirely carried by 
$\rhobar$. Using (\ref{Xi3_Ctilde}) in (\ref{field_flow6}) yields 
\be
\int_{\rho}^{\rhobar} \frac{du}{u} h^{\rm beta}(u,\lb) =
\frac{\lb}{2\pi} \frac{1}{C(\lb)} \ln \mu/\mu_0\,
\quad \Longleftrightarrow \quad 
\mu \frac{d}{d\mu} \ln \rhobar\Big|_{h^{\rm beta}} =
\frac{\lb}{2\pi} 
\frac{\lbar(\mu)}{C(\lb)}\,, 
\label{field_flow8}
\ee
where $C(\lb) = C_0 = p/\zeta_1$ corresponds to the minimal 
$h^{\rm beta}(\rho)$ in (\ref{betaminsol}). Comparing with 
(\ref{RGdist1}) one sees that $\lbar(\mu)$ also parameterizes the 
leading order of the distance squared traveled along the 
total renormalization flow. $\rhobar|_{h^{\rm beta}}$ now has 
somewhat nicer properties, but not nice enough: The integral in 
(\ref{lbarflow3}) now does define a $\mu$-dependent harmonic 
function of $x$. Further, comparing (\ref{field_flow8}) with (\ref{3geodesic}) 
one identifies the residual $\rhobar$-flow as geodesic, 
with $\ln \mu/\mu_0$ playing the role of the affine parameter. 
However the same is not true for $\sigmabar$, otherwise the integrated 
distance inferred from (\ref{RGdist1}) would vanish. 
As it doesn't vanish $\lbar$ remains an inessential coupling 
for $h=h^{\rm beta}$.

\newsection{Operator constraints and trace anomaly}

Recall from appendix A that the classical hamiltonian and 
diffeomorphism constraints coincide with the components 
$T_{00}$ and $T_{01}$, respectively, of the energy momentum 
tensor (\ref{constr}). The construction of the operator constraints 
therefore is equivalent to the construction of the renormalized 
energy momentum tensor $\nl T_{\mu\nu} \nr$ of the flat space 
quantum field theory whose (source-extended) renormalized 
Lagrangian was constructed in the previous sections. The 
so-defined energy momentum operator may be expected to 
have a non-vanishing trace anomaly $\nl T^{\mu}_{\; \mu}\nr$. 
At the fixed point (\ref{betaminsol}) of the functional $h$-flow 
however one may hope that the anomaly vanishes. In the following we 
take up these issues consecutively. 

\newsubsection{Improved energy momentum tensor} 

On the bare level the energy momentum operator is uniquely determined 
by the conservation equation up to 
an improvement term. It is thus given by (\ref{abconstr}) 
with constant $a(\rho) = a,\, b(\rho) = b$, modified by the addition of 
a generic improvement term $(\dd_{\mu} \dd_{\nu} - 
\eta_{\mu\nu} \partial^2)\Phi^{\b}$, whose potential $\Phi^{\b}$ replaces 
$f$ in the last term of (\ref{abconstr}). It is not 
hard to see that $\Phi^{\b}$ can only depend on $\sigma_{\b}$ and $\rho_{\b}$ 
and that the counter terms must be $\sigma$-independent. The counter terms 
are in principle determined by the requirement that $T^{\b}_{\mu\nu} =
\nl T_{\mu\nu} \nr$ is a finite composite operator in minimal subtraction  
whose insertion into correlation functions produces answers for 
which the UV cutoff can be removed. Combined with the principle
of ``conformal renormalizability'' this turns out to determine
the counter terms, and eventually the renormalized improvement potential 
$\Phi$ as the solution of a functional flow equation. 
For the actual computation of the counter terms it is useful to 
treat $\Phi$ as an arbitrary function of $\phi^j$, work 
out the counter terms and then impose the additional restrictions.
In this setting one can again take advantage of the 
results available in the literature on Riemannian sigma-models 
because improvement terms with a potential $\Phi$ 
correspond to minimal couplings to the scalar curvature of a fiducial 
background metric in the Lagrangian. We refer to appendix B for 
a compilation of the relevant counter terms.

The restrictive notion of ``conformal renormalizability'' 
adopted here for the Ernst-like systems implies that the 
renormalized improvement potential $\Phi$ can 
only depend on $\rho$ and $\sigma$, and that the dependence on $\sigma$ 
must be {\it linear}. To see this we apply the diffeomorphism 
Ward identity (\ref{diffward}) to the vector $\dd^i \Phi$ and the action 
of the Ernst-like systems. This gives 
\be  
\nl \dd^{\mu} \dd_{\mu} \Phi \nr = \nl \nabla_i \nabla_j \Phi \,
\dd^{\mu} \phi^i \dd_{\mu} \phi^j\nr  - \lb 
\frac{\delta S_{\b}}{\delta \phi^j} \dd^j \Phi\,.
\label{ddPhi_ward}
\ee
The presence of a $\dd^{\mu} \sigma \dd_{\mu} \sigma$ term on the 
right hand side would destroy one of the conformal Killing vectors 
on the target space in which case (\ref{ddPhi_ward}) would not be a viable 
improvement for the trace $\nl T^{\mu}_{\;\mu} \nr$. The absence of 
such a term requires $\Phi$ to be linear in $\sigma$. For later 
convenience we parameterize it as 
\be
\Phi = f(\rho,\lb) + f_0(\lb)\sigma\,,
\label{Phiparam1} 
\ee
where both $f$ and the constant $f_0$ may depend on $\lb$. The linear 
$\sigma$-dependence has the consequence that also no $\dd^{\mu}\sigma
\dd_{\mu} \rho$ term appears in the improvement of the trace.
Moreover in the counter terms (\ref{Rcounter1}) relating the bare potential 
$\Phi_{\b}(\phi_{\b})$ to the renormalized one $\Phi(\phi)$ the function 
$f(\rho)$ drops out. To verify this recall that a typical monomial 
in the differential 
operator $Z(g)-1$ is of the form $z^{i_1\ldots i_n}(g) \nabla_{i_1} \ldots 
\nabla_{i_n}$. By an argument similar to the ones in section 3.2 one 
establishes that only for $n=2$ can this give a non-vanishing 
contribution upon acting on $\Phi$ of the form (\ref{Phiparam1}). 
$f$ only appears in the $3$-$3$ component of $\nabla_i\nabla_j\Phi$ 
and thus disappears upon contraction with a $z^{ij}(g)$ that has 
vanishing covariant $4$-components. We conclude that 
$[Z(g)-1]\Phi$ is a local function of $h$ whose $l$-loop
contribution scales like $\Lambda^l$ under $h \ra \Lambda^{-1} h$. 
The contributions coming from $\Psi(g)$ have a similar structure. 
We write 
\ba
\label{Phiparam2}
&& [Z(g/\lb) -1]\Phi + \lb \Psi(g/\lb) =
\frac{1}{2 -d}k[h](\rho) + O\Big( \frac{1}{(2-d)^2} \Big)\,,
\nonum
&& k(\rho) = -\frac{f_0}{b} k_1(\rho)\rho \dd_{\rho} \ln h + 
h k_2(\rho)\;,
\\[2mm] 
&& k_1(\rho) = \frac{\lb}{2\pi} \frac{1}{2h} + O(\lb^3)\;,\sspace
k_2(\rho) = \frac{\lb}{2\pi} \frac{2}{3 h} + 
\Big( \frac{\lb}{2\pi} \Big)^3 \frac{1}{12 h^3}+ O(\lb^4) \,,
\nonumber
\ea
where the explicit form of the low order contributions follows from 
Eqs.~(\ref{Rcounter3}),(\ref{Rcounter4}). Parameterizing the bare 
potential as $\Phi_{\b} = f_{\b}(\rho_{\b}) + \mu^{d-2} f_0\, \sigma_{\b}$ one 
finds from (\ref{field_flow1}), (\ref{Rcounter1}) that $f_0$ is unrenormalized 
while the bare function $f_{\b}(\,\cdot\,)$ is related to the 
renormalized one, $f(\,\cdot \,)$, by 
\be
f_{\b}(\rho) = \mu^{d-2} f(\rho) + \frac{\mu^{d-2}}{2-d} 
[ k(\rho) - \dd_{\rho} f(\rho) \Xi^3(\rho) - f_0 \Xi^4(\rho) ] 
+  \ldots\,.
\label{fB_fren}
\ee
In particular the function $f_{\b}(\,\cdot\,)$ will in general 
not be the same as $f(\,\cdot\,)$. A strict counterpart of 
the non-renormalization property (\ref{curr_nonren}) valid for the 
currents cannot hold therefore. Of course if one takes into account 
the additional functional change in $f$ the weaker property $T^{\b}_{\mu\nu} = 
\nl T_{\mu\nu} \nr$ holds by construction. 

The $\rho$-dependence of $f$ is constrained by its functional 
flow, which in contrast to that of $h$ is not autonomous. 
The equation governing the flow $\mu \ra \fbar(\,\cdot\,,\mu)$ 
can be obtained either directly from (\ref{fB_fren}) or by 
combining (\ref{Gflow}) with (\ref{field_flow2}), keeping 
$\mu \frac{d}{d\mu} f_0 = (2-d) f_0$ in mind. One finds
either way 
\be
\mu \frac{d}{d \mu} \fbar = (2-d) \fbar + 
\dd_{\rho} \fbar \dot{\Xi}^3[\hbbar] + f_0 \dot{\Xi}^4[\hbbar] 
- K[\hbbar] \,,
\label{fflow}
\ee
where $K[h]$ is obtained from $k$ by substituting $k_1 \ra 
\dot{k}_1$ and $k_2 \ra \dot{k}_2$, and the latter can be interpreted 
either in the sense of (\ref{Xdot}) or (\ref{Odot}). Note that
$K[h] = h\! \cdot\! \frac{\delta}{\delta h} k[h] - h k_2[h]$, 
and $K(\rho) = - \frac{\lb}{2\pi} \frac{2}{3} + O(1/h)$, for 
large $\rho$. For a given $\hbbar$-flow Eq.~(\ref{fflow}) is 
a linear inhomogeneous equation
for $\fbar$. We shall only be interested in the solution corresponding 
to $h^{\rm beta}$, i.e.~to the fixed point of the $\hbbar$-flow.
As $h$ is the essential coupling one expects that 
at its fixed point the form of the improvement potential likewise 
stabilizes. This consistency condition determines $f^{\rm beta}(\rho) := 
f[h^{\rm beta}](\rho)$ to be
\be 
\rho \dd_{\rho} f^{\rm beta} - 
f_0 \frac{a}{2b} = - \frac{2\pi C}{\lb} h 
\Big[K(\rho) + \frac{\lb}{2\pi} \frac{2}{3} \Big] -  
\frac{2\pi}{\lb} \frac{C f_0}{2 b} h  
\int_{\rho}^{\infty} \frac{du}{u} \dot{S}(u,\lb)\,,
\label{ffix1}
\ee
where we set $d=2$ and the right hand side is evaluated for 
$h= h^{\rm beta}$. The $\dot{S}$ term enters through 
Eq.~(\ref{HXidot}) and we re-adjusted 
the lower integration boundary so as to extract the constant 
$-\frac{\lb}{2\pi} \frac{2}{3 f_0}$. The redefined integration boundary 
was then set to infinity, which removes terms proportional to 
$h$ in the $\rho \ra \infty$ limit. Generally the choice of integration 
constants affects $f(\rho)$ merely by a shift proportional to the 
potential of the Noether current (\ref{sigma_curr}) 
\be 
f(\rho) \rra f(\rho)  + d(\lb) \int^{\rho} 
\frac{du}{u} h(u,\lb)\sspace 
\mbox{with} \sspace d(\lb) = \sum_{l \geq 1}d_l 
\Big( \frac{\lb}{2\pi} \Big)^l\,.
\label{Phiambiguity} 
\ee
Since such a term is already present on the classical level 
-- c.f.~Eq.~(\ref{phi_hb}) with constant $b(\rho)$ -- and one is not 
forced to modify it, it is natural to impose the absence of 
ad-hoc $\lb$-dependent corrections to it as a boundary condition.

So far our focus lay on the construction of the renormalized energy
momentum tensor. Apart from the ambiguities stemming from the 
 solution of (\ref{fflow}) it is now fully determined and one can 
proceed to investigate its trace. The key result to be shown in the 
next section is
\be
\mu \frac{d}{d \mu} \hbbar = 0 = \mu \frac{d}{d \mu} \fbar 
\quad \Longleftrightarrow \quad \nl T^{\mu}_{\; \mu} \nr =0 \,.
\label{zerotrace}
\ee
That is, the trace anomaly of the improved energy momentum tensor 
vanishes precisely at the fixed point of the functional flow. 
Of course $h$ and $f$ are not on the same footing though: 
$h$ is the essential coupling while $f$ `merely' defines the
proper improvement term. 

Technically the equivalence is non-trivial because on 
both sides different types of information enter. The left hand side 
contains only information about the basic Lagrangian (without 
sources) and its renormalization. The very definition of 
$\nl T_{\mu\nu} \nr$ as a composite operator, on the other hand, 
requires additional counter terms beyond those needed to 
renormalize the basic Lagrangian. In particular operator mixing 
takes place, i.e.~the counter terms of operators with lower 
engineering dimension enter. The equivalence (\ref{zerotrace}) 
thus requires both types of counter terms to be correlated.   
The fact that they indeed are correlated can be traced back 
to the `non-renormalization' property of the energy 
momentum tensor $T^{\b}_{\mu\nu} = \nl T_{\mu\nu} \nr$. The latter 
gives rise to a precursor of the Curci-Paffuti relation 
\cite{CurciPaff} which is instrumental for the proof of 
(\ref{zerotrace}).

\newsubsection{Trace anomaly}

We begin with the following expression for the trace anomaly 
\be 
\nl T^{\mu}_{\;\;\mu} \nr  = \nl \frac{\lb}{h} \beta_h(h/\lb) \,L \nr
+ \frac{1}{4\pi} \nl (\cL_{K}g_{ij})(\phi) 
\dd^{\mu} \phi^i \dd_{\mu} \phi^i \nr \,,\sspace 
K^j = \frac{2\pi}{\lb}[W^j - \dot{\Xi}^j + \dd^j \Phi]\,,
\label{anomaly2} 
\ee
where $L$ is the Lagrangian. It is obtained by inserting 
Eq.~(\ref{ts_beta}) into (\ref{anomaly1}) and taking the limit of 
a flat base space. We already know that the improvement 
potential $\Phi$ is of the form (\ref{Phiparam1}) with $f(\rho)$ 
subject to (\ref{fflow}). The vector $W^j$ is likewise 
highly constrained. Starting from the definition (\ref{Svector}) 
an argument similar to the one yielding (\ref{xiren3}) shows 
that to all loop orders the covariant vector $W_i$ must be of 
the form $W_i=(0,0,W_3(\rho,\lb),0)$. Equivalently 
\be
W^i = \left(0,0,0, \frac{\rho}{b h}W_3(\rho,\lb) \right)\,,
\quad \mbox{with} \quad  
W_3(\rho,\lb) = \Big(\frac{\lb}{2\pi} \Big)^3 \frac{1}{8} \dd_{\rho} 
\Big( \frac{1}{h^2} \Big) + O(\lb^4) \,.
\label{Wvector}
\ee
An important further constraint arises from the `pre'-Curci-Paffuti
relation (\ref{CurciPaff}). It can be shown to be equivalent to 
\ba   
&& \dd_{\rho}(\dot{\Xi}^3 W_3) 
= P(\dot{H})  -\dd_{\rho}(h \dot{k}_2) - (\dot{Z}^V)_{3j} \dot{\Xi}^j
+  \frac{h}{\rho^2} \dot{\Xi}^3 \dot{S} \,,
\nonum
&& \mbox{with} \quad P(\dot{H}) = \dot{N}_3^{\;jk}(\dot{H}(\rho) g_{jk}) 
-\dot{H}(\rho) \dot{W}_3(\rho)\,.
\label{preCP}
\ea

Before proceeding with the general analysis let us briefly 
check the decoupling limit to the ordinary 
${\rm O}(1,2)$ sigma-model. For a constant $h= 1/\lb_0$ one expects 
to recover the trace anomaly of the ordinary 
${\rm O}(1,2)$ sigma-model because the $\rho,\sigma$ part 
of the action decouples and is non-interacting. 
We already saw that for constant $h$ the $\beta_h(h)$ function 
reduces to the ordinary $\beta_{\lb}$ function, c.f.~(\ref{beta_hlb}). 
When specializing $\dot{\Xi}^3$, $\dot{\Xi}^4$, the integration 
constants $\rho_l$ in (\ref{ren_sol1}), (\ref{ren_sol2})  
have to be taken finite, say $\rho_l = \rho_1, \,l \geq 1$.  
This gives $\dot{\Xi}^3 = 
\frac{1}{\lb\lb_0} \beta_{\lb}(\lb \lb_0) \rho \ln \rho/\rho_1$, 
$\dot{\Xi}^4 = -\frac{a}{2b\rho} \dot{\Xi}^3$. When inserted into 
(\ref{anomaly2}) the $\rho,\sigma$ dependent part in the Lagrangian 
cancels and one ends up with 
\be 
\nl T^{\mu}_{\;\;\mu} \nr \Big|_{h = 1/\lb_0} = 
- \frac{1}{2} \frac{1}{(\lb \lb_0)^2} 
\beta_{\lb}(\lb\lb_0) \nl \dd^{\mu} n \cdot \dd_{\mu} n \nr\,.
\label{O12anomaly} 
\ee
This agrees with the (${\rm O}(1,2)$ analogue of the) result in 
\cite{Montanari} modulo terms proportional to the equations 
of motion operator. Because of the different 
schemes used such terms cannot be compared.

An instructive way to proceed with the general analysis is by trying 
to adjust $\Phi$ such that $K^j$ becomes a conformal Killing 
vector of $g_{ij}(\phi)$, 
\be
\cL_K g_{ij} \stackrel{\displaystyle !}{=} \Omega \,g_{ij}\,. 
\label{anomaly3}
\ee
Off-hand (\ref{anomaly3}) would imply only that the trace anomaly 
(\ref{anomaly2}) is proportional to the Lagrangian. In fact it 
turns out to be equivalent to the vanishing of the anomaly! To show 
this we parameterize $\Phi$ as before, i.e.~$\Phi(\rho,\sigma) = 
f(\rho) + f_0\sigma$. Spelling out the conformal Killing equations 
for $K^j$ gives rise to a pair of differential equations. 
After using (\ref{HXidot}) and (\ref{Wvector}) the first one reads 
\be
\frac{\lb}{2\pi} \rho \dd_{\rho} h = C(\lb) h^2\, 
\frac{h}{\lb} \beta_{\lb}\Big(\frac{\lb}{h} \Big)\quad \mbox{with} 
\quad C(\lb) = - \frac{\lb}{2\pi} \frac{b}{f_0(\lb)}\,. 
\label{traceh1} 
\ee
This exactly coincides with Eq.~(\ref{betaode}) -- which entails the 
vanishing of the $\beta_h(h)$ function -- provided the constants 
are matched as indicated. Thus the first term on the right hand side 
of (\ref{anomaly2}) vanishes. The second differential equation 
deriving from (\ref{anomaly3}) is 
\be 
h \dd_{\rho} \left( \frac{\rho \dd_{\rho} f}{h} \right) =
\frac{\lb}{2\pi} \frac{a}{C} \dd_{\rho} \ln h + b h \dd_{\rho} 
(\dot{\Xi}^4 - W^4)\,,
\label{tracePhi1}
\ee
and defines $f^{\rm trace}$. Here $\dot{\Xi}^4$ is given in 
(\ref{HXidot}) and $h$ refers to $h^{\rm beta}$. We postpone the 
integration of (\ref{tracePhi1}) for a moment and compute the 
conformal factor in (\ref{anomaly3})
\be
\Omega = \frac{1}{4} g^{ij} \cL_{K} g_{ij} = 
- \frac{\lb}{2\pi} \dd_{\rho} \ln h \Big[ \dot{\Xi}^3  + 
\frac{\lb}{2\pi} \frac{\rho}{C h} \Big] = 0 \,.
\label{Omega}
\ee 
In the last step the integrated form (\ref{Xi3_Ctilde}) of 
(\ref{traceh1}) was used. One concludes that $K^j$ is actually 
a proper Killing vector and the second term in (\ref{anomaly2})
vanishes as well.

It remains to integrate equation (\ref{tracePhi1}). One of the 
integrations can be performed trivially and gives rise to a 
$\lb$-dependent integration constant $\widetilde{d}(\lb)$. 
Inserting further $W^4$ from (\ref{Wvector}) and $\dot{\Xi}^4$ 
from (\ref{HXidot}) some of the terms combine due to 
(\ref{Xi3_Ctilde}). We absorb $\widetilde{d}(\lb)$ into the (anyhow 
unspecified) lower integration boundary of the $\dot{S}$ term. 
Eventually one ends up with 
\ba
\dd_{\rho} f^{\rm trace} \is - \frac{\lb}{2\pi} \frac{a}{2 C \rho} +  
\frac{h}{2 \rho} \int^{\rho} \frac{du}{u} \dot{S}(u,\lb) 
- W_3(\rho)\,,
\nonum
\Phi^{\rm trace} \is f^{\rm trace}(\rho) - 
\frac{\lb}{2\pi} \frac{b}{C}\sigma\,, 
\label{tracePhi2} 
\ea 
completing the solution of (\ref{anomaly3}). Of course also      
the vector $K^j$ is determined and comes out to be proportional 
to ${\bf t}_- = ( 0, 0, 0, 1)$. The proportionality constant 
parameterizes an additive ambiguity in $\Phi^{\rm trace}$ of the 
form (\ref{Phiambiguity}). One will naturally set this constant to zero 
in which case $K^j$ vanishes identically. Equivalently $\dot{\Xi}^j$ 
is the gradient of a potential 
\be
\dot{\Xi}^j\big|_{h^{\rm beta}} = 
\dd^j(\Phi^{\rm trace} + \omega)
\quad \mbox{with} \quad \omega(\rho,\lb) = - \int_{\rho}^{\infty} 
du W_3(u,\lb)\,.
\label{Kvanishes} 
\ee

In summary, we find  
\be
\cL_K g_{ij} \stackrel{\displaystyle !}{=} \Omega \,g_{ij} 
\quad \Longrightarrow \quad \beta_h(h) =0 \quad \mbox{and} \quad
K^j =0\quad \Longrightarrow \quad \nl T^{\mu}_{\;\,\mu}\nr =0\,.
\label{no_anomaly}
\ee
Here $\beta_h(h) =0$ and $K^j =0$ are equivalent to Eqs.~(\ref{traceh1}) 
and (\ref{tracePhi2}), respectively. We proceed by showing that a converse 
of the statement (\ref{no_anomaly}) is also true.

Specifically we verify that the vanishing of the tensorial 
Weyl anomaly coefficient $B_{ij}(g)$ in Eq.~(\ref{Weyl1}) 
again implies (\ref{traceh1}) and that the already computed 
$\Phi^{\rm beta}= f^{\rm beta}(\rho) + f_0 \sigma$ is the associated 
dilaton field. In other words it should come out that 
\be
B_{ij}(g)\Big|_{h^{\rm beta},\Phi^{\rm trace}} =0 \,\quad \mbox{and} \quad  
\Phi^{\rm trace} =  \Phi^{\rm beta}\,.
\label{Weyl4}
\ee
In order to verify (\ref{Weyl4}) we first compute
\be
\sum_{l \geq 1}\Big(\frac{\lb}{2\pi} \Big)^l l T_{ij}^{(1,l)}(g) 
= {\rm diag}\left(
\frac{\eps}{\Delta^2} \frac{h^2}{\lb} \beta_{\lb}\Big(\frac{\lb}{h}\Big),\;
\frac{1}{\Delta^2} \frac{h^2}{\lb} \beta_{\lb}\Big(\frac{\lb}{h}\Big),\;
-\frac{h}{\rho^2}\dot{S}(\rho,\lb),\; 0\right)\,,
\label{Tdotlambda} 
\ee
with $\dot{S}(\rho,\lb)$ as in (\ref{HXidot}). Further  
\ba
&\nspace & - \frac{2\pi C}{\lb}\, 
\cL_{W + \partial \Phi} \,g_{ij}  
\nonum
&\nspace & = {\rm diag}\left( 
\frac{\eps}{\Delta^2} \rho \dd_{\rho}\ln h, \;
\frac{1}{\Delta^2} \rho \dd_{\rho}\ln h, \;
\frac{a}{\rho} \dd_{\rho}\ln h - \frac{2 \pi C}{\lb} \frac{2 h}{\rho} 
\dd_{\rho} \Big[ \frac{\rho}{h}( \dd_{\rho} f + W_3) \Big] ,
\;0 \right)\,.
\label{Weyl5} 
\ea
with $W^j$ as in (\ref{Wvector}) and $\Phi$ of the form 
$\Phi(\phi,\lb) = f(\rho,\lb) - \frac{\lb}{2\pi} \frac{b}{C}\sigma$, 
as before. Together the condition $B_{ij}=0$ translates 
into just two differential equations. They can be seen to coincide  
with (\ref{traceh1}) and (\ref{tracePhi2}), as asserted. Thus:
\be
\nl T^{\mu}_{\;\;\mu} \nr  \stackrel{\displaystyle !}{=} 0
\quad \Longrightarrow \quad \beta_h(h) =0 \quad \mbox{and} \quad
K^j =0\,.
\label{no_anomaly_inv}
\ee
In particular the differential equations (\ref{traceh1}) and 
(\ref{tracePhi2}) are necessary and sufficient conditions for 
the vanishing of the trace anomaly. The first one determines
$h= h^{\rm beta}$ and the other one the improvement potential 
$\Phi^{\rm trace}$. An instructive consistency check is obtained 
by starting from the alternative expression (\ref{anomaly5}) for 
$\nl T^{\mu}_{\;\mu}\nr$: Using the non-renormalization of the 
conserved current $h(\rho) \dd_{\mu}\ln \rho$ -- 
c.f.~Eq.~(\ref{curr_nonren}) -- the total divergence term 
$\dd^{\mu} \nl \dd_{\mu} \rho W_3(\rho)\nr$ can be rewritten as 
$\nl \rho^{-1} \dd^{\mu} \rho \dd_{\mu} \rho \; 
h \dd_{\rho} (\rho W_3/h) \nr$. Inserting further (\ref{ddPhi_ward}) 
and (\ref{Tdotlambda}) the vanishing of $\nl T^{\mu}_{\;\mu}\nr$
translates into the previously found differential equations. 

Before proceeding let us briefly note the corresponding results in 
the abelian subsector. We already know from (\ref{ERcounter}), 
(\ref{ERren}) that $h(\cdot)$ and $\rho$ remain unrenormalized,
while $\sigma$ is renormalized (only) at the 1-loop level.    
Using this, one finds that the only way to solve $B_{ij} =0$ 
(with a $\rho$-dependent $h$) is by having 
$\Phi(\phi,\lb) = f(\rho,\lb)$ independent of $\sigma$. 
Then $h(\rho,\lb)$ turns out to be unconstrained and only the 
counterpart of the differential equation (\ref{tracePhi2}) for 
$f(\rho,\lb) = \sum_{l \geq 1}(\frac{\lb}{2\pi})^l 
f_l(\rho)$ has to be solved. Taking $h(\rho) = \rho^p$, 
the 1-loop contribution is $f_1(\rho) = \frac{p}{8} \ln \rho$.
Just as with $\Xi^4$ we expect all higher contributions
to vanish, rendering also the distinction between $f^{\rm beta}$ 
and $f^{\rm trace}$ superfluous. 

In the nonabelian system $f^{\rm beta}$ and $f^{\rm trace}$ are 
defined through very different conditions. Their equivalence, 
as asserted in the second part of (\ref{Weyl4}) is also needed to 
conclude the derivation of (\ref{zerotrace}). We now show that 
indeed 
\be
f^{\rm trace} \simeq f^{\rm beta}\,,
\label{ftrace_beta}
\ee
where `$\simeq$' denotes equality modulo (\ref{Phiambiguity}). 
Matching the expressions in (\ref{tracePhi2}) and (\ref{ffix1}) 
one finds that (\ref{ftrace_beta}) requires the following identity 
\be
\rho W_3(\rho) = \frac{2\pi C}{\lb} h \Big( K[h] + 
\frac{\lb}{2\pi} \frac{2}{3} \Big) - 
h \int_{\rho}^{\infty} \frac{du}{u} 
\dot{S}(u,\lb) \quad \mbox{for} \quad h = h^{\rm beta} \,.
\label{preCP_fix}
\ee   
Luckily this indeed is an identity; it arises from the `pre'-Curci-Paffuti
relation (\ref{preCP}) as follows: From Eq.~(\ref{ZVZY}) we have
$(\dot{Z}^V)_i^j \dd_j V = \dd_i (\dot{Z} V)$ for any scalar $V$. 
Applied to $\omega + \Phi$ in (\ref{Kvanishes}) one finds 
$(\dot{Z}^V)_3^j\, \dot{\Xi}^j = - \frac{f_0}{b} \dd_{\rho} 
[ \dot{k}_1(\rho) \rho \dd_{\rho} \ln h]$. Keeping Eq.~(\ref{Xi3_Ctilde})
and the definition of $K$ in (\ref{fflow}) in mind one arrives at 
(\ref{preCP_fix}). As an explicit check one can verify that the right 
hand side vanishes up to and including $O(\lb^2)$,
consistent with $W_3(\rho) = O(\lb^3)$. 

In summary the improvement potential stationary at the 
fixed point coincides -- to all loop orders -- with the one that 
cancels the trace anomaly. The improved energy momentum tensor 
differs from that in (\ref{constr}) by an improvement term 
\be 
\Delta T_{\mu\nu} = \left( 
\dd_{\mu} \dd_{\nu} - \eta_{\mu\nu} 
\dd^{\kappa} \dd_{\kappa}\right) \Phi^{\rm trace}\,,
\label{Timpr}
\ee
with $\Phi^{\rm trace} = f(\rho) - \frac{\zeta_1 b}{p} \sigma = 
\Phi^{\rm beta}$ determined by (\ref{tracePhi2}) or (\ref{ffix1}).  
It is separately conserved and cancels the unwanted quantum corrections 
to the trace of (\ref{constr}).

A well-known consequence of the Curci-Paffuti relation  
is that the Weyl anomaly coefficient $B^{\Phi}$ is a constant
whenever $B_{ij}$ vanishes. In view of (\ref{Weyl4}) this 
should now come out automatically. $B^{\Phi}$ can be 
parameterized in terms of $K[h]$ and the improvement potential 
(\ref{tracePhi2}) as 
\be
\lb B^{\Phi}(\Phi,g/\lb) = - K[h] + \frac{f_0}{b h} \Big[ 
2 \rho \dd_{\rho} f - \frac{a}{b} f_0 + \rho W_3\Big]\,. 
\label{BPhi_Ernst} 
\ee 
Combining (\ref{preCP_fix}) with (\ref{ftrace_beta}) one finds indeed 
\be
\lb B^{\Phi}(\Phi,g/\lb)\Big|_{h^{\rm beta}, \Phi^{\rm trace}} = 
\frac{\lb}{2\pi} \frac{2}{3} \,,\quad \mbox{i.e.} \quad c =4\,.
\label{Bphi_value}
\ee
Here $c=4$ is the formal central charge of the improved 
energy momentum tensor at the fixed point. Note that we established 
(\ref{Bphi_value}) to all loop orders 
despite the fact that the explicit form of (\ref{BPhi_Ernst}) is 
known only at low orders. The specific 
value $2/3$ for the constant hinges on a `natural' 
choice for the various integration constants involved. This is best 
illustrated by an explicit computation.

Since $B^{\Phi}$ is known explicitly up to and including $O(\lb^2)$ 
one can solve the differential equations
(\ref{traceh1}), (\ref{tracePhi2}) to the same order and verify that 
$B^{\Phi}$ comes out to be a constant to that order 
upon inserting the general solutions. This is what we shall do now. 
One starts by expanding the right hand side of (\ref{tracePhi2}) in 
powers of $\lb$, inserts the general solutions for $h_0^{\rm beta},\,
h_1^{\rm beta}$ from (\ref{hbeta_sol2}) and performs the $\rho$-integration.
Only $C_0 = - \eps p$ and an arbitrary constant $C_1$ enter. 
In order to illustrate its impact we also modify $h^{\rm beta}_1(\rho)$ 
by a trivial additive contribution $t_1 \rho^p$, i.e.~we use
\be
h^{\rm beta}(\rho) = \rho^p + \frac{\lb}{2\pi} \eps 
\Big( 1 - C_1 \rho^p \ln \rho + t_1 \rho^p \Big) + O(\lb^2) \,.
\ee
After some computation one finds the 
following general solution depending on the power $p$ and the 
integration constants $C_1,d_1,d_2$, while $t_1$ drops out:  
\begin{subeqnarray}
\mbox{}\nspace 
\Phi^{\rm trace}(\phi) &=& \frac{\lb}{2\pi}\Phi_1(\rho,\sigma) + 
\Big(\frac{\lb}{2\pi}\Big)^2 \Phi_2(\rho,\sigma) + O(\lb^3) \,,
\\
\Phi_1(\rho,\sigma) &=& \frac{1}{4p}[2 \eps a + p^2]\, 
\ln \rho + d_1\, \rho^p + \frac{\eps b}{p}\,\sigma\;,
\\
\Phi_2(\rho,\sigma) &=& \frac{3 \eps}{8} \rho^{-p} + 
\left[ \frac{C_1}{4}\Big(\frac{2 a}{p^2} - \eps\Big) + 
\eps d_1 p\right] \ln \rho
+ \rho^p[d_2 - \eps C_1 d_1 \ln \rho] + \frac{b C_1}{p^2} \sigma\;.
\label{WeylPhi_2loop}
\end{subeqnarray}
Both in $\Phi_1$ and $\Phi_2$ an irrelevant additive constant has been 
omitted. Evaluated on this solution $B^{\Phi}$ is 
field independent as it should and comes out as 
\be
\lb B^{\Phi}(\Phi,g/\lb)\bigg|_{h^{\rm beta},\Phi^{\rm trace}} 
= \frac{\lb}{2\pi} \frac{2}{3} + \Big(\frac{\lb}{2\pi} \Big)^2 2 \eps d_1
+ \Big(\frac{\lb}{2\pi} \Big)^3 2(\eps d_2 - d_1 t_1) +O(\lb^4)\,.
\label{WeylPhi_2loop3}
\ee 
Note that the result is independent of the parameters $p=\zeta_1 C_0$ 
and $C_1$ entering $h^{\rm beta}$. The parameter
$t_1$ introduced for illustration merely changes the 
overall normalization of $h_0(\rho)$, and hence of the 
tree-level Lagrangian. It should clearly be set to zero. 
The constants $d_1,d_2$, etc enter through the integration 
of (\ref{tracePhi2}) and modify the large $\rho$ asymptotics of 
$\rho \dd_{\rho} \Phi$. If they are put to zero 
$\frac{2\pi}{\lb} \rho \dd_{\rho} \Phi$ approaches the constant 
$- a/(2 C(\lb)) + p/4$ for $\rho \ra \infty$, while 
switching them on produces a power-like asymptotics of the form 
$\rho^p$ or $\rho^{p-1} \ln \rho$, etc. Again, this is not enforced 
by the counter terms and one would naturally stipulate that 
the large $\rho$ asymptotics of $\frac{2\pi}{\lb} \rho \dd_{\rho} \Phi$ 
is given by the $\lb$-independent 1-loop constant $\eps a/2p + p/4$.  
This is the same as removing additive contributions of the form 
(\ref{Phiambiguity}). Doing so one recovers the `canonical' 
value $\lb B^{\Phi}(\Phi,g/\lb) = \frac{2}{3} \frac{\lb}{2\pi}$, 
i.e.~$c=4$.   

Based on the results of \cite{Osb88} one then expects that for the 
energy momentum tensor improved via (\ref{Timpr}) the  
combinations $\nl T_{\pm\pm} \nr \sim \nl \cH_0 \pm \cH_1 \nr$ 
generate a 2D conformal algebra with formal central charge 
$c=4$. The value $c=4$ in itself has little significance because
the state space generated by  $\rho,\dd_{\mu} \sigma$ and e.g.~the 
Noether currents has indefinite metric. The latter feature is not 
an artifact, it is reflects the notorious positivity problem related to 
the ``conformal factor problem'' of 4D quantum Einstein gravity. 
As stressed in \cite{CJZ96}, even 
for free field doublets of opposite signature 
inequivalent quantizations exist which affect the value of $c$ 
through the choice of vacuum. In the case at hand quantum counterparts  
of the non-local charges mentioned after Eq.~(\ref{i1})
together with the quantities found in \cite{qernst} are candidates 
for quantum observables which collectively should generate the physical 
state space. A complete construction of these observables in 
a Lagrangian-based formulation is likely to be very difficult.
As with other integrable field theories, however, one can try to 
accumulate evidence that both of these vastly different formulations
actually describe the same system. For the 2-Killing vector reduction 
already the successful construction of one nonlocal observable 
would presumably entail the factorization properties characteristic for 
`integrability', -- and hence would strongly indicate that the 
non-perturbative bootstrap formulation of \cite{qernst} and the 
present Lagrangian-based quantum theory coincide.

\newpage 
\newsection{Conclusions}
\vspace{-3mm}

The main result obtained here is that truncations of quantum Einstein gravity
corresponding to geometries with two Killing vectors are asymptotically 
safe. The truncated systems can be renormalized at the expense 
of introducing infinitely many essential couplings that are combined into
the function $h(\,\cdot\,)$. The renormalization flow then has a unique
ultraviolet stable fixed point at which the trace anomaly vanishes. 
This holds irrespective of the signature of the Killing vectors, although 
in the stationary sector a dualization 
is needed in order to see this. The significance of the result for the full 
4D theory is that the asymptotic safety scenario passes an important 
self-consistency test: If it is true for the full theory it should be 
true in every truncation that preserves the presumed `ferromagnetic' 
nature of the selfcoupling (but not vice versa). Heuristically one would 
expect that simply `freezing' the fluctuations transversal to the Killing 
orbits meets this criterion.     
 
We may also offer some comments on future directions. 
For pure Einstein gravity as considered here an important open problem 
is to link the present Lagrangian-based Dirac quantization 
to the bootstrap formulation of \cite{qernst}. Concretely 
this can be done by studying the conditions under which the 
first non-local charge survives quantization. Following L\"{u}scher's
strategy in the ${\rm O}(3)$ nonlinear sigma-model (without 
coupling to gravity) \cite{Lusch} one will check for the 
absence of unwanted dimension two operators in the operator 
product expansion of two Noether currents, which ideally would be the 
case at the fixed point. An early example for such a match 
between integrability and the vanishing of a beta function was found 
in \cite{Bonneau83b}. Understanding the relation between 
Dirac- and reduced phase space quantization 
is also required to identify the origin of the spontaneous 
${\rm O}(1,2)$ symmetry-breaking found in \cite{qernst}: Is it induced 
by the projection onto the physical state space or already present 
on the enlarged state space providing the `arena' for the constraints?

Another strand to be taken up is the inclusion of matter. 
Classical integrability is known to be preserved in the 2-Killing
vector reduction of a wide class of matter extensions \cite{BMG88}. 
They range from Einstein-Maxwell over dilaton-axion gravity 
\cite{IBakas94,IBakas96} to $N\!=\!16$ supergravity \cite{Nic87}. 
The renormalizability and fixed point structure of all these systems 
can be investigated by the techniques developed here. In particular
a potential relation to \cite{PolyDam94a,PolyDam94b} should be worth
exploring. 

Finally, since already one Killing vector is enough 
to produce the coset structure \cite{BMG88} which has been 
instrumental here, one might hope that similar ideas apply 
in these yet larger sectors, eventually helping to `tame' gravity's 
non-renormalizability. 
\medskip

{\bf  Acknowledgments:}
I wish to thank P.~Forg\'acs for the enjoyable collaboration in part 
of this project and important discussions throughout. 
In addition I profited from conversations and correspondence with 
J.~Balog, T.~Jacobson, D.~Maison, N.~Mohammedi, M.~Reuter, 
C.~Torre, and P.~Weisz. Part of this work was done at the 
Max-Planck-Institut f\"{u}r Physik, whose hospitality is gratefully 
acknowledged. 
\newpage

\appendix
\newappendix{The 2-Killing vector reduction of 
general relativity}
 
The solutions of Einstein's equations with two Killing vectors cover 
a variety of physically interesting situations: If one of the 
Killing vectors is timelike these are stationary axisymmetric 
spacetimes, among them in particular all the prominent black hole 
solutions. If both Killing vectors are spacelike the subsector comprises 
(depending on the sign of $\dd^{\mu} \rho \dd_{\mu} \rho$ with $\rho$ defined below) 
cylindrical gravitational waves, colliding plane gravitational waves, as 
well as generalized Gowdy 
cosmologies. In contrast to the spherical reduction and the 
matter-coupled systems based on it, here one is dealing with 
infinitely many nonlinearly self-interacting gravitational degrees of 
freedom. In a Hamiltonian formulation this results (with and 
without matter) in a `kinematical' diffeomorphism and a `dynamical' 
hamiltonian constraint, very much like in the full theory. 
Unlike in the full theory, however, an infinite set of (nonlocal) 
observables Poisson commuting with all the constraints can be 
constructed explicitly; see \cite{qernst,KorSam98,Torre91} and the 
references therein.  

In order to fix notations and conventions we recall here 
the main steps of the reduction procedure; see also \cite{Maison00} 
for a recent review. Our spacetime conventions are that of 
Landau-Lifshitz, The classical theory of fields, editions after 1971. 
In the classification of Misner-Thorne-Wheeler these are $(-,+,+)$ 
conventions for the metric, Riemann tensor, Einstein tensor, respectively.   
In particular the spacetime metric has signature $(+,-,-,-)$, and is 
denoted by $G_{MN}$ (in abstract index notation). 
As usual it is convenient to adopt a coordinate system, say 
$(x^0,x^1,y^1,y^2)$, in which the Killing vector 
fields act as coordinate derivatives, $\frac{\partial}{\partial y^1},\,   
\frac{\partial}{\partial y^2}$. An ansatz for the 4D line element 
in these coordinates will then be parameterized by functions
depending on $x =(x^0,x^1)$ only. Further it is convenient to 
treat both possible signatures of the Killing vectors simultaneously. 
We distinguish both cases by a sign, such that $\eps=+1$
corresponds to both Killing vectors being spacelike and 
$\eps =-1$ corresponds to one being spacelike and the other being 
timelike. For the reduction one then has two options: {\bf (a)} direct 
reduction, and {\bf (b)} reduction and dualization. For $\eps = +1$ 
both procedures lead to identical actions. For $\eps =-1$ 
the reduced actions are classically equivalent but differ by a
crucial signature which (most likely) leads to distinct quantum 
theories.   

{\bf (a) Direct reduction:} The ansatz for the line element reads
\cite{Ernst,MatzMis67}  
\be
\eps dS^2 = \gamma_{\mu\nu}(x) dx^{\mu} dx^{\nu} -
\rho(x) M_{ab}(x) dy^a dy^b\,.
\label{lineel}  
\ee
Here $\gamma_{\mu\nu}(x)$ is a 2D metric with Lorentzian signature 
if $\eps =+ 1$ and with Euclidean signature if $\eps =-1$. $M_{ab}(x)$ 
is a symmetric $2 \times 2$ matrix normalized to have determinant 
$\det M = \eps$ (so that $\det G <0 $ always). In particular with 
these conventions we can assume $\rho \geq 0$ for the degree of freedom 
parameterizing the determinant.  
In the axisymmetric case $y^2$ is the time variable and the overall 
$\eps =-1$ sign on the right hand side of (\ref{lineel}) is needed 
to restore the $(1,-1,-1,-1)$ signature. In order to minimize the 
number of sign flips in the 2D theory we shall base the block 
decomposition on the 4D metric $\eps G_{MN}$. Since the 4D Ricci 
scalar changes sign under a sign flip of the metric this can be 
compensated simply by multiplying the reduced action by $\eps$. 

We shall assume throughout that the metric $\gamma_{\mu\nu}(x)$ is 
conformally flat, i.e.~that by a diffeomorphism in the $(x^0,x^1)$ 
coordinates it can be brought into the form 
\be
\gamma_{\mu\nu} = \eta_{\mu\nu} \,e^{\sigma}\;,\quad 
\eta_{\mu\nu} = \left(\begin{array}{cc} 1 & 0\\ 0 &-\eps \end{array} \right)
\sspace \mbox{with}\;\;\eps = \pm 1\,.
\label{confgauge}
\ee
For $\eps =1$ this is no restriction, for $\eps =-1$ there could be 
topological obstructions to achieving (\ref{confgauge}) globally. 
For the matrix $M$ often a parameterization in terms of `hyperbolic spins' 
$n^j,\,j =0,1,2$, is useful, where $(n^0)^2 - (n^1)^2 -(n^2)^2 =: n\cdot n =
\eps$. Explicitly
\be
M = \left(\begin{array}{cc} n^0 + n^2 & - n^1 \;,\\
- n^1 & n^0 - n^2 \end{array} \right) \;,\sspace \det M = \eps \,,
\sspace  n^j \tau_j = M \, \tau_0 \;, 
\label{Mn}
\ee
where $\tau_j,\,j=0,1,2$, is a basis of $sl(2,\R)$. Frame rotations in 
the (differentials of the) Killing coordinates then induce O$(1,2)$ rotations 
of the $n^j$, via
$$
{dy^1 \choose dy^2 } \rra A^T {dy^1 \choose dy^2}\;,
\quad A \in SL(2,\R)\quad 
\Longrightarrow \quad n^j \tau_j \rra A(n^j \tau_j)A\inv = 
(n^j \Lambda _j^k) \tau_k\,,
$$
with an O$(1,2)$ matrix $\Lambda$. By definition O$(1,2)$ preserves
the constraint $n \cdot n =\eps$ and thus is the symmetry group of a 
2D hyperboloid $H_{\eps}$. The sign $\eps$ determines whether the 
hyperboloid is one- or two-sheeted; in the latter case we restrict 
attention to one branch (leaving only SO$(1,2)$ as the invariance group). 
Explicitly 
\ba
&& H_+ = \{n \in \R^{1,2}\,|\, n \cdot n = 1,\,n^0 > 0\}
\quad   \mbox{two-sheeted hyperboloid}\;,\nonum
&& H_- = \{n \in \R^{1,2}\,|\, n \cdot n = -1\}
\sspace\quad\;  \mbox{one-sheeted hyperboloid}\;.
\label{hyp}
\ea
Both hyperboloids are (pseudo-) Riemannian spaces of constant 
negative curvature, normalized to $-2$ in our conventions. We write 
$ds^2_{H_{\eps}} = (d\Delta^2 + \eps dB^2)/\Delta^2$ for the 
metric in canonical coordinates $(\Delta,B)$. The latter are related 
to the hyperbolic spins by 
\be
n^0 = \frac{1 + \eps \Delta^2 + B^2}{2\Delta}\;,\quad
n^1 =  -\frac{B}{\Delta}\;,\quad
n^2 = \frac{-1 + \eps \Delta^2 + B^2}{2\Delta}\;.
\label{nspins}
\ee
For the matrix $M$ it amounts to a Gauss decomposition
$$
M =  \left( \begin{array}{cc} 1 & B \\
0 & 1  \end{array}\right) 
\left( \begin{array}{cc} \eps \Delta & 0 \\
0 & \Delta^{-1}  \end{array}\right)
\left( \begin{array}{cc} 1 & 0\\
B & 1  \end{array}\right) \,,
$$
and the line element (\ref{lineel}) takes the form
\be
dS^2 = e^{\sigma}[\eps (dx^0)^2 - (dx^1)^2] -\eps \frac{\rho}{\Delta}\,
(dy^2 + B dy^1)^2 -\rho \Delta(dy^1)^2\,.
\label{Elineel}
\ee
The complex combination $E = \Delta + i B$ is the Ernst potential 
\cite{Ernst}. The signature of the Killing vectors poses the constraints
$\rho/\Delta > 0$ and $\Delta^2 + \eps B^2 >0$.

Inserting the ansatz (\ref{Elineel}) into the Einstein equations 
a system of partial differential equations is obtained which
on general grounds (see \cite{TorreFels01} for a survey) coincide 
with the Euler-Lagrange equations following from the reduced action
(\ref{Lgamma}). Importantly also the symplectic structure following from 
the 2D action coincides with the restriction of the symplectic 
structure one has on the full phase space of general relativity. 
This is crucial with regard to quantization. Strictly speaking 
this equivalence of the symplectic structures has been shown only 
for the $\eps =1$ case, but most likely it also holds in the axisymmetric 
sector;~see e.g.~\cite{TorreRom96,TorreFels01}  
for a discussion.  A complete description of the phase space 
in addition requires the specification of boundary or fall-off 
conditions for the fields. For the cylindrical waves and the 
generalized Gowdy cosmologies this is available in the literature 
\cite{TorreRom96}; for the other sectors the results are incomplete. 
For the development of a perturbative quantum theory, however, not all 
of the subtle differences in signatures and boundary terms are important. 
The essential dynamical information about the phase space 
(as embedded into the full phase space of general relativity) 
is contained in the 2D reduced action. In the bulk of the article  
we therefore distinguish only the two main situations -- one or both 
Killing vectors spacelike -- by a sign ($\eps = -1,1$, respectively) 
and try to develop the framework for all subsectors 
simultaneously, starting directly from the reduced action.

The constraints associated with the 2D diffeomorphism invariance 
of (\ref{Lgamma}) are obtained by the familiar ADM prodecure,
see e.g.~\cite{TorreRom96} in the present context. A technically 
convenient shortcut is to compute the ``would-be'' energy momentum 
tensor by varying the action $S_{\eps} = \int \!d^2 x L_{\eps}$ with respect 
to the metric. One finds 
\ba
\label{constr}
\lb \,T_{\mu\nu} \is  
\frac{\lb}{\sqrt{\gamma}} 
\frac{\delta S_{\eps}}{\delta \gamma^{\mu\nu}}\Bigg|_{\gamma = e^{\sigma} \eta}
= 
-\rho \bigg( \dd_{\mu}n\cdot \dd_{\nu} n - \eta_{\mu\nu}
\frac{1}{2} \dd^{\kappa} n \cdot \dd_{\kappa} n \bigg) 
\\[2mm] 
&\!-\!&\!\!\eps \bigg(\dd_{\mu}\rho \dd_{\nu} + \dd_{\nu} \rho \dd_{\mu}
- \eta_{\mu\nu}\,\dd^{\kappa} \rho \dd_{\kappa} \bigg) 
\Big(\sigma + \frac{1}{2} \ln \rho \Big) 
+ 2\eps (\dd_{\mu}\dd_{\nu} - \eta_{\mu\nu} 
\dd^{\kappa} \dd_{\kappa})\rho\;.
\nonumber
\ea
Re-expressed in terms of the momenta $T_{00}$ coincides with 
the hamiltonian constraint $\cH_0$ and $T_{01}$ coincides with 
the 1D diffeomorphism constraint $\cH_1$. Due to the last 
(`anti-improvement') term in (\ref{constr}) the trace is non-zero, 
$\lb\,T^{\mu}_{\;\;\mu}= -2 \eps \dd^{\mu}\dd_{\mu} \rho$. Its form 
however complies with conformal invariance of the flat space system;
see e.g.~\cite{Polch}. More directly the trace can also be obtained by 
varying with respect to $\sigma$: 
$$ 
T^{\mu}_{\;\;\mu} = -2 \frac{\delta}{\delta \sigma} 
\int \!d^2x\,L_{\eps}(n,\rho, \sigma )\,.
\label{trace}
$$ 
The overall sign in (\ref{constr}) has been chosen 
such that upon reduction $\rho,\,\sigma=const$, $T_{\mu\nu}$ becomes the 
energy-momentum tensor of the O$(1,2)$ nonlinear sigma-model. Its 
energy density then is positive semi-definite 
for $n\cdot n =1$ and indefinite for $n \cdot n =-1$. Among the 2D Weyl 
transformations in Eq.~(\ref{Lconf}) with $\nabla^2 \omega =0$ are 
those induced by diffeomorphisms of the form 
$x^{\pm} \ra f^{\pm}(x^{\pm})$, where $x^{\pm}$ are lightcone 
coordinates with respect to the flat metric $\eta_{\mu\nu}$. 
Our conventions are 
$x^{\pm} = (x^0 \pm x^1)/2$ for $\eps =1$ 
and $x^{\pm} = (x^0 \pm i x^1)/2$ for $\eps =-1$.
We write $\dd_{\pm} = \dd_0 \pm \sqrt{\eps} \dd_1$, with 
$\sqrt{\eps} = 1,-i$, for $\eps = 1,-1$, respectively.
The transformations $x^{\pm} \ra f^{\pm}(x^{\pm})$ are 
the usual 2D conformal transformations that preserve the conformally
flat form of the metric with $\sigma \ra \sigma - \ln \dd_+ f^+ \dd_-f^-$.
The lightcone components $T_{\pm\pm}$ transform covariantly 
as second rank tensors since the non-covariance of $\sigma$ 
cancels that of $\dd_+^2 \rho$. The trace $T_{+-}$, non-zero off-shell, 
could be canceled by switching to an improved
$T_{\mu\nu}$ in the usual way. Technically it is often simpler to put 
$\rho$ on shell, $\dd_+\dd_-\rho =0$, and express $T_{\pm\pm}$ in terms 
of the conformal scalar $\sigma - \frac{1}{2}
\ln (\dd_+ \rho \dd_-\rho)^2$ \cite{NKS96}.

In conformal gauge $\gamma_{\mu\nu} = e^{\sigma} \eta_{\mu\nu}$ 
the action based on (\ref{Lgamma}) becomes that of a flat space 
sigma-model 
\be
S_{\eps} =  -\frac{1}{2\lb} \int d^2 x 
\left[\rho \dd^{\mu} n \cdot \dd_{\mu} n
+ \eps \dd^{\mu}\rho \dd_{\mu}(2 \sigma + \ln \rho)\right]\;. 
\label{action2}
\ee
The $T_{\pm\pm}$ constitute its energy-momentum tensor and 
the gravitational origin of the system enters only through the 
vanishing conditions $T_{\pm\pm} \approx 0$. They can be verified to be 
first class constraints and to generate two commuting copies of a 
(centerless) Virasoro-Witt algebra with respect to the Poisson structure 
induced by (\ref{action2}). On general grounds the Einstein equations 
for the metrics (\ref{lineel}) will coincide with those obtained 
by variation of the reduced action (\ref{action2}). 
The equations of motion are 
\ba
&& \dd_{\mu}\dd^{\mu} \rho =0\;,\sspace 
\dd^\mu \dd_{\mu}(2 \sigma + \ln \rho)
= \eps \dd^{\mu}n \cdot \dd_{\mu} n\;,\nonum
&& \dd^{\mu}(\rho \dd_{\mu} n^j) + 
\eps \rho(\dd^{\mu}n\cdot \dd_{\mu} n) n^j =0\;.
\label{motion}
\ea
In particular they ensure the consistency conditions $\dd_- T_{++} =0 
= \dd_+ T_{--}$. 
\bigskip

{\bf (b) Dual action:} Recall that in canonical coordinates $(\Delta,B)$ 
one has $\dd^{\mu} n \cdot \dd_{\mu} n = -
(\eps \dd^{\mu} \Delta \dd_{\mu} \Delta + \dd^{\mu} B \dd_{\mu} B)/\Delta^2$.
The action (\ref{action2}) thus has a manifest $B \ra B + {\rm const}$ invariance.
It is instructive to perform an abelian T-duality transformation (in the 
sense of Buscher \cite{Buscher}) with respect to this symmetry. Denoting the 
dual field by $B_{\d}$ the resulting Lagrangian is 
\ba
L^{\d}_{\eps} \is \frac{1}{2\lb}
\Big[ \eps \frac{\rho}{\Delta^2} \dd^{\mu} \Delta \dd_{\mu} \Delta + 
\eps \frac{\Delta^2}{\rho} \dd^{\mu}B_{\d} \dd_{\mu} B_{\d}
-\eps \rho^{-1} \dd^{\mu} \rho \dd_{\mu} \rho - 2 \eps \rho^{-1} \dd^{\mu} \rho
\dd_{\mu} \sigma \Big]\,,
\nonum
&+& \eps \dd^{\mu} [\rho \dd_{\mu} 
( \sigma + \ln(\rho^{1/2}/\Delta)) + 2\dd_{\mu} \rho]\,. 
\label{Ldual}
\ea
The total derivatives are introduced for later use. By means of an 
involutive field redefinition the dual Lagrangian can {\it almost} be 
brought back into the original form 
\ba
&& L^{\d}_{\eps}(\phi_{\d})) = \eps L_+(\eta(\phi_{\d}))\,,
\sspace \eta^2 = {\rm id} \,,  
\nonum
&& \eta (\Delta, B_{\d}, \rho, \sigma) = \Big(\frac{\rho}{\Delta},
B_{\d}, \rho, \sigma + \frac{1}{2} \ln \rho - \ln \Delta\Big) 
=: (\Delta_{\d}, B_{\d}, \rho, \sigma_{\d} ) \,.
\label{LDD}
\ea
Here $\phi_{\d} = (\Delta, B_{\d}, \rho,\sigma)$ and $L_+$ is the Lagrangian 
(\ref{Lgamma}) in conformal gauge ($\gamma_{\mu\nu} = e^{\sigma} \eta_{\mu\nu}$),
including total derivatives. In our conventions 
$\dd_{\mu} B_{\d} = - \eps \rho \Delta^{-2} \eps_{\mu\nu} 
\dd^{\nu} B$, on shell. One notices an important (convention-independent) 
difference between the situation when both Killing vectors are spacelike 
($\eps =1$) and when one is timelike ($\eps =-1$). For $\eps =+1$ 
the signature of the target space remains unchanged under dualization
while for $\eps =-1$ it changes into $(-,-,+,-)$.%
\footnote{The same result is obtained by first performing the reduction 
from 4 to 3 dimensions, dualizing the 3D theory and then reducing with 
respect to the second Killing vector \cite{BMG88,Maison00}. The additional sign flip 
as compared to the usual dualization formulas can also be understood 
in terms of the familar (formal) functional integral argument; c.f.~section 2.2.} 
As a consequence the action (\ref{Ldual}) can only for $\eps = +1$ be 
obtained directly from a local and real 4D line element, namely 
\be
dS_{\d}^2 = \frac{\rho^{1/2}}{\Delta} e^{\sigma} [\eps (dx^0)^2 - (dx^1)^2] 
- \eps \Delta\,(dy^2 + \psi dy^1)^2 - \frac{\rho^2}{\Delta} (dy^1)^2\,,
\label{Dlineel}
\ee
with $\eps =+1$. For $\eps =-1$ the line element (\ref{Dlineel}) 
describes stationary axisymmetric solutions; the spacelike/timelike 
nature of the two Killing vectors $\dd/\dd y^1$ and $\dd/\dd y^2$ 
poses the conditions $\Delta >0$ and $\rho^2 + \eps \psi^2 \Delta^2 >0$.
Evaluating the Einstein-Hilbert Lagrangian on 
(\ref{Dlineel}) gives
\ba
\nspace -\sqrt{-\det G_{\d}}\, R(G_{\d}) \is 
\frac{1}{2} \left[ \eps
\frac{\rho}{\Delta^2}\dd^{\mu} \Delta \dd_{\mu}\Delta + 
\frac{\Delta^2}{\rho} \dd^{\mu} \psi \dd_{\mu} \psi
- \eps \frac{\dd^{\mu} \rho \dd_{\mu} \rho}{\rho}  
-2 \eps \dd^{\mu} \rho \dd_{\mu} \sigma\right]
\nonum
&+& \eps \dd^{\mu} [\rho \dd_{\mu} 
( \sigma + \ln(\rho^{1/2}/\Delta)) + 2\dd_{\mu} \rho]\,. 
\label{LDdirect}
\ea
One sees that for $\eps =1$ the identification 
$\psi = B_{\d}$ reproduces (\ref{Ldual}). In contrast for $\eps =-1$ a purely 
imaginary dual field would be needed $\psi = \pm i B_{\d}$
to match (\ref{LDdirect}) and (\ref{Ldual}), which would however make the line 
element (\ref{Dlineel}) complex. The mapping $(\Delta, B_{\d}, \rho, \sigma) 
\ra ( \Delta_{\d}, \sqrt{\eps} B, \rho, \sigma_{\d})$ which in modern terminology 
brings the T-dual action back into identically the original form (with 
fields local with respect to the line element) is known as the 
Kramer-Neugebauer involution \cite{KN}. Its existence 
is closely related to the classical integrability of the system.  
Because of the signature change no local real line element can reproduce 
(\ref{Ldual}) in the Einstein-Hilbert Lagrangian. Nevertheless also for 
$\eps =-1$ is the dual Lagrangian classically  equivalent to the original one. 
For the equations of motion this holds by construction, for the Poisson structure 
one can check directly that the transition from $L_{\eps}$ in (\ref{Lgamma}) to 
$L^{\d}_{\eps}$ in (\ref{Ldual}) is a canonical transformation. 

In summary, after dualization both sectors $\eps =1$ and $\eps = -1$ 
are described by reduced actions whose `matter' part corresponds to 
a sigma-model on a noncompact Riemannian (rather than pseudo-Riemannian) 
symmetric space, i.e.~the hyperboloid $H_+$ in (\ref{hyp}).

\newpage
\newappendix{Renormalization of Riemannian sigma-models}
 
For convenient reference we review here some aspects of the 
renormalization of Riemannian sigma-models. We largely follow
the thorough treatment of Osborn \cite{Osb87}. As usual we adopt 
the covariant background field expansion, dimensional regularization 
and minimal subtraction. For the purposes of renormalization 
it is useful to consider an extended Lagrangian of the 
form 
\ba
\lb L(G; \phi) = \frac{1}{2} \gamhat^{\mu\nu} g_{ij} 
\dd_{\mu} \phi^i \dd_{\nu} \phi^j + 
\gamhat^{\mu\nu} \dd_{\mu} \phi^i V_{\nu i} + 
\frac{1}{2} R^{(2)}(\gamhat)\Phi + F\,. 
\label{Lsource}
\ea    
Here $G = \{g,V,\Phi,F\}\,,\;G = G(\phi;x)$ is a collection 
of generalized couplings/sources (of the tensor type indicated 
by the index structure) that depend both on the fields $\phi^j$ 
and explicitly on the point $x$ in the ``base space''. The
latter is a 2-dimensional Riemannian space with metric 
$\gamhat_{\mu\nu}(x)$, extended to $d$ dimensions in 
the sense of dimensional regularization, and $R^{(d)}(\gamhat) = 
R^{(2)}(\gamhat)/(d-1)$. The action functional 
is $S[G;\phi] = \int d^d x \sqrt{\gamhat} L(G;\phi)$. 
The explicit $x$-dependence of the sources $G$ allows one to define
local composite operators via functional differentiation after 
renormalization. In addition the scalar source $F$ provides 
an elegant way to compute the nonlinear renormalizations 
of the quantum fields in the background expansion \cite{HPS88}.

The background field method involves decomposing the fields $\phi^j$ into 
a classical background field configuration $\varphi^j$ and a 
formal power series in the quantum fields $\xi^j$ whose coefficients 
are functions of $\varphi^j$. The series is defined in terms of the 
unique geodesic from the point $\varphi$ in the target manifold 
to the (nearby) point $\phi$, where $\xi^j$ is the tangent 
vector at $\varphi$. We shall write $\phi^j(\varphi;\xi)$ for 
this series, and refer to $\phi,\,\varphi$, and $\xi$ as the full 
field, the background field, and the quantum field, respectively. 
On the bare level one starts with $\phi^j_{\b} := \phi^j(\varphi_{\b}; \xi_{\b})$ 
which upon renormalization are replaced by 
$\phi^j = \phi^j(\varphi; \xi)$. The transition function 
$\xi_{\b}(\xi)$ can be computed from the differential operator 
$Z -1$ in Eq.~(\ref{Rcounter1}) below. For our purposes we in addition 
have to allow for a renormalization $\varphi_{\b}(\varphi)$ of the background 
fields. As usual we adopt the convention that the fields
$\phi_{\b}^j$ remain dimensionless for base space dimension $d\neq 2$. 
Then the bare couplings/sources $G^{\b}(\phi_{\b};x)$ have dimension 
$[\mu]^{d-2}$ and are expressed as a dimensionless sum of the 
renormalized $G(\phi;x)$ and covariant counter tensors 
built from $G(\phi;x)$. A suitable parameterization is 
\ba
g^{\b}_{ij} \is \mu^{d-2}\left[ g_{ij} + T_{ij}(g)\right]\;,
\nonum
V_{\mu i}^{\b} \is  \mu^{d-2}\left[ Z^V(g)_i^j V_{\mu j} + N_i^{\;\;jk}(g)
\overline{\dd}_{\mu} g_{jk} \right]\;,
\nonum
\Phi^{\b} \is  \mu^{d-2}\left[ Z(g) \Phi + \Psi(g) \right]\;,
\nonum
F^{\b} \is \mu^{d-2}\left[ Z(g) F + Y \right]\;.
\label{Rcounter1}
\ea
Here $\overline{\dd}_{\mu}$ denotes differentiation with 
respect to $x$ at fixed $\phi$. The quantities $T_{ij},
N_i^{\;\;jk}, \Psi, Y$ and $Z^V\!\! -1, Z-1$ contain poles and 
only poles in $(2-d)$ whose coefficients are defined by 
minimal subtraction. Except for $Y$ they depend on $g_{ij}$
only; $Y$ in addition depends quadratically on $V_{\mu i}$ 
and $\overline{\dd}_{\mu} g_{jk}$, but the quadratic forms 
with which they are contracted again only depend on $g_{ij}$.
All purely $g$-dependent counter tensors are algebraic 
functions of $g_{ij}$, its covariant derivatives and its 
curvature tensors. $Z-1$ and $Z^V\!\! -1$ specifically are linear 
differential operators acting on scalars and vectors on the target 
manifold, respectively. The combined pole and loop expansion 
takes the form 
\be
\cO = \sum_{\nu \geq 1} \sum_{l \geq \nu} \frac{1}{(2-d)^{\nu}} 
\Big( \frac{1}{2\pi} \Big)^l \;\cO^{(\nu,l)}\,,   
\label{nulpieces}
\ee 
for any of the quantities $T_{ij},N_i^{\;\;jk}, \Psi, Z^V\!\! -1, 
Z-1,Y$. The residue of the simple pole is denoted by $\cO^{(1)}$. 
We do not include explicitly powers of the loop counting
parameter $\lb$ in (\ref{nulpieces}). For the purely $g$-dependent 
counter terms of interest here they are easily restored by inserting 
$g/\lb$ and utilizing the scaling properties listed below. 
However once $g$ is `deformed' into a nontrivial function of $\lb$ 
the `scaling decomposition' (\ref{nulpieces}) no longer coincides with 
the expansion in powers of $\lb$ and the former is the fundamental one.
Under a constant rescaling of the metric the purely $g$-dependent 
counter term coefficients transform homogeneously as follows
\ba
&& \cO^{(\nu,l)}(\Lambda^{-1} g) = \Lambda^{l-1} \cO^{(\nu,l)}(g)\,\quad 
\mbox{for} \quad \cO = T_{ij},\,\Psi\,,
\nonum
&& \cO^{(\nu,l)}(\Lambda^{-1} g) = \Lambda^l \;\cO^{(\nu,l)}(g)\,\quad 
\mbox{for} \quad \cO = Z,\,Z^V,\,N\,.
\label{Rcounter5}
\ea
In principle the higher order pole terms 
$\cO^{(\nu,l)},\,l \geq \nu \geq 2$, are determined recursively by 
the residues $\cO^{(1,l)}$ of the first order poles via 
``generalized pole equations''. The latter can be worked out 
in analogy to the quantum field theoretical case; see   
\cite{AGFM81,CurciPaff,Osb87}. 
Taking the consistency of the cancellations for granted one can 
focus on the residues of the first order poles, which we shall do 
throughout. 

Explicit results for them are typically available up to 
and including two loops \cite{Frie85,Tseytlin87,CurciPaff,HPS88,Osb87}. 
For the metric and $\Phi$ also the three-loop results are known:%
\footnote{Our conventions are: 
$\nabla_i v^k = \dd_i v^k + \Gamma^k_{\;\;ij}\,v^j$, 
with $\Gamma^k_{\;\;ij} = 
\frac{1}{2}g^{kl}[\dd_j g_{il} + \dd_i g_{jl} - \dd_l g_{ij}]$.
The Riemann tensor is defined by 
$(\nabla_i \nabla_j - \nabla_j \nabla_i)v^k = 
R^k_{\;\;lij}\,v^l$, so that $R^k_{\;\;lij} = \dd_i \Gamma^k_{\;\;lj}- 
\dd_j \Gamma^k_{\;\;li} + 
\Gamma^k_{\;\;im} \Gamma^m_{\;\;\;lj} - 
\Gamma^k_{\;\;jm} \Gamma^m_{\;\;\;li}$.
The Ricci tensor is $R_{ij} = R^m_{\;\;imj}$. 
For the computation of curvature tensors the maple tensor package is 
useful. Compared with our conventions one has 
$R_{ijkl} = (R_{ijkl})^{\rm maple}$, $R_{ij} = - (R_{ij})^{\rm maple}$.}
\ba
T_{ij}^{(1,1)}(g) \is R_{ij}\;,
\nonum
T_{ij}^{(1,2)}(g) \is \frac{1}{4} R_{iklm}\,R_j^{\;\;klm}\;.
\nonum
T_{ij}^{(1,3)}(g) \is \frac{1}{6} R_{imn}^{\;\;\;\;\;\;\;k} 
R_{jpqk} R^{pnmq} - \frac{1}{8} R_{iklj} R^k_{\;\;mnp}R^{lmnp} 
- \frac{1}{12} \nabla_n R_{iklm} \nabla^k R_{j}^{\;\, lmn}\,, 
\label{Rcounter2}
\ea
where the three-loop term has been computed independently by
Fokas-Mohammedi \cite{FokasMohamm87} and Graham \cite{Graham}.
For $\Phi$ the results are \cite{HullTown86,Osb87,CurciPaff,Tseytlin87} 
\be
\Psi^{(1,1)}(g) = \frac{c_T}{6} \,,\quad \Psi^{(1,2)}(g) = 0 \,,\quad
\Psi^{(1,3)}(g) = \frac{1}{48} R_{ijkl} R^{ijkl} \,,
\label{Rcounter3}
\ee 
where $c_T$ is the dimension of the target manifold. 
For the other quantities one has \cite{Osb87, HPS88, Tseytlin87,JJM89}: 
\begin{subeqnarray}
[Z^V(g)_i^j]^{(1,1)} &=& \frac{1}{2}[ - \nabla^2 \delta_i^j + R_i^{\;j}]\,,
\\ \nonumber
[Z^V(g)_i^j]^{(1,2)} &=& \frac{1}{4} R_i^{\;\,klj} \nabla_k \nabla_l\;,
\\[1mm] 
Z(g)^{(1,1)} &=& -\frac{1}{2} \nabla^2\,,\quad Z(g)^{(1,2)} = 0\,,
\\ \nonumber
Z(g)^{(1,3)}  &=& - \frac{3}{16} R^{iklm}R^j_{\;\,klm} \nabla_i \nabla_j\,.
\\[1mm]
[N_i^{\;jk}(g)]^{(1,1)}
&=& \frac{1}{2} \delta_i^j \nabla^k - \frac{1}{4} g^{jk} \nabla_i\;,
\\ \nonumber
[N_i^{\;jk}(g)]^{(1,2)}\, &=& \frac{1}{2} R_i^{\;\,jkl}\nabla_l\,.
\label{Rcounter4}
\end{subeqnarray} 
The expressions for $Y^{(1,1)}$ and $Y^{(1,2)}$ are likewise 
known \cite{Osb87} but are not needed here.

Some explanatory comments should be added. First, in addition 
to the minimal subtraction scheme the above form of the 
counter tensors refers to the background field expansion in terms of 
Riemannian normal coordinates. If a different covariant expansion 
is used the counter tensors change (see e.g.~\cite{HullTown86} for 
a one-loop illustration). Likewise the standard form 
of the higher pole equations \cite{AGFM81,CurciPaff,Osb87}
is only valid in a preferred scheme. E.g.~for the metric counter 
terms in this scheme additive contributions to $T_{ij}(g)$ 
of the form $\cL_V g_{ij}$ are absent \cite{HPS88}. 
Note that adding such a term for $\nu=1$ leaves the metric beta 
function in Eq.~(\ref{Gflow}) below unaffected, provided $V^j$ 
is functionally independent of $g_{ij}$.

So far only the full fields entered, $\phi^j_{\b}$ on the bare and  
$\phi^j$ on the renormalized level. Their split into 
background and quantum contributions is however likewise 
subject to renormalization. A convenient way to determine the
transition function $\xi_{\b}^j(\xi)$ from the bare to the 
renormalized quantum fields was found by Howe, Papadopolous and   
Stelle \cite{HPS88}. In effect one considers the inversion
$\xi^j(\varphi; \phi- \varphi)$ of the normal coordinate expansion 
$\phi^j = \phi^j(\varphi; \xi)$ of the renormalized fields. 
If Z in (\ref{Rcounter1}) is regarded as a differential operator 
acting on the second argument of this function, i.e.~on $\phi$,  
\be 
\xi^j(\xi_{\b}) = Z\,\xi^j(\varphi; \phi- \varphi)\,,
\label{xiren1}
\ee
one obtains the desired $\xi^j_{\b}(\xi)$ relation by inversion.
To lowest order $Z^{(1,1)} = -\frac{1}{2} \nabla^2$ yields 
\be
\xi_{\b}^i = \xi^i + \frac{1}{2 -d} \frac{\lb}{2\pi} 
\left[ \frac{1}{3} R^i_{\;j} \xi^j + \frac{1}{4} 
\nabla_k R^i_{\;j} \xi^k \xi^j - \frac{1}{24} 
\nabla^i R_{kj} \xi^k \xi^j + O(\xi^3) \right]\,.
\label{xiren2} 
\ee
At each loop order the coefficient is a power series in $\xi$
whose coefficients are covariant expressions built from 
the metric $g_{ij}(\varphi)$ at the background point.  

With all these renormalizations performed the result can be 
summarized in the proposition \cite{HPS88,Osb87} 
that the source-extended background functional 
\be
\exp \Gamma[G;\varphi] = \int [\cD \xi] \exp\left\{ 
- S[G_{\b},\phi_{\b}] + \frac{1}{\lb} \int \!d^dx J_i(\varphi) \xi^i \right\}   
\label{Gamma}
\ee
defines a finite perturbative measure to all orders of the loop
expansion. The additional source $J_i(\varphi)$ here is constrained 
by the requirement that $\bra \xi^j \ket =0$. The key properties 
of $\Gamma(G;\varphi)$ are: 
\begin{itemize}
\item It is invariant under reparameterizations 
of the background fields $\varphi$.
\item It obeys a simple renormalization group equation (which 
would not be true without the F-source). 
\item A generalized action principle holds that allows one 
to construct local composite operators of dimension $0,1,2$, by 
variation with respect to the renormalized sources.
\end{itemize}     
Let $V(\phi),\,V_i(\phi),\,V_{ij}(\phi)$ be a scalar, a vector, 
and a symmetric tensor on the target manifold, respectively. `Pull-back'
composite operators of dimension 0,1,2 are defined by \cite{Osb87}
\ba
\nl V(\phi)\nr \is \lb V \cdot \frac{\dd}{\dd F} L_{\b} 
=\mu^{d-2} Z(g) V\,,
\nonumber \\
\nl V_i(\phi)  \dd^{\mu} \phi^i \nr \is 
\lb V_i\cdot \frac{\dd}{\dd V_{\mu i}} L_B = 
\mu^{d-2}\left[ \dd^{\mu} \phi^i Z^V(g)_i^{j} V_j 
+ V_i \cdot\frac{\dd}{\dd V_{\mu i}} Y \right]\,,
\\
\nl \frac{1}{2} V_{ij}(\phi) \dd^{\mu} \phi^i \dd_{\mu} \phi^j \nr
\is \lb V\cdot \frac{\dd}{\dd g} L_B  
-\frac{\mu^{d-2}}{\sqrt{\gamhat}} \dd_{\mu} \left[\sqrt{\gamhat} 
\dd^{\mu} \phi^i N_i^{jk}(g) \, V_{jk} +\sqrt{\gamhat} V_{ij} \cdot 
\frac{\dd}{\dd (\overline{\dd}_{\mu} g_{ij})} Y \right]. 
\nonumber
\label{normalproducts}
\ea
The functional derivatives here act on functionals on the target 
manifold at fixed $x$, e.g.~$V \!\cdot \!\frac{\dd}{\dd F} = 
\int d^D\phi \sqrt{g} 
V(\phi;x) \frac{\dd}{\dd F(\phi;x)}$. For $g_{ij}$ in addition 
the dependence of the counter terms on $\overline{\dd}_{\mu}g_{ij}$
has to be taken into account, so that 
$V\cdot \frac{\dd}{\dd g} := V_{ij} \cdot \frac{\dd}{\dd g_{ij}}
+ \overline{\dd}_{\mu} V_{ij} \cdot 
\frac{\dd}{\dd (\overline{\dd}_{\mu} g_{ij})}$. 
Further $L_{\b} = L(G_{\b},\phi_{\b})$
is the bare Lagrangian regarded as a function of the renormalized 
quantities. The contractions on the base space are with respect to 
the backgound metric $\gamhat_{\mu\nu}$. The additional total 
divergence in (\ref{normalproducts}c) reflects the effect of 
operator mixing. The normal products as given in (\ref{normalproducts})  
still refer to the functional measure as defined by the source-extended
Lagrangian. After all differentiations have been performed the 
sources should set to zero or rendered $x$-independent again to get 
the composite operators e.g.~for the purely metric sigma-model.

The definition (\ref{normalproducts}) of the normal products is 
consistent with redefinitions of the couplings/sources 
that change the Lagrangian only by a total divergence. 
The operative identities are
\be
(Z^V)_i^j \dd_j V = \dd_i (Z V) \,,\sspace 
(\overline{\dd}^{\mu} Z)V = \dd_i V \cdot 
\frac{\dd Y}{ \dd V_{\mu i}}\,,  
\label{ZVZY} 
\ee
for a scalar $V(\phi;x)$. They entail 
\be
\dd_{\mu} \nl V \nr = \nl \dd_i V \dd_{\mu} \phi^i \nr + 
\nl \overline{\dd}_{\mu} V \nr\,.
\label{dnormalproduct}
\ee 
Moreover the invariance of the regularization under reparameterizations 
of the target
manifold allows one to convert the reparameterization invariance 
of the basic Lagrangian (\ref{Lsource}) into a ``diffeomorphism 
Ward identity'' \cite{Shore,Osb87}: 
\ba
&\nspace & \frac{1}{\sqrt{\gamhat}} \dd^{\mu} 
\nl\sqrt{\gamhat} \lb J_{\mu}(v) \nr 
= \nl \frac{1}{2} \cL_v g_{ij} \dd^{\mu} \phi^i \dd_{\mu} \phi^j + 
\dd^{\mu} \phi^i \cL_v V_{\mu i} + \frac{1}{2} R^{(2)}(\gamhat) 
\cL_v \Phi + \cL_v F \nr - \lb v^i \cdot \frac{\delta S_{\b}}{\delta \phi^i},  
\nonum
&\nspace & \bspace \quad \mbox{with} \quad \lb J_{\mu}(v) = 
\dd_{\mu} \phi^i v_i + v^i V_{\mu i}\,. 
\label{diffward}
\ea
The Lie derivative terms on the right hand side are the response 
of the couplings/sources under an infinitesimal diffeomorphism 
$\phi^j \ra \phi^j + v^j(\phi)$. Thus $J_{\mu}(v)$ may be 
viewed as a ``diffeomorphism current''. The last term on the right 
hand side is the (by itself finite) ``equations of motion operator''.     
In deriving (\ref{diffward}) the following useful consistency conditions
arise
\ba
g^{\b}_{ij} v^j \is \mu^{d-2}\left[ Z^V(g)_i^j v_j + N_i^{\;jk}(g) 
\cL_v g_{jk} \right]\,,
\nonum
v^i V_{\mu i}^{\b} \is \mu^{d-2}\left[ \cL_v g_{ij} \cdot 
\frac{\dd Y}{ \dd (\overline{\dd}^{\mu} \!g_{ij})} + 
v_i \cdot \frac{\dd Y}{ \dd V_i^{\mu}} + Z( v^i V_{\mu i}) \right]\,.
\label{diffconsist} 
\ea

So far the renormalization was done at a fixed normalization scale 
$\mu$. The scale dependence of the renormalized couplings/sources 
$G = \{g_{ij}, V_{\mu i}, \Phi, F\}$ is governed by a 
set of renormalization functions which follow from (\ref{Rcounter1}).
For a counter tensor of the form (\ref{nulpieces}) it is convenient 
to introduce 
\be
\dot{\cO} = - \sum_{l \geq 1} \Big(\frac{\lb}{2\pi} \Big)^l \,l\, 
\cO^{(1,l)} \,,
\label{Odot}
\ee
which in view of (\ref{Rcounter5}) can be regarded as a parametric 
derivative of $\cO^{(1)}$. Then
\ba 
\mu \frac{d}{d \mu} g_{ij} \is \beta_{ij} := (2 -d) g_{ij} - 
\dot{T}_{ij}\,,
\nonum 
\mu \frac{d}{d \mu} V_{\mu i} \is \gamma^V := 
(2 -d) V_{\mu i} - (\dot{Z}^V)_i^j
V_{\mu j} - \dot{N}_i^{\;jk} \overline{\dd}_{\mu} g_{jk} \,,
\nonum
\mu \frac{d}{d \mu} \Phi \is \gamma^{\Phi} := 
(2 -d - \dot{Z}) \Phi - \dot{\Psi} \,,
\nonum
\mu \frac{d}{d \mu} F  \is \gamma^F := (2-d - \dot{Z}) F - \dot{Y}\,.
\label{Gflow} 
\ea 
The associated renormalization group operator is 
\be 
\cD = \mu \frac{\dd}{\dd \mu} + \beta \cdot  \frac{\dd}{\dd g} 
+  \gamma^V_{\mu}  \cdot  \frac{\dd}{\dd V_{\mu}}  
+  \gamma^{\Phi}  \!\cdot  \frac{\dd}{\dd \Phi} 
+  \gamma^F  \!\cdot  \frac{\dd}{\dd F}\,.
\label{RGoperator}
\ee
For example the dimension 0 composite operators in (\ref{normalproducts}) 
obey 
\be 
\cD \nl V(\phi)\nr = \nl (d - 2 + \dot{Z} + \cD) V\nr \,,
\label{Dnormalproduct}
\ee
and similar equations hold for the dimension 1,2 composite operators.

An important application of this framework is the determination 
of the Weyl anomaly as an ultraviolet finite composite operator.
We shall only need the version without vector and scalar functionals.
The result then reads  \cite{Shore,Tseytlin87,Osb87}
\be
\gamhat^{\mu\nu}\nl T_{\mu\nu} \nr = 
\frac{1}{2} \nl B_{ij}(g/\lb) \gamhat^{\mu\nu} 
\dd_{\mu} \phi^i \dd_{\nu} \phi^j \nr + \frac{1}{2} R^{(2)}(\gamhat) 
\nl B^{\Phi} \nr \,.
\label{anomaly1}
\ee
Here the so-called Weyl anomaly coefficients enter:
\ba
&& \lb B_{ij}(g/\lb) := \lb \beta_{ij}(g/\lb)\big|_{d=2} + 
\cL_S g_{ij}\,,\nonum
&& \lb B^{\Phi}(\Phi,g/\lb) := \lb \gamma^{\Phi}(g/\lb)\big|_{d =2} 
+ S^j \dd_j \Phi\,, 
\label{Weyl1} 
\ea
where $\beta_{ij}$ and $\gamma^{\Phi}$ are the renormalization group 
functions of Eq.~(\ref{Gflow}) and 
\ba
S_i \!&:=&\! W_i + \dd_i \Phi \quad \mbox{with}
\nonum
W_i \!&:=&\! N^{(1)}(g)_i^{\;jk} g_{jk} = \Big(\frac{\lb}{2\pi}\Big)^3 
\frac{1}{32} \dd_i (R_{klmn}R^{klmn}) + O(\lb^4)\;.
\label{Svector} 
\ea
These expressions hold in dimensional regularization, minimal 
subtraction, and the backgound field expansion in terms of normal
coordinates. Terms proportional to the equations of motion 
operator $\frac{\delta S_{\b}}{\delta \phi^j}$ have been omitted. 
The normal-products (\ref{normalproducts}) are normalized such 
that the expectation value of an operator contains as its leading 
term the value of the 
corresponding functional on the background, $\bra {\cal O}(\phi) \ket =
{\cal O}(\varphi) + \ldots$, where the subleading terms are in general 
nonlocal and depend on the scale $\mu$. For the expectation value of the 
trace anomaly this produces a rather cumbersome expression, see 
e.g.~\cite{Tseytlin87}. As stressed in \cite{Shore} the result 
(\ref{anomaly1}), in contrast, allows one to use $B_{ij}(g) =0$ 
as a simple criterion to select functionals which `minimize' the 
conformal anomaly.

The Weyl anomaly coefficients (and the anomaly itself) can be shown 
to be invariant under field redefinitions of the form  
\be
\phi^j_{\b} \rra \phi_{\b}^j + \frac{1}{2-d} V^j(\phi,\lb) \;,\quad 
\label{diff10}
\ee
with $V^j(\phi,\lb) = \sum_{l \geq 1} (\frac{\lb}{2\pi})^l V^j_l(\phi)$ 
functionally {\it independent} of the metric. 
Roughly speaking (\ref{diff10}) changes the beta function by a
Lie derivative term that is compensated by a contribution of the 
diffeomorphism current to the anomaly which amounts to 
$W_j \ra W_j - V_j$ \cite{Shore}. It is important to 
distinguish these diffeomorphisms from field renormalizations
like (\ref{field_flow1}) that depend on the metric. 
Although formally (\ref{diff10}) amounts to 
$\Xi^j(\phi,\lb) \rra \Xi^j(\phi,\lb) + V^j(\phi,\lb)$ in (\ref{field_flow1}),
clearly one cannot cancel one against the other. The distinction 
is also highlighted by considering the change in the 
metric counter terms 
\be
T^{(1)}_{ij}(g) \rra T^{(1)}_{ij}(g) - \cL_V g_{ij}\;,
\label{diff11}
\ee
under (\ref{diff10}). Without further specifications this would not 
be legitimate for a $g$-dependent vector. Although the Lie derivative 
term in (\ref{diff11}) drops out when recomputing $\beta_{ij}$ 
directly as a parametric derivative, in combinations like 
\be
\beta_{ij}(\phi_{\b}) \dd^{\mu} \phi_{\b}^i \dd_{\mu} \phi_{\b}^j = 
\beta_{ij}(\phi) \dd^{\mu} \phi^i \dd_{\mu} \phi^j + 
\frac{1}{2-d} \cL_V \beta_{ij}(\phi) \dd^{\mu} \phi^i \dd_{\mu} \phi^j 
+ \ldots \,,
\ee 
the term $(2 -d) g_{ij}$ in the metric beta function of (\ref{Gflow}) 
induces an effective shift 
\be
\beta_{ij}(g) \rra \beta_{ij}(g) + \cL_V g_{ij}\,.
\label{betadiff}
\ee
Similarly $W_i$ is shifted to $W_i - V_i$ and the Weyl 
anomaly coefficients are invariant. 

In the context of Riemannian sigma-models $\Phi$ is usually 
interpreted as a ``string dilaton'' for the systems (\ref{Lsource}) 
defined on a curved base space. If one is interested in the 
renormalization of (\ref{Lsource}) on a flat base space, $\Phi$ 
on the other hand plays the role of a potential for the improvement 
term of the energy momentum tensor. This role of $\Phi$ can be made 
manifest by rewriting (\ref{anomaly1}) by means 
of the diffeomorphism Ward identity. Returning to a flat base 
space one finds \cite{Shore,Osb87} 
\be
\nl T^{\mu}_{\;\mu} \nr  = \frac{1}{2} 
\nl \beta_{ij}(g/\lb) \dd^{\mu} \phi^i \dd_{\mu} \phi^j \nr
+ \dd^{\mu} \dd_{\mu} \nl \Phi \nr + 
\dd^{\mu} \nl \dd_{\mu} \phi^i W_i\nr \,,
\label{anomaly5}
\ee
where again terms proportional to the equations of motion 
operator have been omitted. Here $\dd^2\nl \Phi \nr$  is the `naive' 
improvement term while the additional total divergence is induced by 
operator mixing.

The functions $B^{\Phi}$ and $B_{ij}(g)$ are linked by an important 
consistency condition, the Curci-Paffuti relation \cite{CurciPaff}.
We present it in two alternative versions
\begin{subeqnarray}
\dd_i \dot{\Psi} \is \dot{N}_i^{\;jk} \,\dot{T}_{jk} - 
\dot{T}\cdot \frac{\dd}{\dd g} W_i + (\dot{Z}^V)_i^j W_j\,,
\\
\dd_i B^{\Phi} \is \dot{N}_i^{\;jk} \,B_{jk} - 
B \cdot \frac{\dd}{\dd g} S_i + B_{ij} S^j\,.
\label{CurciPaff}
\end{subeqnarray} 
The first version displays the fact that the identity relates  
various $g$-dependent counter terms without $\Phi$
entering. In the second version $\Phi$ is introduced    
in a way that yields an identity among the Weyl anomaly 
coefficients. It has the well-known consequence that $B^{\Phi}$ 
is constant when $B_{ij}$ vanishes:
\be
B_{ij} =0 \quad \Longrightarrow \quad 
B^{\Phi} = c_T/6 \;,
\label{Weyl3}
\ee 
where $c_T$ is the central charge of energy momentum tensor derived 
from (\ref{Lsource}).

\newpage
\newappendix{Strict renormalizability yields no fixed point}

Although the fixed point $h^{\rm beta}$ was constructed in a 
loopwise expansion it can be regarded as ``non-Gaussian'' in the 
following sense: 
There exists a function $h^{\rm ren}(\,\cdot\,,\lb)$ for which 
$h^{\rm ren}_{\b}(\,\cdot\,,\lb) = h^{\rm ren}(\,\cdot\,,\lb)$, so that 
one almost recovers conventional renormalizability.%
\footnote{I thank P.~Forg\'acs for pointing this out. 
A similar concept of recovering renormalizability by finite 
quantum deformations was recently employed in the context of 
T-duality \cite{Bonneau01}; see also \cite{Bonneau83b}.}  
However beyond one loop $h^{\rm ren}$ {\it differs} from  $h^{\rm beta}$,
in particular $\beta_h(h^{\rm ren}) \neq 0$. Thus if one was to 
explore the vicinity of this conventionally renormalizable theory 
the fixed point $h^{\rm beta}$ could not be seen and in view of 
Eq.~(\ref{zerotrace}) their was little chance to impose the operator 
constraints.

In order to find $h^{\rm ren}$ we return to the general solution 
of the finiteness condition (\ref{Hlbfinite}) with $h(\rho,\lb)$ 
of the form (\ref{hlb}). On can then ask whether there exist special 
choices for the functions 
$h_l(\rho),\,l \geq 0$, such that $H(\rho,\lb)$ is $\rho$-independent. 
This requirement translates into a system of second 
order differential equations that can be recursively solved 
for $h_0,h_1,\,$etc.  
\ba
&& H(\rho,\lb) = - Z(\lb) = 
-\sum_{l \geq 1} z_l \Big(\frac{\lb}{2\pi} \Big)^l
\quad \Longleftrightarrow 
\nonum
&& \frac{h (\rho \dd_{\rho})^2 h}{(\rho \dd_{\rho} h)^2}
= 1 + \frac{B_{\lb}(\lb/h) - \beta_{\lb}(\lb/h)h/\lb}%
{B_{\lb}(\lb/h) + Z(\lb)}\,.
\label{defren1}
\ea
In this situation one almost recovers conventional renormalizability: 
The bare and the renormalized metric in (\ref{RBlbmetric}) are 
related by a {\it numerical} though singular prefactor, which can be 
attributed in the usual way to a renormalization of the coupling
\be
\lb_{\b} = \mu^{2 -d} \lb \left[1+ \frac{1}{2 -d} Z(\lb) + 
\ldots \right]\,.
\label{defren2}
\ee
Combined with the (still in general nonlinear) field redefinitions
all counter terms can be absorbed. The price to pay is that the 
`good' target space geometry in which the bare and the renormalized 
Lagrangian have the same functional form is not known a-priori. 
The functions $h_1, \,h_2,$ etc, constitute {\it finite quantum 
deformations} of the naive target space metric. Judiciously chosen they 
ensure conventional renormalizability, but the proper choice
has to be determined order by order by solving (\ref{defren1}).
At the one-loop level one finds that the only solutions of 
$H_1[h_0] = - z_1$ are again powers of $\rho$, with the power 
independent of $z_1$. A non-zero $z_1$ can arise either for 
the constant solution, $h_0 = - \zeta_1/z_1$, or when a finite 
integration constant in (\ref{ren_sol1}) is permitted. Then 
$H_1[\rho^p] = \zeta_1 \rho_1^{-p}$, and with the choice 
$\rho_1 = \infty$ advocated before only the solution 
$\rho^p,\,p>0$, (with $z_1=0$) are left. For $l\geq 1$ the solutions 
for $h_l$ will have a nontrivial dependence on $z_1,\ldots , z_l$.
However they can not be chosen so as to make $h_l$ 
vanish, i.e.~the required quantum deformations are always 
non-trivial. Since on the other hand the structure of the 
solutions considerably simplifies if the $z_l$'s are chosen to vanish, 
we consider only this case in the following. The solutions
for $h_l$ will then still depend on $l$ arbitrary deformation parameters;
switching on the $z_l$'s simply enlarges the number of parameters at 
the expense of the cumbersome coupling renormalization (\ref{defren2}).

The case $H(\rho,\lb)=0$ has the bonus that the 
field renormalization vectors $\Xi^j(\rho,\lb)$  are gradients of 
a potential: 
\ba
&&  H(\rho,\lb) =0 \quad \Longleftrightarrow  \quad 
h(\rho,\lb) \Xi^3(\rho,\lb) = \rho \frac{\lb}{2\pi}C(\lb)^{-1} 
\quad \Longleftrightarrow
\nonum
&& \Xi^j = - \dd^j \Phi\,,\quad \mbox{with} \quad
\Phi = \sum_{l \geq 1} \Big(\frac{\lb}{2\pi} \Big)^l \Phi_l\,, 
\label{XiPhi}
\ea 
for some $C(\lb) = \sum_{l \geq 0} C_l (\frac{\lb}{2\pi})^l$ with
constant $C_l$. The potential is given by 
\be
\Phi^{\rm ren}(\rho,\sigma,\lb) = 
-\frac{\lb}{4\pi} C(\lb)^{-1}(a \ln \rho + 2 b \sigma) - 
\frac{1}{2} \int^{\rho} \frac{du}{u} h(u,\lb) \int^u 
\frac{dv}{v} S(v,\lb)\,.
\label{Philb}
\ee
The ambiguity stemming from the various integration constants 
is again of the form (\ref{Phiambiguity}).

The condition (\ref{XiPhi}) converts into the first order 
differential equation
\be
\frac{\lb}{2\pi} \rho \dd_{\rho} h(\rho,\lb) = 
C(\lb) h(\rho,\lb)^2 B_{\lb}
\left(\frac{\lb}{h(\rho,\lb)}\right)\,,
\label{Hode}
\ee 
from which upon insertion of (\ref{hlb}) the functions 
$h_l,\,l \geq 0$, are determined recursively. We denote the
solutions by $h^{\rm ren}_l$. There exists a minimal solution 
corresponding to a $\lb$-independent $C(\lb) = C_0$. It comes out as 
\be 
h^{\rm ren}(\rho,\lb) = 
\rho^{p} - \frac{\lb}{2\pi} \frac{\zeta_2}{\zeta_1} 
- \Big(\frac{\lb}{2\pi}\Big)^2 \frac{\zeta_3}{2 \zeta_1} \rho^{-p}
+ \ldots\,,
\label{Hminsol}
\ee
where $C_0 = p/\zeta_1$. Switching on the parameters $C_1,C_2, \ldots$,
leads to a deformation of the functions $h_1,h_2, \ldots$,
similar to the one in (\ref{hbeta_sol2}). Indeed, in view of 
(\ref{Blb}), the solutions of (\ref{Hode}) must be related to those
of (\ref{betaode}) by substituting $\zeta_l \ra \zeta_l/l$.

Let us now compare $h^{\rm ren}$ with $h^{\rm beta}$ defined through 
the vanishing of the $\beta_h(h)$ functional. Already for the minimal 
solution one sees from (\ref{betaminsol}) that  
\be
h^{\rm ren}_l(\rho) \neq h^{\rm beta}_l(\rho)\,,\sspace l \geq 2\,.
\label{ren_vs_beta}  
\ee
Note that the disagreement starts at the two-loop level where the 
coefficients $\zeta_1,\zeta_2$ are still universal. In particular 
\be
\lb \beta_h(h^{\rm ren}/\lb) = \Big(\frac{\lb}{2\pi}\Big)^2 
\frac{\zeta_2}{2 \rho^p} + O(\lb^3) \,,
\label{ren_betah}
\ee  
cannot be made to vanish by a change of scheme. In section 6 we 
related the $h_l^{\rm beta}$ class to a physical requirement
-- the vanishing of the trace anomaly -- which is a necessary condition 
for conformal invariance. The mismatch (\ref{ren_vs_beta}) therefore 
implies that one cannot have both desirable properties, 
conformal invariance and strict renormalizability, at the same time. 
Since this is an important conclusion, we made sure that
it is not an artifact of some inessential assumption. For example 
at first sight it seems that by using the more general general target space 
metric (\ref{Tabmetric}) (where the residual freedom in choosing 
adapted coordinates has not been used to simplify the functions 
$a(\rho), \,b(\rho)$) the additional freedom to adjust 
$a(\rho)$ and $b(\rho)$ as functions of $\lb$ could be used 
to compensate a mismatch $h_l^{\rm ren}(\rho)\neq 
h^{\rm beta}_l(\rho)$. However one can check that this is 
not the case by repeating the entire construction in this more general 
setting. 

It may be useful to summarize and juxtapose the key relations 
defining the ``renormalizable'' $h$ and $\Phi$ and that satisfying 
the ``beta'' condition. As alternative defining relations one 
may take 
\begin{subeqnarray} 
\cL_{\Xi} g_{ij} &=& - 2 \nabla_i \nabla_j \Phi
\sspace \;\mbox{for ``ren''}  
\\[2mm]
\cL_{\dot{\Xi} -W} g_{ij} &=& 2 \nabla_i \nabla_j \Phi 
\sspace\, \quad \mbox{for ``beta''}  
\label{summary1}
\end{subeqnarray}
In both cases the condition (\ref{summary1}) has an unexpected 
spin-off: 
\begin{subeqnarray} 
H(\rho,\lb) &=& 0\,,\quad \mbox{i.e.}\quad \quad 
\frac{2\pi}{\lb} \frac{\Xi^3}{\rho} =\frac{1}{C(\lb) h}\,, 
\sspace \,\mbox{for ``ren''}  
\\[2mm]
\beta_h(h) &=& 0\,,\quad \mbox{i.e.}\quad  
- \frac{2\pi}{\lb} \frac{\dot{\Xi}^3}{\rho} = \frac{1}{C(\lb) h}\,, 
\sspace \mbox{for ``beta''}  
\label{summary2}
\end{subeqnarray}
Combining (\ref{ddPhi_ward}) with (\ref{summary1}a) and 
(\ref{summary2}a) the finiteness condition (\ref{Hlbfinite}) can 
be rewritten as $2 \dd^{\mu} \dd_{\mu} \Phi 
= \lb T^{(1)}_{ij}(g/\lb) \dd^{\mu} \phi^i \dd_{\mu} \phi^j$, 
modulo terms proportional to the equations of motion operator.
For $h=h^{\rm ren}$ the counter terms for the Lagrangian can 
therefore to all loop orders be written as $\dd^2 \Phi$. At 1-loop 
this also entails the vanishing of the trace anomaly in agreement 
with the criterion in \cite{HullTown86}. At higher loops however this 
equivalence breaks down. 

In summary, the class of $h_l$'s that ensure (almost) conventional 
renormalizability 
is {\em not} the same as the one that ensures the vanishing of the 
$\beta_h(h)$ function. Likewise the improvement potential 
$\Phi^{\rm trace}= \Phi^{\rm beta}$ for the energy momentum tensor 
at the fixed point differs from $\Phi^{\rm ren}$.

\newpage 


\end{document}